\documentclass[twocolumn]{aastex63}
\usepackage{times}
\usepackage{amsmath}
\usepackage{textcomp}
\usepackage{amssymb}
\usepackage{natbib}
\usepackage{float}
\usepackage{color}
\usepackage{graphicx}
\usepackage{epstopdf}
\newcommand{\msun}{\ensuremath{M_{\odot}}}
\newcommand{\lum}{erg\,s$^{-1}$}
\newcommand{\fermi}{{\it Fermi}}

\newcommand{\swift}{{\it Swift}}

\newcommand{\ergflux}{\mbox{${\rm \, erg \,\, cm^{-2} \, s^{-1}}$}}
\newcommand{\gm}{$\gamma$}
\newcommand{\mbh}{$M_{\rm BH}$}
\newcommand{\ld}{$L_{\rm disk}$}
\newcommand{\km}{\mbox{${\rm \, km \,\,  s^{-1}}$}}
\newcommand{\ledd}{$L_{\rm Edd}$}

\newcommand{\hbeta}{H{$\beta$}}
\newcommand{\halpha}{H{$\alpha$}}
\newcommand{\FeII}{Fe{\sevenrm II}}
\newcommand{\CIV}{C{\sevenrm IV}}

\newcommand{\MgII}{Mg{\sevenrm II}}
\newcommand{\OIII}{[O{\sevenrm\,III}]}

\newcommand{\NII}{[N{\sevenrm\,II}]}

\newcommand{\msinterr} {\ensuremath{8.12 \pm 0.08}}
\newcommand{\msslopeerr} {\ensuremath{4.24 \pm 0.41}}

 \font\sevenrm=cmr7 scaled 1000

\received{\today}
\revised{\today}
\accepted{\today}
\submitjournal{ApJS}

\shorttitle{Blazar's Central Engine}
\shortauthors{Paliya et al.}

\begin{document}

\title{The Central Engines of \textbf{\textit {Fermi}} Blazars}

\correspondingauthor{Vaidehi S. Paliya}
\email{vaidehi.s.paliya@gmail.com}

\author[0000-0001-7774-5308]{Vaidehi S. Paliya}
\affiliation{Aryabhatta Research Institute of Observational Sciences (ARIES), Manora Peak, Nainital 263001, India}
\affiliation{Deutsches Elektronen Synchrotron DESY, Platanenallee 6, 15738 Zeuthen, Germany}

\author[0000-0002-3433-4610]{A. Dom{\'{\i}}nguez}
\affiliation{Institute of Particle and Cosmos Physics (IPARCOS), Universidad Complutense de Madrid, E-28040 Madrid, Spain}
\affiliation{Department of EMFTEL, Universidad Complutense de Madrid, E-28040 Madrid, Spain}

\author[0000-0002-6584-1703]{M. Ajello}
\affiliation{Department of Physics and Astronomy, Clemson University, Kinard Lab of Physics, Clemson, SC 29634-0978, USA}

\author[0000-0001-7859-699X]{A. Olmo-Garc\'{i}a}
\affiliation{Departamento de F\'{i}sica de la Tierra y Astrof\'{i}sica, Universidad Complutense de Madrid (UCM, Spain) and Instituto de F\'{i}sica de Part\'{i}culas y del Cosmos (IPARCOS)}

\author[0000-0002-8028-0991]{D. Hartmann}
\affiliation{Department of Physics and Astronomy, Clemson University, Kinard Lab of Physics, Clemson, SC 29634-0978, USA}

\begin{abstract}
We present a catalog of central engine properties, i.e., black hole mass (\mbh) and accretion luminosity (\ld), for a sample of 1077 blazars detected with the \fermi~Large Area Telescope. This includes broad emission line systems and blazars whose optical spectra lack emission lines but dominated by the absorption features arising from the host galaxy. The average \mbh~for the sample is $\langle \log~M_{{\rm BH,all}~\msun} \rangle=8.60$ and there are evidences suggesting the association of more massive black holes with absorption line systems. 
Our results indicate a bi-modality of \ld~in Eddington units (\ld/\ledd) with broad line objects tend to have a higher accretion rate (\ld/\ledd$>$0.01). 
We have found that \ld/\ledd~and Compton dominance (CD, the ratio of the inverse Compton to synchrotron peak luminosities) are positively correlated at $>$5$\sigma$ confidence level, suggesting that the latter can be used to reveal the state of accretion in blazars. Based on this result, we propose a CD based classification scheme. Sources with CD$>$1 can be classified as High-Compton Dominated or HCD blazars, whereas, that with CD$\lesssim$1 are Low-Compton Dominated (LCD) objects. This scheme is analogous to that based on the mass accretion rate proposed in the literature, however, it overcomes the limitation imposed by the difficulty in measuring \ld~and \mbh~for objects with quasi-featureless spectra. We conclude that the overall physical properties of \fermi~blazars are likely to be controlled by the accretion rate in Eddington units. The catalog is made public at \url{http://www.ucm.es/blazars/engines} and Zenodo.
\end{abstract}


\section{Introduction}{\label{sec:Intro}}

Blazars are a subclass of jetted active galactic nuclei (AGN) family that host relativistic jets closely aligned to the line of sight to the observer. Due to their peculiar orientation, the emitted radiation is relativistically amplified thereby making blazars observable at cosmological distances ($z>5$). Typically, blazars are hosted by elliptical galaxies and powered by massive (10$^{\sim 8-10}$ \msun) black holes \citep[e.g.,][]{2000ApJ...532..816U,2012ApJ...748...49S}. The multi-wavelength spectral energy distribution (SED) of a blazar exhibits a typical double hump structure. The low-energy bump peaks at radio-to-X-rays and is well explained with the synchrotron mechanism. On the other hand, the inverse Compton process is found to satisfactorily reproduce the high-energy bump located in the MeV-TeV enetgy range, considering the leptonic radiative models \citep[e.g.,][]{2009ApJ...692...32D,2019ApJ...874...47V}. The properties of the central engine, i.e., black hole mass (\mbh) and accretion disk luminosity (\ld), are found to significantly correlate with the broadband SED of blazars \citep[][]{2014Natur.515..376G,2017ApJ...844...32P,2019ApJ...881..154P,2020ApJ...897..177P}, indicating a close-connection between the accretion process and relativistic jets.

Historically, the beamed AGN have been classified based on the appearance of the broad emission lines in their optical spectra. Blazars exhibiting strong and broad emission lines (rest-frame equivalent width of EW $>$5 \AA) are known as flat spectrum radio quasars or FSRQs. The objects showing quasi-featureless spectra (EW $<$5 \AA), on the other hand, belong to the category of BL Lacertae or BL Lac sources \citep[][]{1991ApJ...374..431S}. It was predicted that the lack of broad emission lines in the optical spectra of the latter might be due to Doppler boosted continuum swamping out any spectral lines even if they exist. However, stellar absorption features originated from the host galaxy are observed in many nearby ($z<1$) BL Lac sources, thus indicating that emission lines are intrinsically weak \citep[][]{2011MNRAS.413..805P}. To make the case more complex, broad emission lines have been detected in the low jet activity states of many BL Lac objects \citep[e.g.,][]{1995ApJ...452L...5V}. Therefore, a blazar classification scheme based on the EW does not reveal the physical distinction between FSRQs and BL Lac sources.

\citet[][]{2011MNRAS.414.2674G} proposed that the criterion to distinguish FSRQs and BL Lac objects should be based on the broad line region (BLR) luminosity ($L_{\rm BLR}$). FSRQs are sources with a luminous BLR ($L_{\rm BLR}\gtrsim10^{-3}$ times Eddington luminosity or \ledd) which, in turn, suggests a radiatively efficient accretion process ($L_{\rm disk}~\gtrsim1\%L_{\rm Edd}$). BL Lac sources, on the other hand, host a radiatively inefficient accretion flow which fails to photo-ionize the BLR clouds \citep[see also,][]{2014MNRAS.445...81S}. The presence of broad and strong emission lines in the optical spectra of FSRQs and absence in BL Lacs, thus, can be explained in this more physically intuitive classification. However, one of the key parameters in this scheme, \ledd, requires the knowledge of the mass of the central black hole. Moreover, an estimation of \ld, or equivalently mass accretion rate, remains a challenge for BL Lac objects.

In the context of the \gm-ray emitting beamed AGN, the research works focused on the central engine properties have so far remained concentrated on the broad emission lines blazars \citep[cf.][]{2014Natur.515..376G}. Apart from a few individual works \citep[e.g.,][]{2020MNRAS.494.6036B}, a population study on the central engine of BL Lac sources is still lacking. In this work, we have attempted to address this outstanding issue by carrying out a detailed optical spectroscopic analysis of a large sample of blazars present in the \fermi~Large Area Telescope (LAT) fourth source catalog data release 2 \citep[4FGL-DR2;][]{2020ApJS..247...33A}. Furthermore, we have also determined that Compton dominance (CD, the ratio of the inverse Compton to synchrotron peak luminosities) can be considered as a good proxy for the accretion rate in Eddington units. This parameter can be crucial for objects whose optical spectroscopic analyses are tedious either due to telescope constraints, intrinsic source faintness, or lack of emission lines in their optical spectrum. As discussed above, the latter could be due to a radiatively inefficient accretion process and/or Doppler boosted jet radiation swamping out emission lines and thus does not allows us to infer the intrinsic nature of the quasar. The CD, therefore, could be the key to connect the FSRQ-BL Lac dichotomy and a classification scheme based on this parameter may fully explain the diverse physical properties of \fermi~blazars. 

In Section~\ref{sec:sample}, we describe the \fermi~blazar sample and the details of the optical spectroscopic data analysis procedures are provided in Section~\ref{sec:spec_analysis}. We elaborate various techniques to determine \mbh~and \ld~in Section~\ref{sec:results} and discuss them in Section~\ref{sec:central_engine}. Section~\ref{sec:discussion} and \ref{sec:sequence} are devoted to the physical interpretation of the estimated central engine parameters and CD. We summarize our findings in Section~\ref{sec:summary}. Throughout, we adopt the flat cosmology parameters: $h=\Omega_\Lambda = 0.7$.

\section{Sample}\label{sec:sample}

\begin{table}[t!]
\caption{The \fermi~blazar sample studied in this work.\label{tab:sample}}
\begin{center}
\begin{tabular}{lll}
\hline
\hline
Sample Classification & Number of Sources \\
\hline
Emission line blazars     & 258 (SDSS)   \\
                                        & 416 (literature) \\
                                        & 674 (all) \\
\hline
Absorption line blazars  & 200 (SDSS) \\
                                        & 146 (literature) \\
                                        & 346 (all) \\
\hline
Blazars with bulge magnitude from literature & 47 \\
Blazars with \mbh~and \ld~from literature & 10\\
\hline
\end{tabular}
\end{center}
\end{table}

We have considered all \gm-ray emitting blazars or blazar candidates present in the 4FGL-DR2 catalog to prepare the base sample of 5250 objects. Since the main goal of this work is to derive \mbh~and \ld~from the optical spectrum, we carried out an extensive search for the optical spectroscopic information for as many blazars as possible. This was done by: (i) cross-matching the 4FGL-DR2 catalog with the 16th data release of the Sloan Digital Sky Survey \citep[SDSS-DR16;][]{2020ApJS..249....3A}, (ii) searching the published optical spectrum of all remaining blazars in the literature using NASA Extragalactic Database and SIMBAD Astronomical Database, and (iii) searching the published \mbh~and \ld~values in the literature for objects leftover after completing steps (i) and (ii). In all of the cases, we used the positional information of the low-frequency counterpart of the \gm-ray source given in the 4FGL-DR2 catalog or in other works and visually examine the quality of the spectra. This exercise led to a collection of 458 SDSS spectra including 258 emission line objects and 200 blazars showing prominent absorption lines. From research articles, we were able to collect optical spectra of 562 blazars either in tabular format or simply as plots which were digitized using {\tt WebPlotDigitizer}\footnote{\url{https://automeris.io/WebPlotDigitizer/}} \citep[][]{Rohatgi2020}. Among them, 416 objects are broad emission line systems and the optical spectra of 146 are dominated by absorption lines arising from the host galaxy. Furthermore, we were able to find host galaxy bulge magnitude for 47 sources and \mbh~and/or \ld~values for 10 blazars from the literature. Altogether, our final sample consists of 1020 objects with available optical spectra and 57 others with bulge magnitude or \mbh~and \ld~directly taken from the literature. Note that emission line blazars mainly belong to FSRQ class of AGN, whereas BL Lacs dominate the absorption line sample. However, since many of the BL Lac objects have exhibited broad emission lines in their optical spectra taken during a low-jet activity state, i.e., akin to FSRQs, we divide the whole sample in emission and absorption line systems, respectively, rather than distinguishing them as FSRQ/BL Lacs. Table~\ref{tab:sample} summarizes our blazar sample.

\begin{figure}[t!]
\hbox{
\includegraphics[scale=0.5]{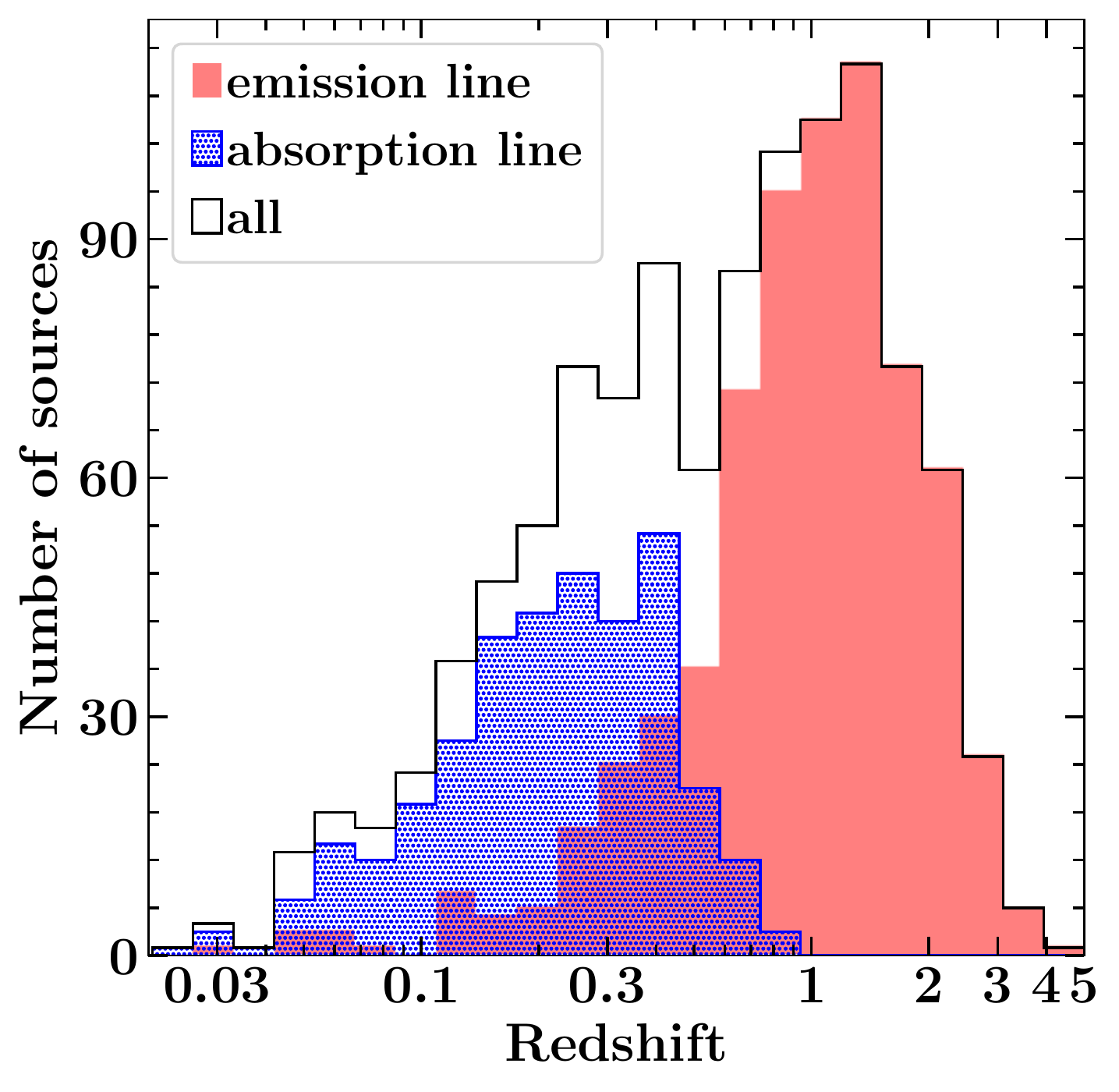} 
}
\caption{The redshift distribution of \fermi~blazars (black) present in our sample. Red solid and blue dotted histograms refer to objects whose optical spectra are dominated by broad emission lines and host galaxy absorption features, respectively.}
\label{fig:redshift}
 \end{figure}

\begin{figure*}[t!]
\hbox{
\includegraphics[scale=0.45]{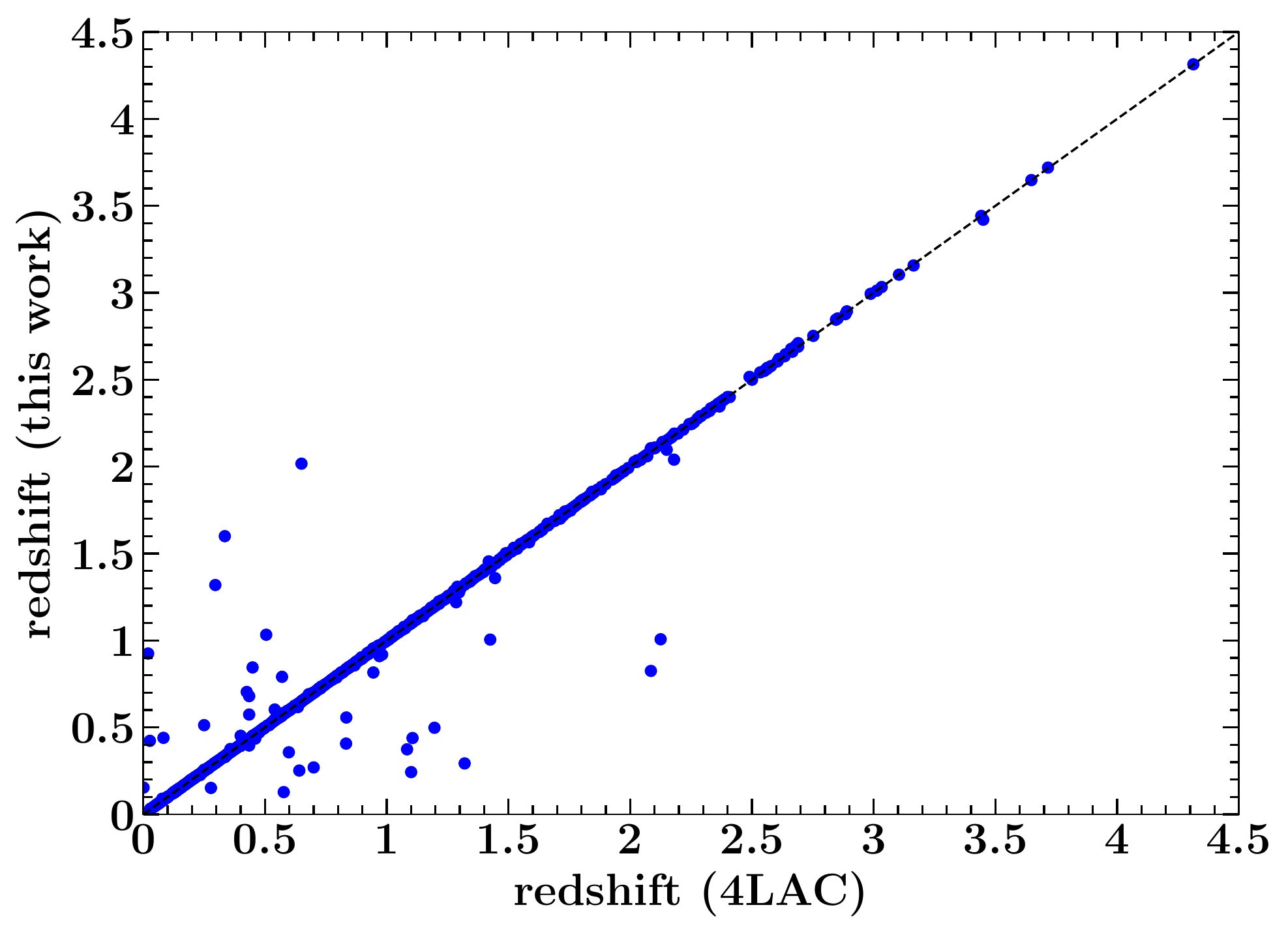} 
\includegraphics[scale=0.45]{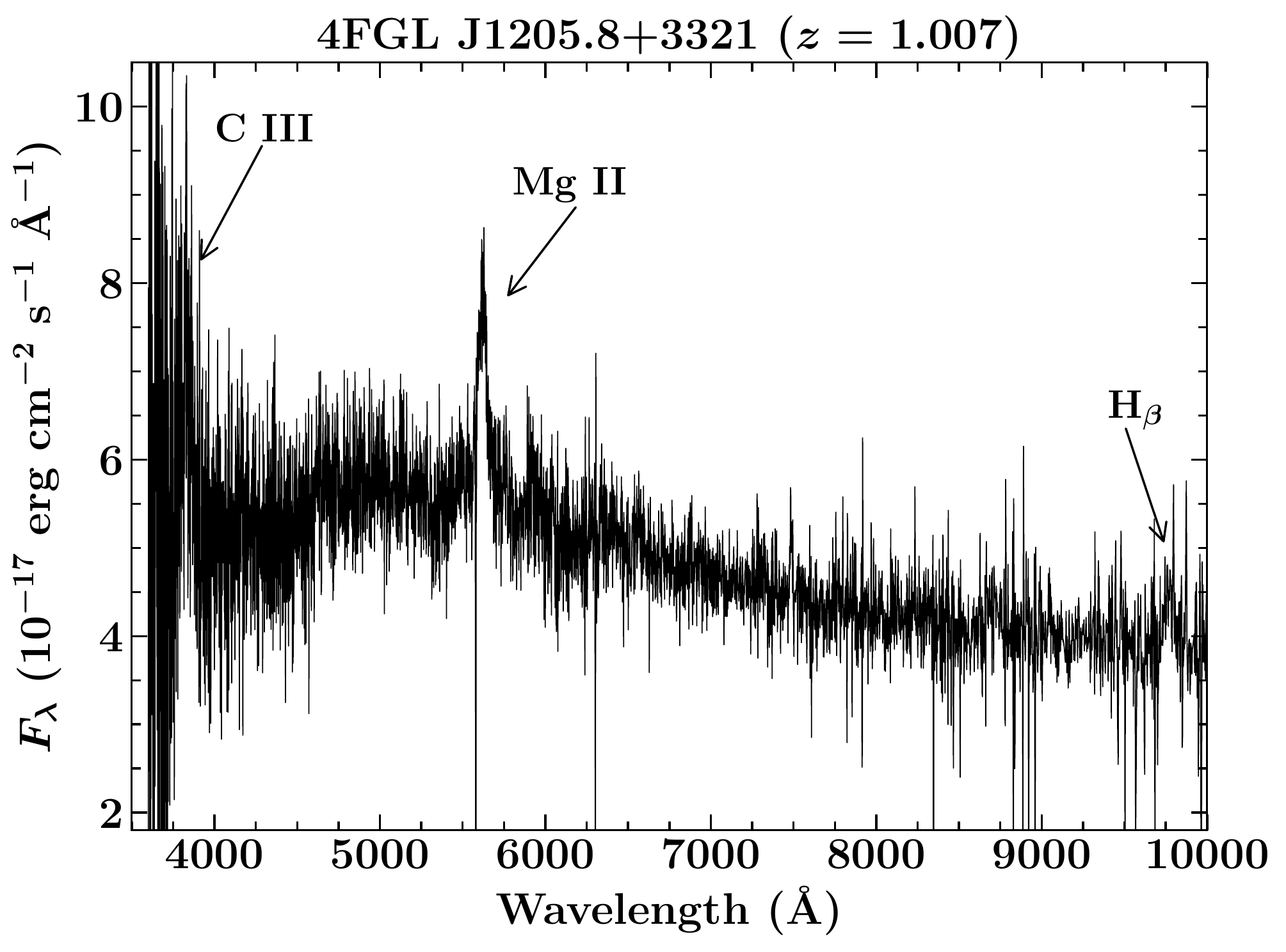} 
}
\caption{Left: A comparison of the redshift values reported in the 4LAC catalog and this work. The dashed line denotes the equality of the plotted quantities. Right: The SDSS-DR16 spectrum of the \gm-ray source 4FGL J1205.8+3321 whose redshift reported in the 4LAC catalog is $z=2.085$. However, based on the detection of various broad emission lines, its correct redshift is $z=1.007$.}
\label{fig:redshift_2}
 \end{figure*}

The redshift distribution of the sample is shown in Figure~\ref{fig:redshift}. Instead of directly using the redshift information from the fourth catalog of the \fermi-LAT detected AGN \citep[4LAC;][]{2020ApJ...892..105A}, we have visually inspected the optical spectrum of all of the sources to determine the accurate source redshift. Though most of our measurements agree with redshifts published in the 4LAC catalog, a small fraction ($<5\%$) of sources do show differences (Figure~\ref{fig:redshift_2}, left panel). This is expected since 4LAC itself is a compilation of information taken from various databases. In the right panel of Figure~\ref{fig:redshift_2}, we show the optical spectrum of the blazar 4FGL J1205.8+3321 as an illustration. The 4LAC catalog reports the redshift of this object as $z=2.085$ which is likely to be photometric in nature  and probably adopted from \citet[][]{2009ApJS..180...67R}. On the other hand, the SDSS-DR16 spectrum clearly shows many broad emission lines securing the spectroscopic redshift as $z=1.007$.

We acquired optical spectra of a few ($<$50) sources from the 6-degree Field Galaxy Survey \citep[][]{2009MNRAS.399..683J}. Since these spectra are not flux calibrated, we adopted the wavelength dependent conversion factor reported in \citet[][]{2018A&A...615A.167C} to convert counts spectra in energy flux units.

\begin{figure*}[t!]
\hbox{
\includegraphics[scale=0.27]{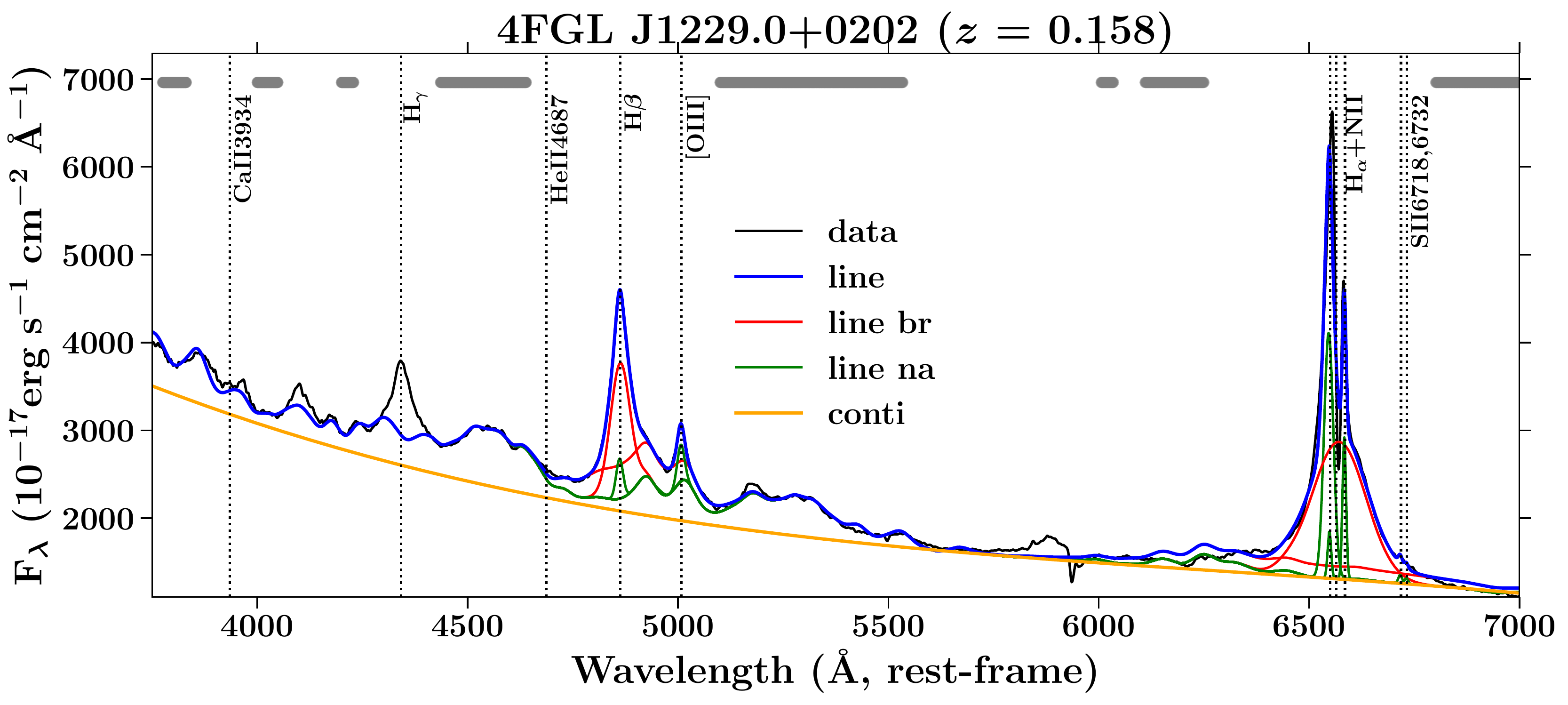} 
\includegraphics[scale=0.27]{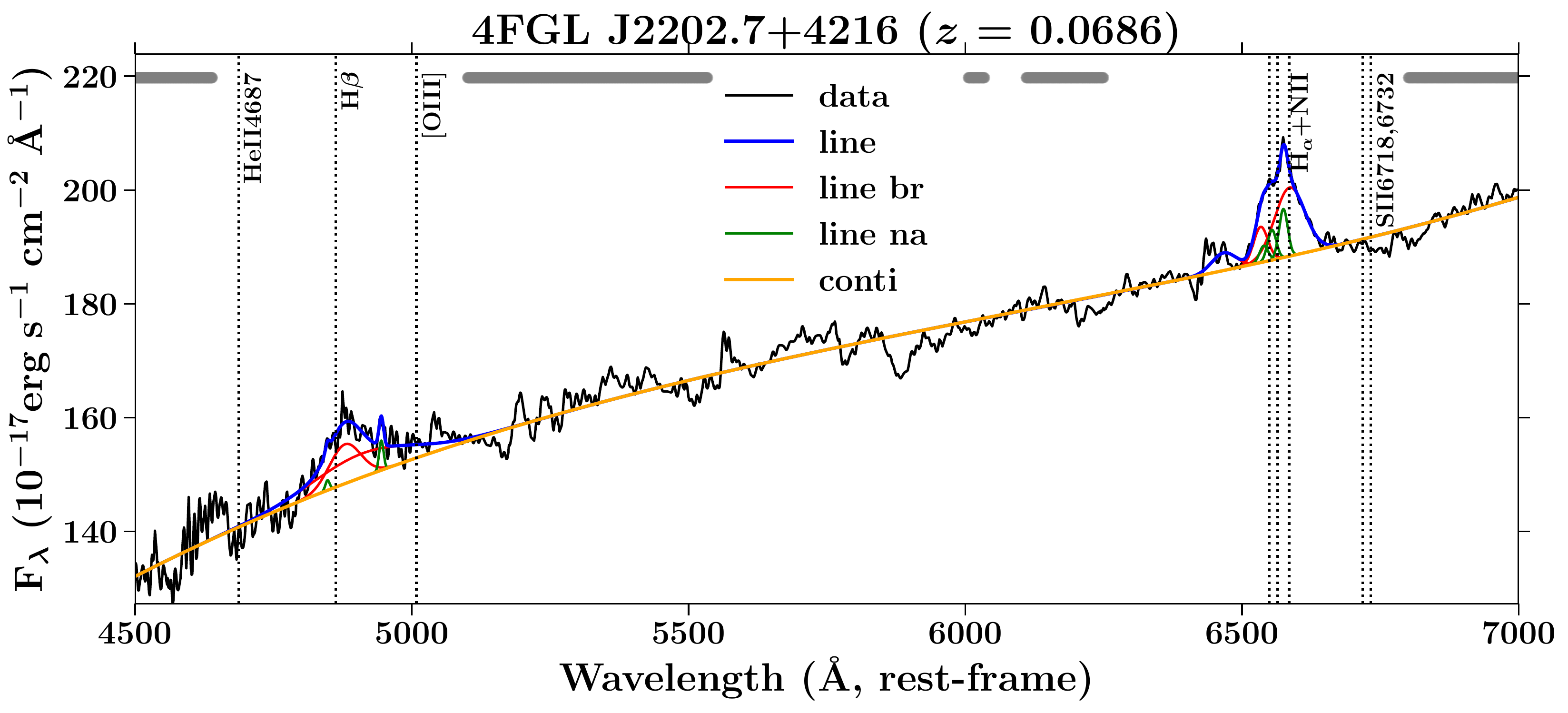} 
}
\caption{The optical spectra of prototype blazars 3C 273 or 4FGL J1229.0+0202 (left) and BL Lacertae or 4FGL J2202.7+4216 modeled with {\tt PyQSOFit}. The spectral data is shown with the black line. Broad and narrow components of the emission line are represented by red and green lines and the modeled continuum is plotted with the orange line. Blue line is the sum of all the components. Horizontal grey dahses at the top of the plots denote the line-free wavelength regions selected to model the continuum emission. The data are adopted from \citet[][for 3C 273]{2012RMxAA..48....9T} and \citet[][for BL Lacertae]{1995ApJ...452L...5V}.}
\label{fig:line_fit}
\end{figure*}
\section{Optical Spectroscopic Analysis}\label{sec:spec_analysis}
A good fraction of sources in our sample has optical spectrum taken from the SDSS-DR16, however, we do not use the emission/absorption line measurements obtained with the automatic SDSS data reduction pipeline. A more sophisticated approach was needed to decompose the broad and narrow components of the emission lines, to model and subtract the host galaxy emission, and to consider the Fe complexes since these are likely not to be properly taken into account by the SDSS pipeline. Moreover, a significant number of blazars do not have SDSS spectra and a separate spectroscopic analysis was required. During the data analysis, every optical spectrum was individually analyzed rather than by running any automatic pipeline.

\subsection{Emission Line Spectrum}
There are 674 \gm-ray sources in our sample whose optical spectra exhibit at least one of the broad emission lines \halpha, \hbeta, \MgII, and \CIV. To derive the spectral parameters associated with these lines, e.g., full-width at half-maximum or FWHM,  and continuum luminosities at 1350 \AA, 3000 \AA, and 5100 \AA~($L_{\rm 1350}$, $L_{\rm 3000}$, and $L_{\rm 5100}$, respectively), we have adopted the publicly available software {\tt PyQSOFit}\footnote{\url{https://github.com/legolason/PyQSOFit}} \citep[][]{2018ascl.soft09008G}. The tool applies the spectral models and templates to data following a $\chi^2$-based fitting technique \citep[see also,][]{2019MNRAS.482.3288G,2019ApJS..241...34S}. Here we briefly describe the adopted steps.

The quasar spectrum was brought to rest-frame and corrected for Galactic reddening following the extinction curve from \citet[][]{1989ApJ...345..245C} and dust map of \citet[][]{1998ApJ...500..525S}. This correction was applied only to the SDSS data. For spectra taken from the published articles, we do not apply Galactic extinction correction since it is usually done as a part of the data reduction prior to the publication. The spectrum was decomposed into the quasar and host galaxy components following the principal component analysis method presented in \citet[][]{2004AJ....128..585Y,2004AJ....128.2603Y}.  We have considered a power-law and a third-order polynomial along with optical and UV \FeII~templates \citep[][]{1992ApJS...80..109B,2001ApJS..134....1V} to fit the line-free continuum over the entire spectrum. The best fitted continuum was then subtracted from the spectrum to acquire a line-only spectrum which was then used to extract the spectral properties of \halpha, \hbeta, \MgII, and \CIV~emission lines.

The \halpha~line was fitted in the wavelength range [6400,~6800] \AA. The broad component of \halpha~was modeled with 3 Gaussians \citep[FWHM $>$1200 \km;][]{2005AJ....129.1783H}, whereas, we adopted a single Gaussian with the FWHM upper limit of 1200 \km~to reproduce the narrow \halpha~line. Furthermore, the flux ratio of the \NII~$\lambda\lambda$6549, 6585 doublet was fixed to 3 \citep[][]{2011ApJS..194...45S}.

\hbeta~line fitting was carried out in the wavelength range [4640$-$5100] \AA. We used 2 Gaussians (FWHM $>$1200 \km) and a single Gaussian (FWHM $<$1200 \km) fit to model the broad and narrow components of the \hbeta~line, respectively. We also fitted the narrow \OIII~$\lambda\lambda$ 4959, 5007 lines with 2 Gaussians and their flux ratio was allowed to vary. However, we tied the velocity and width of the narrow \hbeta~line to the core \OIII~component. 

We fitted \MgII~and \CIV~emission lines in the wavelength range [2700$-$2900] \AA~and [1500$-$1700] \AA, respectively. The broad component of both lines were modeled with 2 Gaussians (FWHM $>$1200 \km) and we used a single Gaussian function (FWHM $<$1200 \km) to fit their narrow emission line components.

{\tt PyQSOFit} determines the uncertainties in the derived parameters employing a Monte-Carlo technique. In particular, a random Gaussian fluctuation with a zero mean and dispersion equals to the uncertainty measured at the given pixel was added to the observed spectral flux at every pixel and 50 mock spectra were created and fitted with the same strategy as that adopted to model the real data. This exercise was repeated 20 times and uncertainties were calculated as the semi-amplitude of the range covering the 16th$-$84th percentiles of the parameter distribution from the trials.

\begin{deluxetable}{llll}
\tabletypesize{\footnotesize}
\tablecaption{The spectral parameters associated with the \halpha~emission line.\label{tab:line_ha}}
\tablewidth{0pt}
\tablehead{
\colhead{4FGL name} & \colhead{redshift} & \colhead{FWHM} & \colhead{$L_{\rm H\alpha}$}\\
\colhead{[1]} & \colhead{[2]} & \colhead{[3]} & \colhead{[4]}}
\startdata
J0001.5+2113   &   0.439   &   1629  $\pm$    84    &  42.942   $\pm$   0.016     \\
J0006.3$-$0620   &   0.347   &   8991  $\pm$    5494  &    42.782 $\pm$     0.189   \\
J0013.6+4051   &   0.256   &   2049  $\pm$    1739  &    41.277 $\pm$     0.205   \\
J0017.5$-$0514   &   0.227   &   2073  $\pm$    982   &   42.926  $\pm$    0.090    \\
J0049.6$-$4500   &   0.121   &   4298  $\pm$    3904  &    41.658 $\pm$     0.117   \\
\enddata
\tablecomments{Column information are as follows: Col.[1]: 4FGL name; Col.[2]: redshift; Col.[3]: FWHM of \halpha~line, in \km; and Col.[4]: log-scale \halpha~line luminosity, in \lum. (This table is available in its entirety in a machine-readable form in the online journal. A portion is shown here for guidance regarding its form and content.) 
}
\end{deluxetable}

\begin{deluxetable}{lllll}
\tabletypesize{\footnotesize}
\tablecaption{The spectral parameters associated with the \hbeta~emission line.\label{tab:line_hb}}
\tablewidth{0pt}
\tablehead{
\colhead{4FGL name} & \colhead{redshift} & \colhead{$L_{\rm 5100\AA}$} & \colhead{FWHM} & \colhead{$L_{\rm H\beta}$}\\
\colhead{[1]} & \colhead{[2]} & \colhead{[3]} & \colhead{[4]} & \colhead{[5]}}
\startdata
J0001.5+2113  &    0.439   &   44.577  $\pm$    0.003   &   2114  $\pm$    483   &   42.029   $\pm$   0.054    \\
J0006.3$-$0620  &    0.347   &   44.757  $\pm$    0.004   &   7051  $\pm$    4699  &    42.004  $\pm$    0.227   \\
J0010.6+2043  &    0.598   &   44.867  $\pm$    0.003   &   2593  $\pm$    207   &   43.047   $\pm$   0.048    \\
J0014.3$-$0500  &    0.791   &   44.951  $\pm$    0.008   &   1619  $\pm$    653   &   42.446   $\pm$   0.060    \\
J0017.5$-$0514  &    0.227   &   44.347  $\pm$    0.005   &   3294  $\pm$    487   &   42.302   $\pm$   0.115    \\
\enddata
\tablecomments{Column information are as follows: Col.[1]: 4FGL name; Col.[2]: redshift; Col.[3]: log-scale continuum luminosity at 5100 \AA; Col.[4]: FWHM of \hbeta~line, in \km; and Col.[5]: log-scale \hbeta~line luminosity, in \lum. (This table is available in its entirety in a machine-readable form in the online journal. A portion is shown here for guidance regarding its form and content.) 
}
\end{deluxetable}

\begin{deluxetable}{lllll}
\tabletypesize{\footnotesize}
\tablecaption{The spectral parameters associated with the \MgII~emission line.\label{tab:line_mg}}
\tablewidth{0pt}
\tablehead{
\colhead{4FGL name} & \colhead{redshift} & \colhead{$L_{\rm 3000\AA}$} & \colhead{FWHM} & \colhead{$L_{\rm \MgII}$}\\
\colhead{[1]} & \colhead{[2]} & \colhead{[3]} & \colhead{[4]} & \colhead{[5]}}
\startdata
J0001.5+2113  &    0.439   &   44.706   $\pm$   0.002   &   1719  $\pm$    165   &   42.503  $\pm$    0.032    \\
J0004.4$-$4737  &    0.880   &   45.345   $\pm$   0.003   &   2965  $\pm$    913   &   42.885  $\pm$    0.099    \\
J0010.6+2043  &    0.598   &   44.799   $\pm$   0.005   &   2222  $\pm$    99    &  43.027   $\pm$   0.017     \\
J0011.4+0057  &    1.491   &   45.589   $\pm$   0.003   &   3382  $\pm$    245   &   43.365  $\pm$    0.045    \\
J0013.6$-$0424  &    1.076   &   44.898   $\pm$   0.003   &   2383  $\pm$    221   &   42.818  $\pm$    0.080    \\
\enddata
\tablecomments{Column information are as follows: Col.[1]: 4FGL name; Col.[2]: redshift; Col.[3]: log-scale continuum luminosity at 3000 \AA; Col.[4]: FWHM of \MgII~line, in \km; and Col.[5]: log-scale \MgII~line luminosity, in \lum. (This table is available in its entirety in a machine-readable form in the online journal. A portion is shown here for guidance regarding its form and content.) 
}
\end{deluxetable}

\begin{deluxetable}{lllll}
\tabletypesize{\footnotesize}
\tablecaption{The spectral parameters associated with the \CIV~emission line.\label{tab:line_c4}}
\tablewidth{0pt}
\tablehead{
\colhead{4FGL name} & \colhead{redshift} & \colhead{$L_{\rm 1350\AA}$} & \colhead{FWHM} & \colhead{$L_{\rm \CIV}$}\\
\colhead{[1]} & \colhead{[2]} & \colhead{[3]} & \colhead{[4]} & \colhead{[5]}}
\startdata
J0004.3+4614  &    1.810   &   45.615   $\pm$   0.001   &   2652  $\pm$    300   &   44.126   $\pm$   0.031     \\
J0011.4+0057  &    1.491   &   45.558   $\pm$   0.001   &   4430  $\pm$    437   &   43.901   $\pm$   0.014     \\
J0016.2$-$0016  &    1.577   &   45.445   $\pm$   0.001   &   4648  $\pm$    216   &   43.809   $\pm$   0.073     \\
J0016.5+1702  &    1.721   &   45.304   $\pm$   0.000   &   5722  $\pm$    275   &   43.849   $\pm$   0.021     \\
J0028.4+2001  &    1.553   &   45.499   $\pm$   0.001   &   2631  $\pm$    128   &   43.927   $\pm$   0.012     \\
\enddata
\tablecomments{Column information are as follows: Col.[1]: 4FGL name; Col.[2]: redshift; Col.[3]: log-scale continuum luminosity at 1350 \AA; Col.[4]: FWHM of \CIV~line, in \km; and Col.[5]: log-scale \CIV~line luminosity, in \lum. (This table is available in its entirety in a machine-readable form in the online journal. A portion is shown here for guidance regarding its form and content.) 
}
\end{deluxetable}

The tool {\tt PyQSOFit} is primarily developed to analyze the SDSS data and hence requires the pixel scale of the optical spectrum to be same as that of SDSS, i.e., $10^{-4}$ in log-space\footnote{\url{https://www.sdss.org/dr16/spectro/spectro\_basics/}}. Since a significant fraction of the optical spectra in our sample is collected from the literature and does not have the requisite binning, we rebinned them using a simple linear interpolation to have the pixel scale same as that of the SDSS data. Furthermore, since we do not have the flux uncertainty measurements in such cases, we conservatively assumed an uncertainty of 25\% of the measured fluxes, a typical value associated with the ground-based optical observations \citep[cf.][]{2008ApJS..175...97H,2012ApJ...748...49S}. 

In Figure~\ref{fig:line_fit}, we show an example of the emission line fitting done on the optical spectra of the prototype blazars 3C 273 (4FGL J1229.0+0202, $z=0.158$) and BL Lacertae (4FGL J2202.7+4216, $z=0.0686$). The spectral parameters derived for \halpha, \hbeta, \MgII, and \CIV~emission lines are provided in Table~\ref{tab:line_ha}, \ref{tab:line_hb}, \ref{tab:line_mg}, and \ref{tab:line_c4}, respectively.

\begin{figure*}[t!]
\hbox{
\includegraphics[scale=0.37]{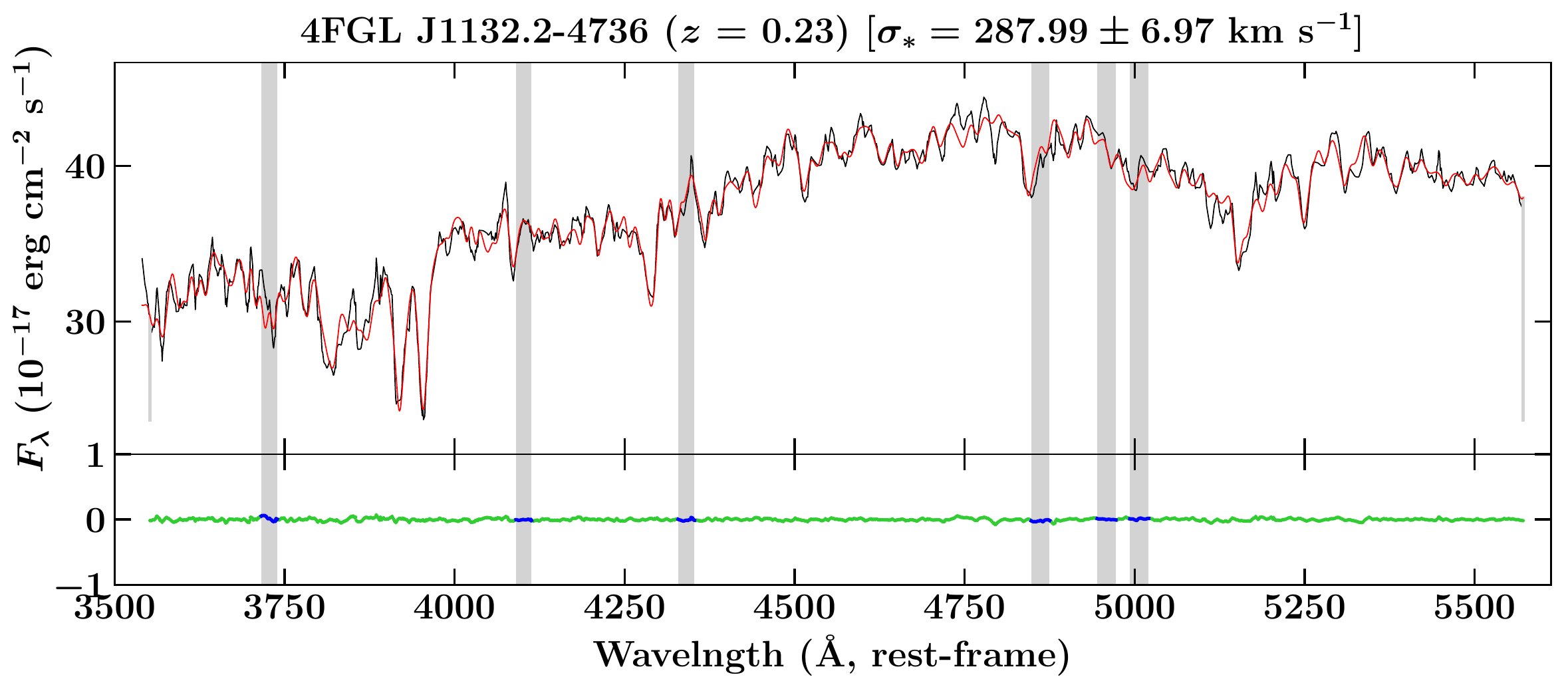} 
\includegraphics[scale=0.37]{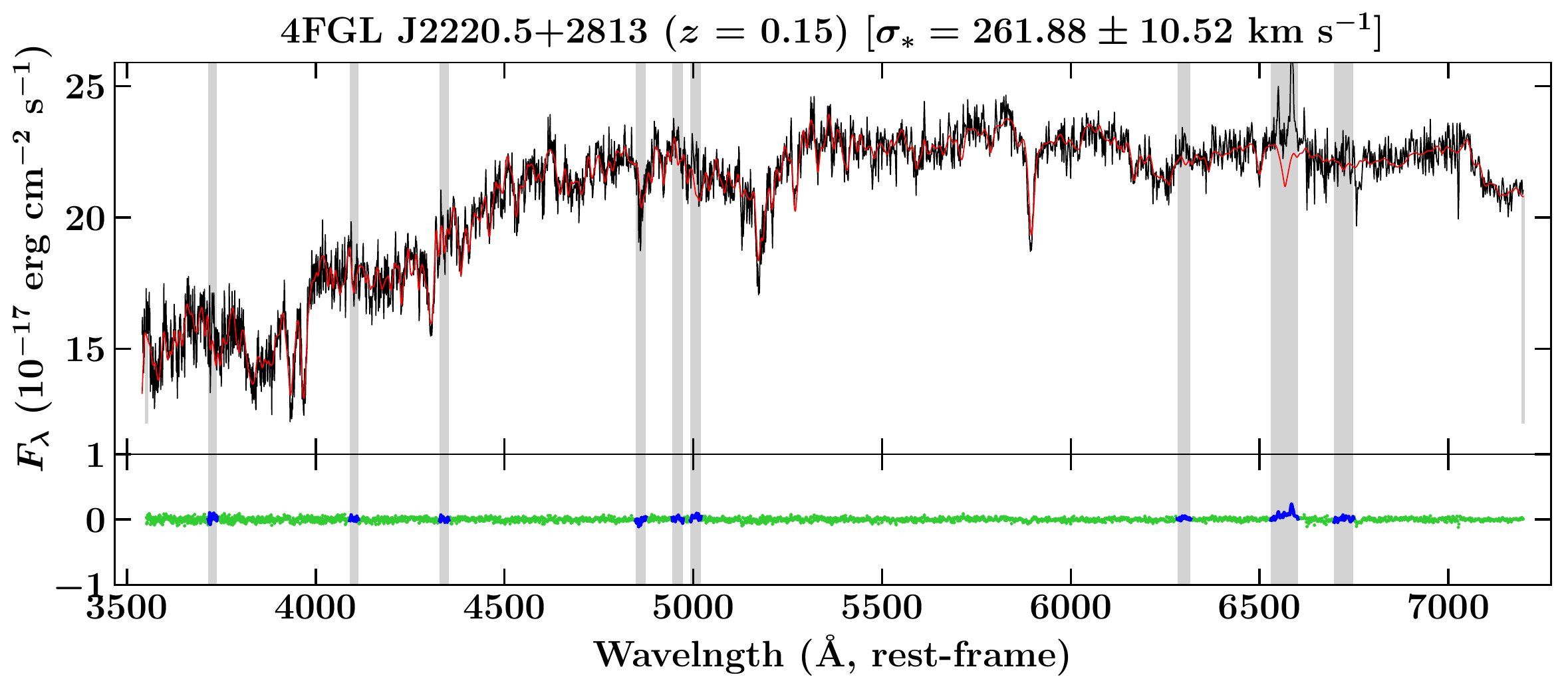} 
}
\caption{The optical spectra of two \fermi-LAT detected blazars (black) along with the results of the stellar population synthesis done using {\tt pPXF} (red). In both plots, the bottom panel refers to the residual of the fit. The wavelength regions excluded from the fit to mask emission lines are highlighted with the grey shaded columns. The spectroscopic data of 4FGL J1132.2$-$4736 is taken from \citet[][]{2017ApNSS.362..228P}, whereas, that of 4FGL J2220.5+2813 is adopted from SDSS-DR16 \citep[][]{2020ApJS..249....3A}.}
\label{fig:line_abs}
\end{figure*}

\subsection{Absorption Line Spectrum}
The optical spectra of 346 \gm-ray blazars show prominent absorption lines, e.g., Ca H\&K doublet, primarily originating from the stellar population in the host galaxy. It has been shown in various works that the mass of the central black hole significantly correlates with the stellar velocity dispersion \citep[$\sigma_*$, cf.][]{2000ApJ...539L...9F,2009ApJ...698..198G,2013ARA&A..51..511K}. Therefore, we have used the penalized PiXel Fitting tool \citep[{\tt pPXF};][]{2004PASP..116..138C} to derive $\sigma_*$ for blazars present in the sample. This software
works in pixel space and uses a maximum penalized likelihood approach to calculate the line-of-sight velocity distribution (LOSVD) from kinematic data \citep[][]{1997AJ....114..228M}. A large set of stellar population synthesis models \citep[adopted from][]{2010MNRAS.404.1639V} with spectral resolution of FWHM = 2.5 \AA~and the wavelength coverage of [3525, 7500] \AA~\citep[][]{2006MNRAS.371..703S} were used in this work. The {\tt pPXF} code first creates a template galaxy spectrum by convolving the stellar population models with the parameterized LOSVD. To mimic the non-thermal power-law contribution from the central nucleus, we further added a fourth-order Legendre polynomial to the model. The regions of bright emission lines were masked prior to the fitting. The model was then fitted on the rest-frame galaxy spectrum and $\sigma_*$ and associated 1$\sigma$ uncertainty were derived from the best-fit spectrum. The whole analysis was carried out on a case-by-case basis to achieve the best results. This is because the optical spectra of many sources reveal strong telluric lines which may overlap with the absorption lines arising from the host galaxy and thus needed to be avoided. Therefore, although we attempted a fit in the full [3525, 7500] \AA~wavelength range, it was not always possible. The examples of this analysis are shown in Figure~\ref{fig:line_abs} and the derived $\sigma_*$ values are provided in Table~\ref{tab:svd}.

\begin{figure}[t!]
\hbox{
\includegraphics[width=\linewidth]{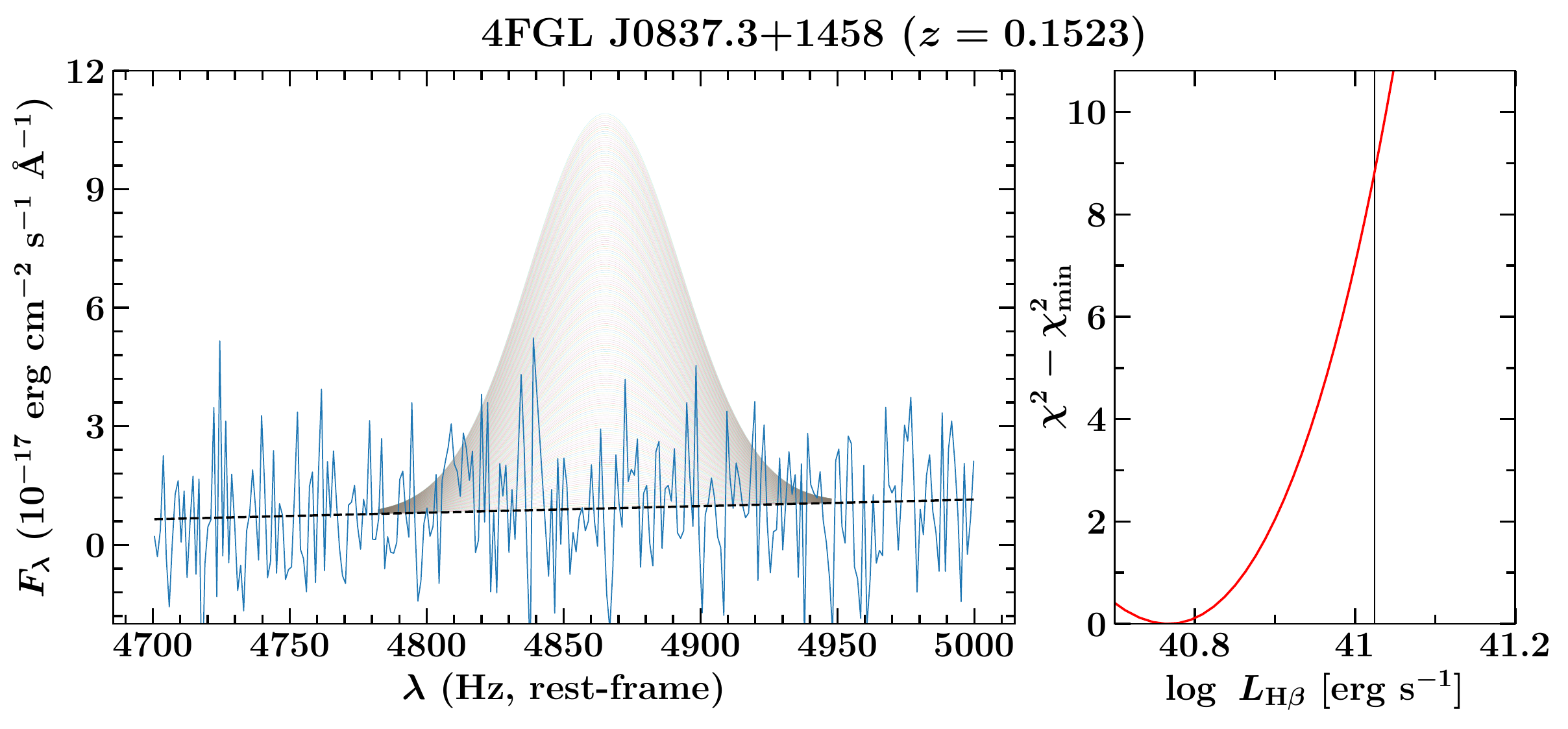} 
}
\caption{A demonstration of the technique adopted to derive the 3$\sigma$ upper limit on the emission line luminosity. Blue thin and black dotted lines refer to the observed spectrum and power-law continuum, respectively. We also plot Gaussian functions with variable line luminosity. The right panel shows the derived $\chi^2$ as a function of the line (\hbeta~in this case) luminosity. The vertical line corresponds to the line luminosity beyond which $\chi^2>\chi^2$ (99.7\%).}
\label{fig:lien_ul}
\end{figure}

Unlike broad line blazars where emission line luminosities can be used to infer the BLR luminosity, it is not possible estimate the latter for absorption line systems since broad emission lines are not detected. Therefore, we have derived 3$\sigma$ upper limit in the \hbeta~(or \MgII, depending on the source redshift and wavelength coverage) line luminosity by adopting the following steps. The observed spectrum was first brought to the rest-frame and the host galaxy component was subtracted using {\tt PyQSOFit}. In the wavelength range [4700$-$5000] \AA~or [2650$-$2950] \AA~for \hbeta~or \MgII, respectively, we then fitted a power-law (reproducing continuum) plus a Gaussian function (mimicking the emission line) with a variable luminosity while keeping the FWHM fixed to 4000 \km, a value typically observed in blazars \citep[e.g.,][]{2012ApJ...748...49S}. Based on the derived $\chi^2$, we considered the upper limit to the line luminosity when $\chi^2>\chi^2$ (99.7\%), i.e., at 3$\sigma$ confidence level. We show an example of the adopted method in Figure~\ref{fig:lien_ul} and report the upper limits in Table~\ref{tab:svd}.

\begin{deluxetable}{llll}
\tabletypesize{\normalsize}
\tablecaption{The stellar velocity dispersion and line luminosity upper limit measurements for 346 blazars.\label{tab:svd}}
\tablewidth{0pt}
\tablehead{
\colhead{4FGL name} & \colhead{redshift} & \colhead{$\sigma_*$} & \colhead{$L_{\rm line,UL}$}\\
\colhead{[1]} & \colhead{[2]} & \colhead{[3]} & \colhead{[4]}}
\startdata
J0003.2+2207   &   0.100   &   197.61  $\pm$    7.87   &   40.34   \\
J0006.4+0135   &   0.787   &   385.99  $\pm$    61.79  &    42.41  \\
J0013.9$-$1854   &   0.095   &   459.46  $\pm$    19.50  &    40.87  \\
J0014.2+0854   &   0.163   &   296.52  $\pm$    10.53  &    40.97  \\
J0015.6+5551   &   0.217   &   467.26  $\pm$    22.05  &    41.65  \\
\enddata
\tablecomments{Column information are as follows: Col.[1]: 4FGL name; Col.[2]: redshift; Col.[3]: stellar velocity dispersion ($\sigma_*$) in \km; and Col.[4]: 3$\sigma$ upper limit in \hbeta~line luminosity (log-scale, in \lum), except for sources 4FGL J0006.4+0135, 4FGL J0204.0-3334, and 4FGL J1146.0-0638, for which we quote the \MgII~line luminosity upper limit. (This table is available in its entirety in a machine-readable form in the online journal. A portion is shown here for guidance regarding its form and content.) 
}
\end{deluxetable}

\section{Black Hole Mass and Disk Luminosity Measurements}\label{sec:results}

\subsection{Emission Line Black Hole Mass}

We have derived \mbh~of emission line blazars from the single-epoch optical spectra assuming that BLR is virialized, broad line FWHM represents the virial velocity, and continuum luminosity can be considered as a proxy for the BLR radius. The virial \mbh~can be calculated using the following equation \citep[][]{2011ApJS..194...45S}:

\begin{equation}\label{eqn:virial_estimator}
\log \left({M_{\rm BH} \over M_\odot}\right)
=\alpha+\beta\log\left({\lambda L_{\lambda} \over 10^{44}\,{\rm
erg\,s^{-1}}}\right)+2\log\left({\rm FWHM\over km\,s^{-1}}\right)
\end{equation}

\begin{figure*}[t!]
\hbox{\hspace{1.cm}
\includegraphics[scale=0.4]{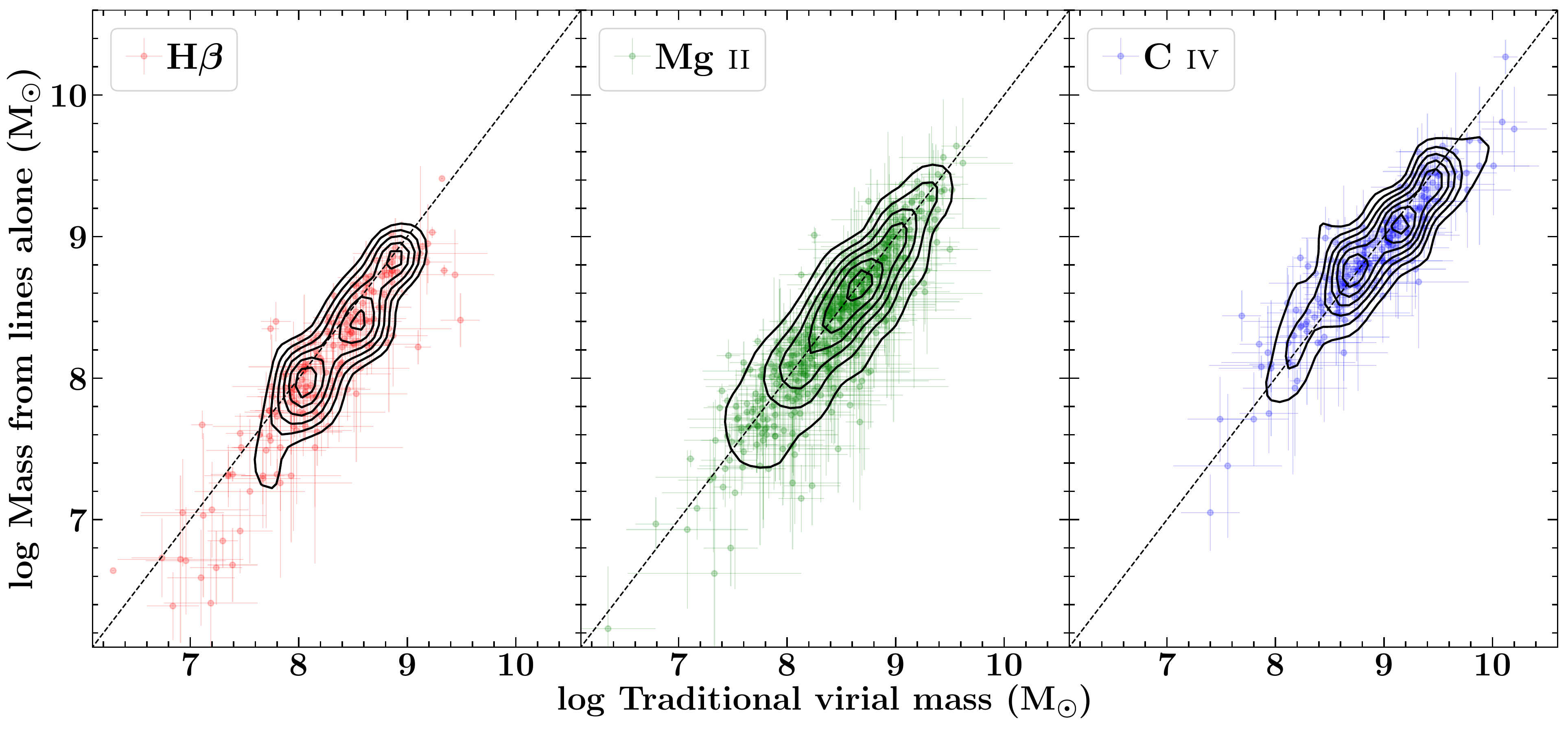} 
}
\caption{A comparison of the \mbh~values derived from the traditional virial technique using line FWHM and continuum luminosity with that estimated from the emission line alone. The dashed line represents the equality of the plotted quantities.}\label{fig:bh_all}
\end{figure*}

\noindent where $\lambda L_{\lambda}$ is the continuum luminosity at 5100 \AA~(for \hbeta), 3000 \AA~(for \MgII), and 1350 \AA~(for \CIV). The calibration coefficients $\alpha$ and $\beta$ are taken from \citet[MD04;][]{2004MNRAS.352.1390M} for \hbeta~and \MgII~and \citet[VP06;][]{2006ApJ...641..689V} for \CIV~lines and have the following values

\begin{equation}
(\alpha,\beta) = \left\{ \begin{array}{ll}
	(0.672,0.61), &  {\textrm \hbeta\ {\rm (MD04)}}\\
	(0.505,0.62), &  {\textrm \MgII\ {\rm (MD04)}}\\
	(0.660,0.53), &  {\textrm \CIV\ {\rm (VP06)}}
	\end{array}
	\right. \ ,
\end{equation}

\citet[][]{2011ApJS..194...45S} also provided the following empirical relation to estimate \mbh~from the \halpha~line FWHM and luminosity:

\begin{eqnarray}\label{eqn:vir_halpha}
\log \left({M_{\rm BH} \over M_\odot}\right)_{\rm H\alpha}=&&0.379 + 0.43\log\left(\frac{L_{\rm H\alpha}}{\rm 10^{42}\, erg\,s^{-1}}\right)\nonumber\\
 &+& 2.1\log\left(\frac{{\rm FWHM_{H\alpha}}}{{\rm km\,s^{-1}}}\right)\ ,
\end{eqnarray}

\noindent where $L_{\rm H\alpha}$ is the total \halpha\ line luminosity. For objects with more than one \mbh~measurements, we have taken the geometric mean. 

The optical spectral continuum of blazars could be significantly contaminated from the non-thermal jetted emission which can affect the \mbh~estimation. To address this issue, we also derived \mbh~values using emission line parameters alone. This was done by replacing continuum luminosity in Equation~\ref{eqn:virial_estimator} with the line luminosity and adopting the calibration coefficients ($\alpha,\beta$) as (1.63, 0.49), (1.70, 0.63), and (1.52, 0.46) for \hbeta, \MgII, and \CIV~lines, respectively, following \citet[][]{2012ApJ...748...49S}. We compared the masses calculated from two approaches in Figure~\ref{fig:bh_all}. There are indications that \mbh~computed from virial approach is slightly higher than that calculated from \hbeta~line alone (Figure~\ref{fig:bh_all}, left panel) which could be due to contamination from the host galaxy emission as also suggested by \citet[][]{2012ApJ...748...49S}. For \MgII~and \CIV~lines, the overall impact of the jetted emission on the \mbh~measurement is negligible. Therefore, the \mbh~values for emission line blazars computed from the traditional virial technique (Equation~\ref{eqn:virial_estimator}) are used in this work as this may also allow a comparison study with other non-blazar class of AGN.  We report them in Table~\ref{tab:em_mbh_ld}.

\subsection{Absorption Line Black Hole Mass}

We used the following empirical relation to compute $M_{\rm BH}$ from the measured stellar velocity dispersion \citep[][]{2009ApJ...698..198G}

\begin{equation}
\log{\left(\frac{M_{\rm BH}}{\msun}\right)} = (\msinterr) + (\msslopeerr) 
\log{\left(\frac{\sigma_*}{200\km}\right)},
\label{e:noulmsigmafit}
\end{equation}

The estimated masses of 346 blazars are provided in Table~\ref{tab:abs_mbh_ld}.

\subsection{Bulge Luminosity Black Hole Mass}

From the literature, we were able to obtain apparent/absolute $R$- or $K$-band magnitudes of the host galaxy bulge for 47 blazars. The following equations waere used to derive \mbh~from the bulge luminosity \citep[][]{2007MNRAS.379..711G}

\begin{equation}
\log{\left(\frac{M_{\rm BH}}{\msun}\right)} = \left\{ \begin{array}{l}
	(-0.38\pm0.06)(M_{\rm R} + 21) + (8.11\pm0.11) \\
	(-0.38\pm0.06)(M_{\rm K} + 24) + (8.26\pm0.11)
	\end{array}
	\right.
\end{equation}

\noindent where $M_{\rm R}$ and $M_{\rm K}$ are the absolute magnitudes of the host galaxy bulge in $R$- and $K$-bands, respectively. The derived \mbh~values are reported in Table~\ref{tab:bulge}.

\begin{deluxetable*}{llllllll}
\tabletypesize{\normalsize}
\tablecaption{The central engine and SED properties of emission line blazars.\label{tab:em_mbh_ld}}
\tablewidth{0pt}
\tablehead{
\colhead{4FGL name} & \colhead{\mbh} & \colhead{\ld} & \colhead{$\nu^{\rm peak}_{\rm syn}$} & \colhead{$\nu F^{\rm peak}_{\nu,\rm syn}$} & \colhead{$\nu^{\rm peak}_{\rm IC}$} & \colhead{$\nu F^{\rm peak}_{\nu,\rm IC}$} & \colhead{CD} \\
\colhead{[1]} & \colhead{[2]} & \colhead{[3]} & \colhead{[4]} & \colhead{[5]} & \colhead{[6]} & \colhead{[7]} & \colhead{[8]} }
\startdata
J0001.5+2113   &   7.54   $\pm$   0.07   &   44.65   $\pm$   0.02  &    13.81  &    -11.97   &   20.64   &   -10.48   &   30.90    \\
J0004.3+4614   &   8.36   $\pm$   0.10   &   46.07   $\pm$   0.03  &    12.35  &    -12.49   &   21.35   &   -11.70   &   6.17     \\
J0004.4$-$4737   &   8.28   $\pm$   0.27   &   45.10   $\pm$   0.10  &    13.01  &    -11.62   &   21.37   &   -11.24   &   2.40     \\
J0006.3$-$0620   &   8.93   $\pm$   0.40   &   44.52   $\pm$   0.15  &    12.92  &    -11.09   &   19.66   &   -12.04   &   0.11     \\
J0010.6+2043   &   7.86   $\pm$   0.04   &   45.35   $\pm$   0.03  &    12.42  &    -12.12   &   22.60   &   -11.97   &   1.41     \\
\enddata
\tablecomments{Column information are as follows: Col.[1]: 4FGL name; Col.[2]: log-scale mass of the central black hole, in \msun; Col.[3]: log-scale accretion disk luminosity, in \lum; Col.[4] and [5]: log-scale synchrotron peak frequency and corresponding flux, in Hz and \ergflux, respectively; Col.[6] and [7]: log-scale inverse Compton peak frequency and corresponding flux, in Hz and \ergflux, respectively; and Col.[8]: Compton dominance. (This table is available in its entirety in a machine-readable form in the online journal. A portion is shown here for guidance regarding its form and content.) 
}
\end{deluxetable*}

\subsection{Accretion Disk Luminosity}

From the emission line luminosities or 3$\sigma$ upper limits, we computed the BLR luminosity (or 3$\sigma$ upper limits) as follows: we assigned a reference value of 100 to Ly$\alpha$ emission and summed the line ratios (with respect to Ly$\alpha$) reported in \citet[][]{1991ApJ...373..465F} and \citet[][for \halpha]{1997MNRAS.286..415C} giving the total BLR fraction $\langle L_{\rm BLR} \rangle  = 555.77\sim 5.6 {\rm Ly\alpha}$. The BLR luminosity can then be derived using the following equation

\begin{equation}
L_{\rm BLR} = L_{\rm line} \times {\langle L_{\rm BLR} \rangle  \over 
                      L_{\rm rel. frac.}} 
\end{equation}

\noindent where $L_{\rm line}$ is the emission line luminosity and $L_{\rm rel. frac.}$ is the line ratio, 77, 22, 34, and 63 for \halpha, \hbeta, \MgII, and \CIV~lines, respectively \citep[][]{1991ApJ...373..465F,1997MNRAS.286..415C}. When more than one line luminosity measurements were available, we took their geometric mean to derive the average $L_{\rm BLR}$ and then calculated \ld~from the $L_{\rm BLR}$ assuming 10\% BLR covering factor. The computed \ld~values and 3$\sigma$ upper limits are provided in Table~\ref{tab:em_mbh_ld} and \ref{tab:abs_mbh_ld}, respectively.

\begin{deluxetable*}{llllllll}
\tabletypesize{\normalsize}
\tablecaption{The central engine and SED properties of absorption line blazars.\label{tab:abs_mbh_ld}}
\tablewidth{0pt}
\tablehead{
\colhead{4FGL name} & \colhead{\mbh} & \colhead{\ld} & \colhead{$\nu^{\rm peak}_{\rm syn}$} & \colhead{$\nu F^{\rm peak}_{\nu,\rm syn}$} & \colhead{$\nu^{\rm peak}_{\rm IC}$} & \colhead{$\nu F^{\rm peak}_{\nu,\rm IC}$} & \colhead{CD} \\
\colhead{[1]} & \colhead{[2]} & \colhead{[3]} & \colhead{[4]} & \colhead{[5]} & \colhead{[6]} & \colhead{[7]} & \colhead{[8]} }
\startdata
J0003.2+2207   &   8.10   $\pm$   0.11   &   42.74   &   15.15   &   -12.23   &   22.16    &  -12.91   &   0.21    \\
J0006.4+0135   &   9.33   $\pm$   0.33   &   44.62   &   15.96   &   -12.49   &   22.93    &  -12.65   &   0.69    \\
J0013.9$-$1854   &   9.65   $\pm$   0.19   &   43.27   &   17.43   &   -11.39   &   23.95    &  -12.47   &   0.08    \\
J0014.2+0854   &   8.85   $\pm$   0.12   &   43.37   &   15.64   &   -12.21   &   22.25    &  -12.58   &   0.43    \\
J0015.6+5551   &   9.68   $\pm$   0.19   &   44.05   &   17.11   &   -11.61   &   24.89    &  -12.07   &   0.35    \\
\enddata
\tablecomments{Column information are same as in Table~\ref{tab:em_mbh_ld} except Column~3 where 3$\sigma$ upper limits on the \ld~are reported. (This table is available in its entirety in a machine-readable form in the online journal. A portion is shown here for guidance regarding its form and content.)
 }
\end{deluxetable*}

 \startlongtable
\begin{deluxetable}{lll}
\tabletypesize{\normalsize}
\tablecaption{The mass of the central black holes for 47 blazars derived from the host galaxy bulge luminosity.\label{tab:bulge}}
\tablewidth{0pt}
\tablehead{
\colhead{4FGL name} & \colhead{redshift} & \colhead{\mbh}\\
\colhead{[1]} & \colhead{[2]} & \colhead{[3]}}
\startdata
J0037.8+1239   &     0.089   &     8.64   $\pm$     0.14    \\
J0050.7$-$0929   &     0.635   &     8.85   $\pm$     0.18    \\
J0109.1+1815   &     0.145   &     9.09   $\pm$     0.23    \\
J0123.1+3421   &     0.272   &     8.68   $\pm$     0.14    \\
J0159.5+1046   &     0.195   &     8.51   $\pm$     0.14    \\
J0202.4+0849   &     0.629   &     8.85   $\pm$     0.16    \\
J0208.6+3523   &     0.318   &     8.62   $\pm$     0.14    \\
J0214.3+5145   &     0.049   &     8.55   $\pm$     0.13    \\
J0217.2+0837   &     0.085   &     8.40   $\pm$     0.13    \\
J0238.4$-$3116   &     0.233   &     8.73   $\pm$     0.15    \\
J0303.4$-$2407   &     0.266   &     8.95   $\pm$     0.17    \\
J0319.8+1845   &     0.190   &     8.66   $\pm$     0.14    \\
J0340.5$-$2118   &     0.233   &     8.29   $\pm$     0.11    \\
J0349.4$-$1159   &     0.188   &     8.60   $\pm$     0.13    \\
J0422.3+1951   &     0.512   &     8.57   $\pm$     0.14    \\
J0424.7+0036   &     0.268   &     8.76   $\pm$     0.16    \\
J0507.9+6737   &     0.416   &     9.00   $\pm$     0.18    \\
J0509.6$-$0402   &     0.304   &     8.88   $\pm$     0.16    \\
J0617.7$-$1715   &     0.098   &     8.83   $\pm$     0.20    \\
J0623.9$-$5259   &     0.513   &     8.85   $\pm$     0.16    \\
J0712.7+5033   &     0.502   &     8.49   $\pm$     0.15    \\
J0757.1+0956   &     0.266   &     8.53   $\pm$     0.13    \\
J0814.6+6430   &     0.239   &     8.23   $\pm$     0.14    \\
J1103.6$-$2329   &     0.186   &     9.01   $\pm$     0.18    \\
J1136.4+7009   &     0.045   &     8.62   $\pm$     0.14    \\
J1217.9+3007   &     0.130   &     8.75   $\pm$     0.15    \\
J1257.2+3646   &     0.530   &     8.92   $\pm$     0.19    \\
J1315.0$-$4236   &     0.105   &     8.56   $\pm$     0.13    \\
J1359.8$-$3746   &     0.334   &     8.77   $\pm$     0.16    \\
J1501.0+2238   &     0.235   &     8.81   $\pm$     0.16    \\
J1517.7+6525   &     0.702   &     10.11  $\pm$      0.31   \\
J1535.0+5320   &     0.890   &     9.64   $\pm$     0.24    \\
J1548.8$-$2250   &     0.192   &     8.69   $\pm$     0.16    \\
J1643.5$-$0646   &     0.082   &     8.43   $\pm$     0.14    \\
J1728.3+5013   &     0.055   &     8.28   $\pm$     0.11    \\
J1748.6+7005   &     0.770   &     10.02  $\pm$      0.30   \\
J1757.0+7032   &     0.407   &     8.78   $\pm$     0.18    \\
J1813.5+3144   &     0.117   &     8.02   $\pm$     0.11    \\
J2005.5+7752   &     0.342   &     8.77   $\pm$     0.16    \\
J2009.4$-$4849   &     0.071   &     8.90   $\pm$     0.17    \\
J2039.5+5218   &     0.053   &     8.11   $\pm$     0.11    \\
J2042.1+2427   &     0.104   &     8.76   $\pm$     0.16    \\
J2055.4$-$0020   &     0.440   &     8.38   $\pm$     0.15    \\
J2143.1$-$3929   &     0.429   &     8.68   $\pm$     0.15    \\
J2145.7+0718   &     0.237   &     8.70   $\pm$     0.15    \\
J2252.0+4031   &     0.229   &     8.72   $\pm$     0.15    \\
J2359.0$-$3038   &     0.165   &     8.67   $\pm$     0.14    \\
\enddata
\tablecomments{Column information are as follows: Col.[1]: 4FGL name; Col.[2]: redshift; and Col.[3]: log-scale black hole mass, in \msun.}
\end{deluxetable}

\begin{figure*}[t!]
\hbox{
\includegraphics[scale=0.37]{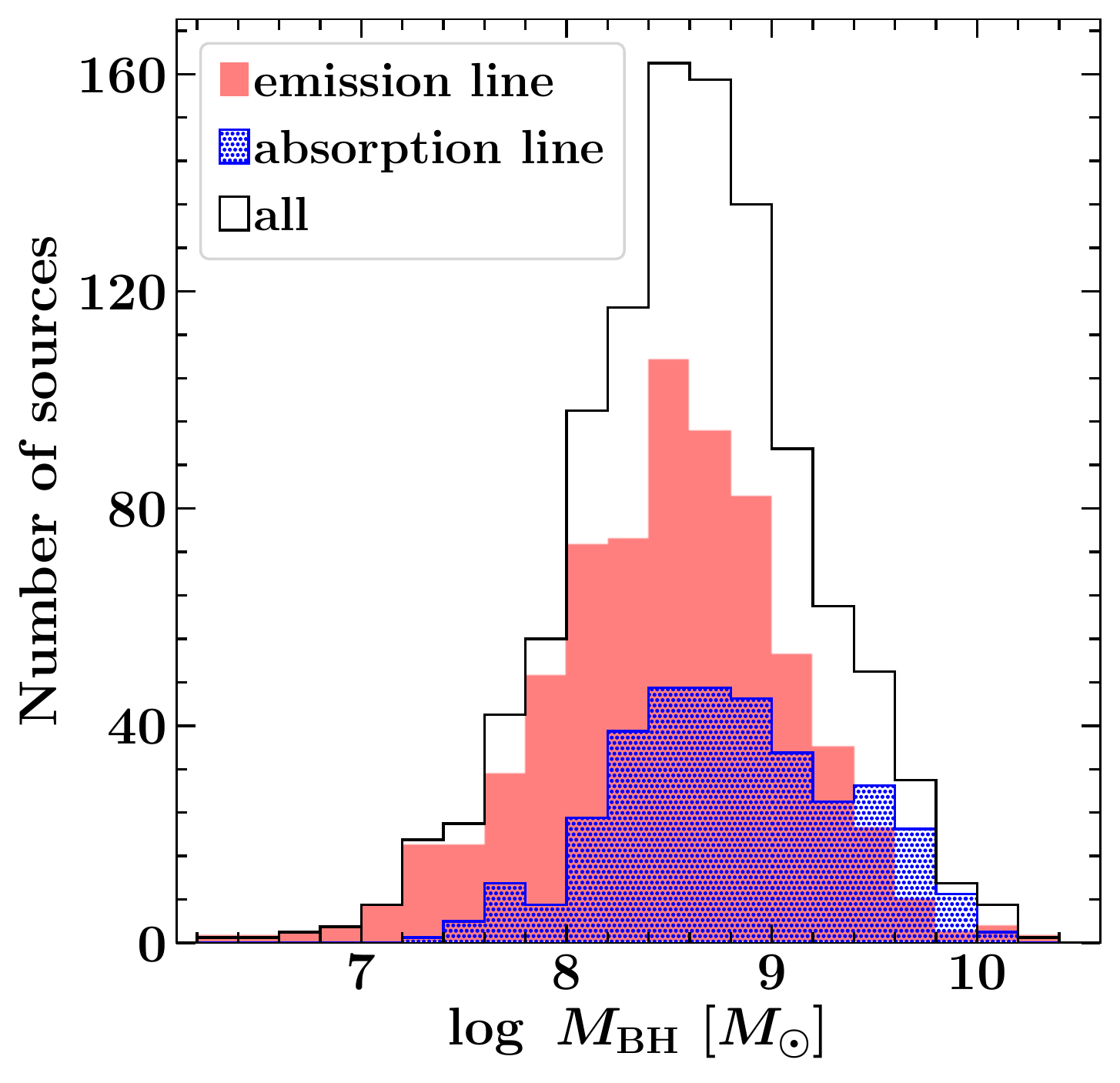} 
\includegraphics[scale=0.37]{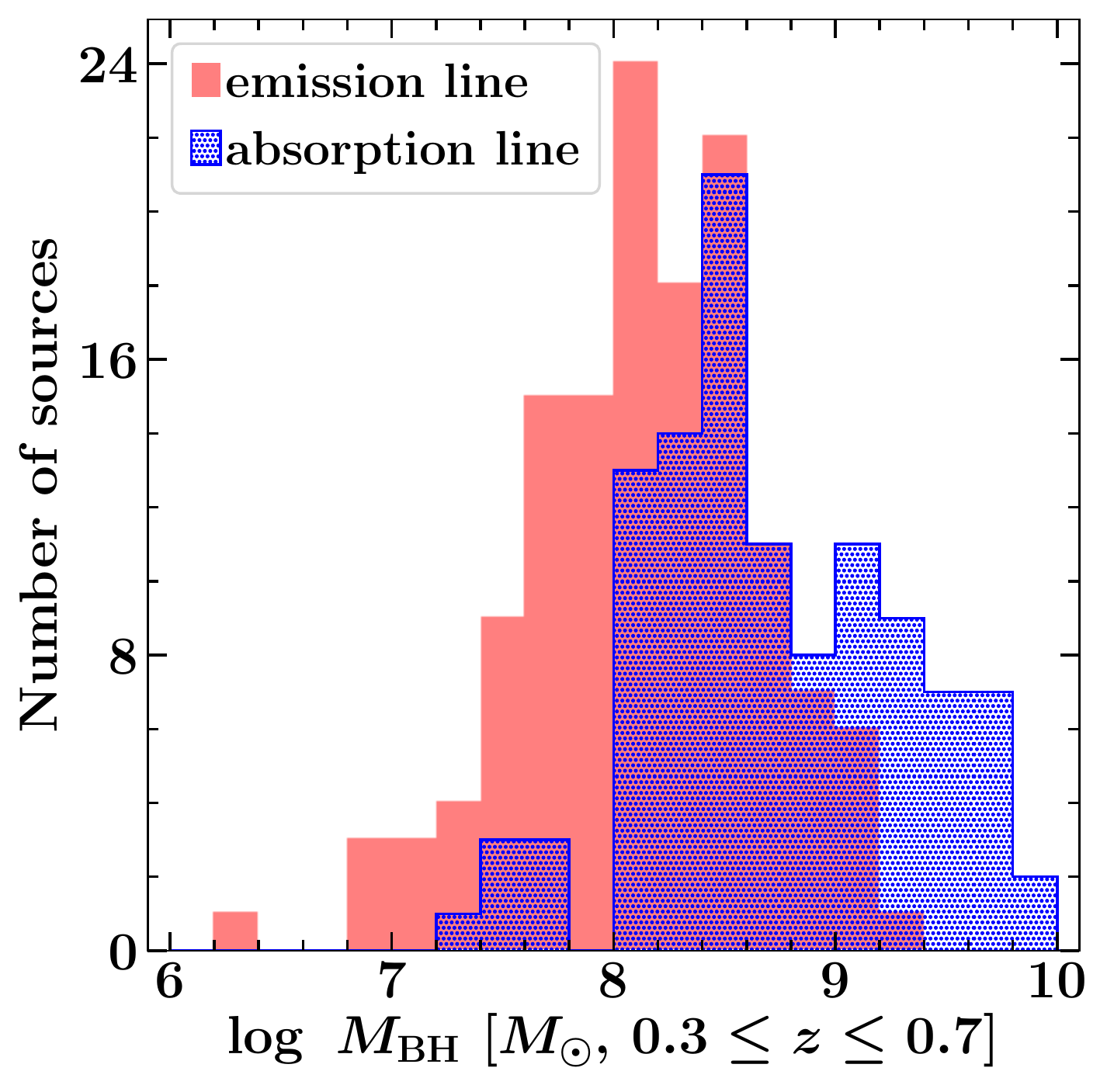} 
\includegraphics[scale=0.37]{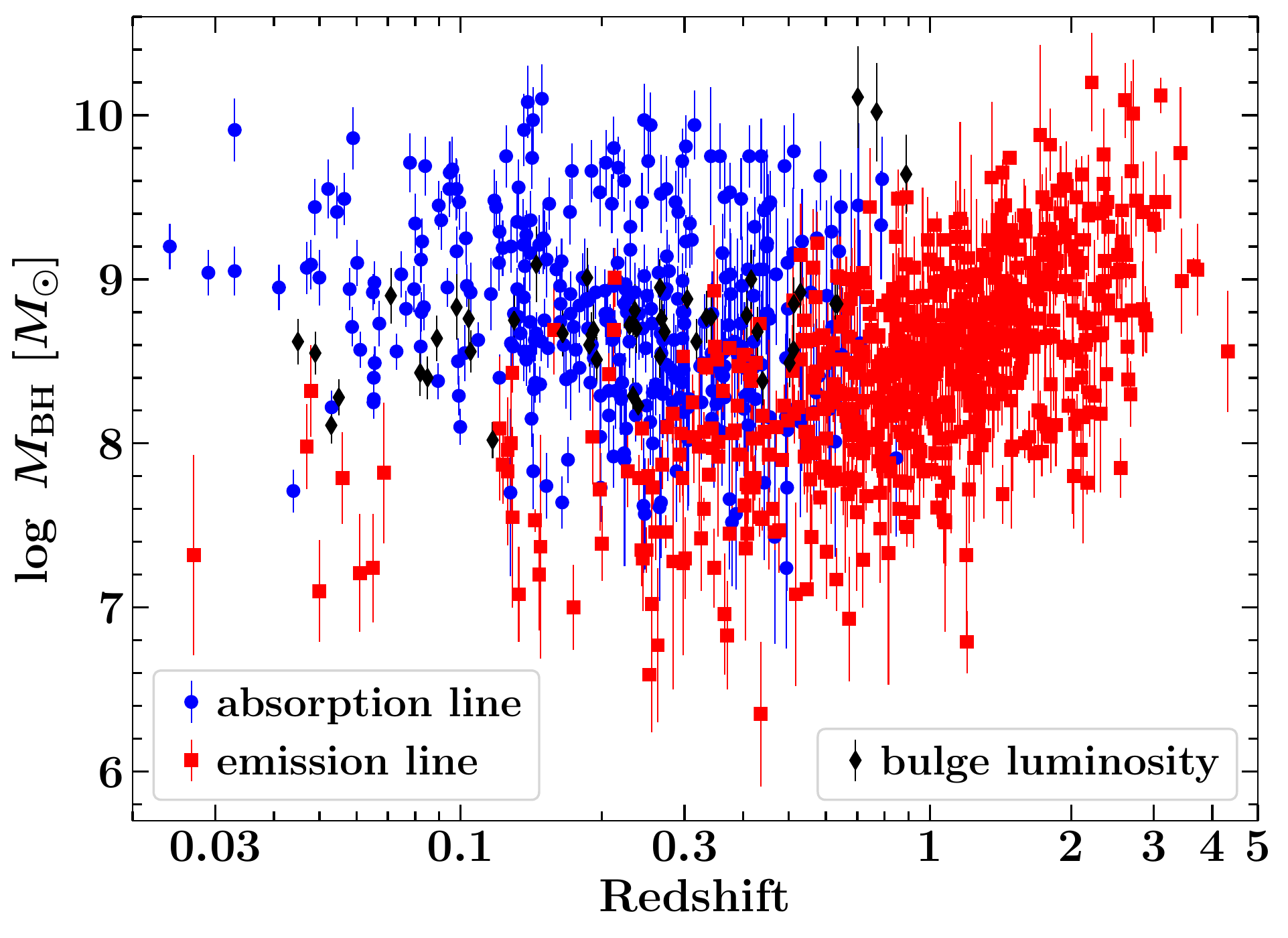} 
}
\caption{Left: The histograms of the \mbh~for broad emission line blazars (red filled), absorption line systems (blue dotted), and the whole sample (black empty) including \mbh~derived from the host galaxy bulge luminosity. Middle: The distributions of \mbh~for blazars in the redshift range $0.3\leq z \leq0.7$. Right: The redshift dependence of the \mbh~derived from various methods as labeled.}\label{fig:bh_mass}
\end{figure*}
\begin{figure*}[t!]
\hbox{\hspace{1.cm}
\includegraphics[scale=0.5]{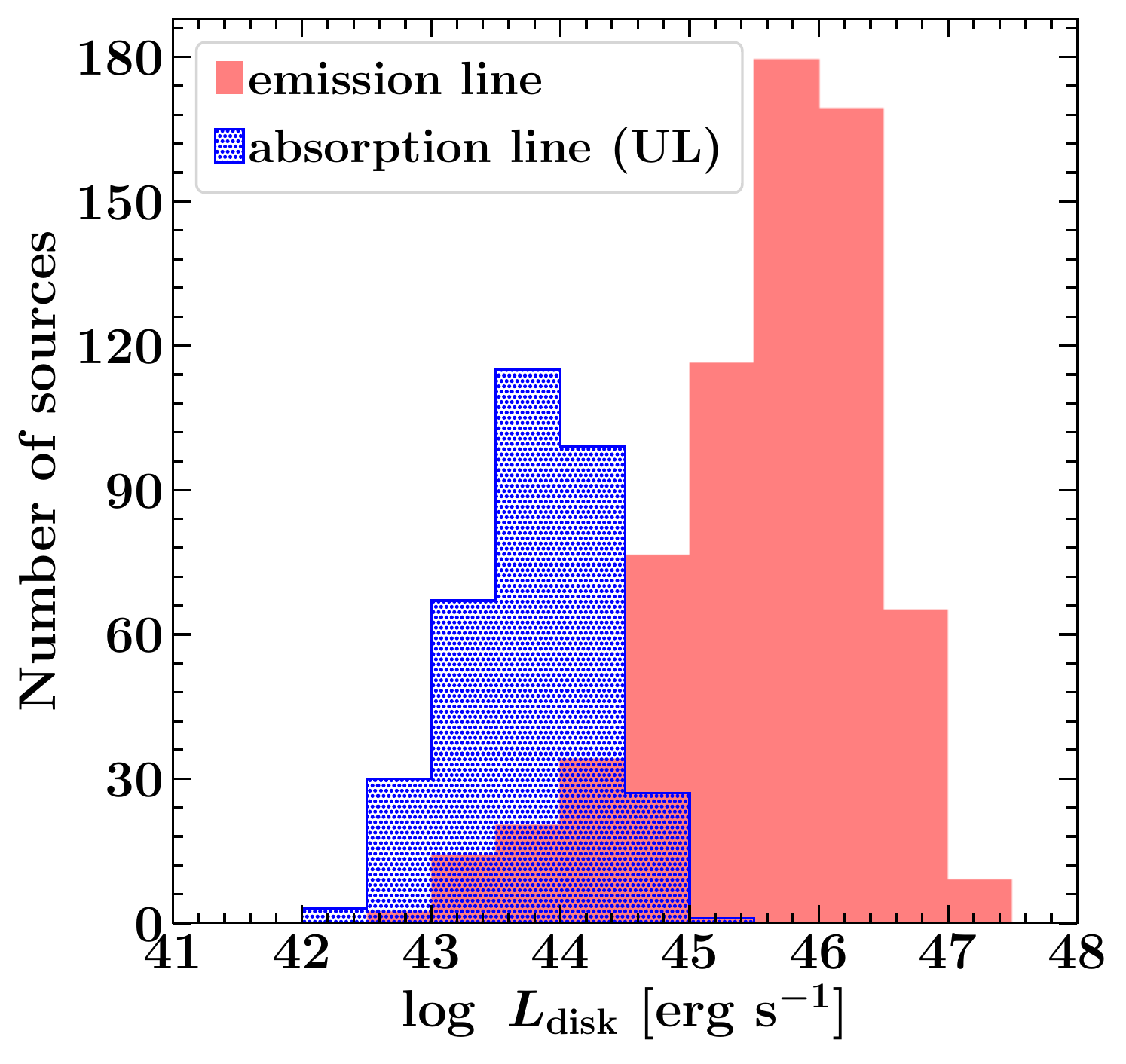} 
\includegraphics[scale=0.5]{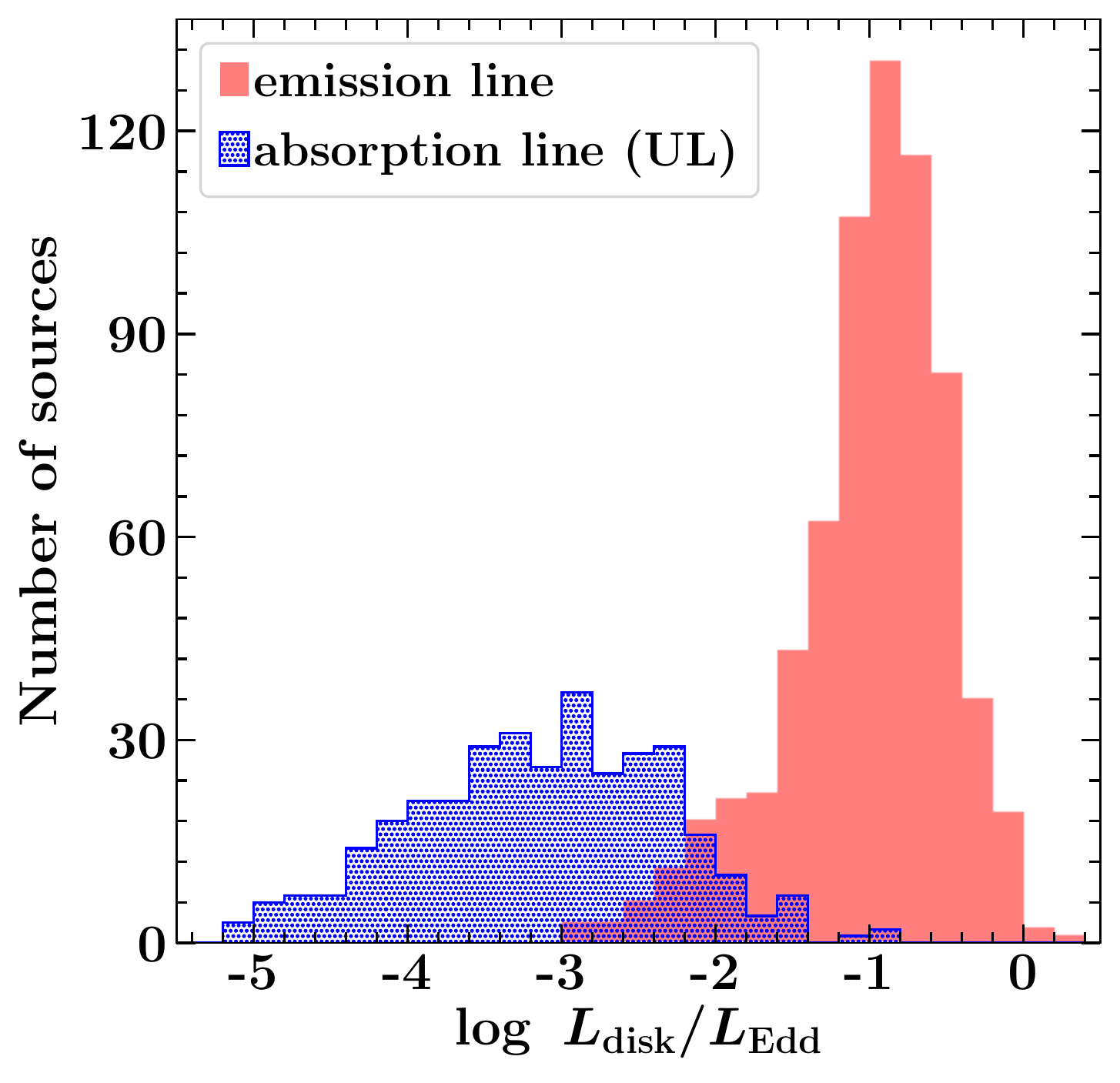} 
}
\caption{Left: The luminosity of the accretion disk for emission line blazars is shown with the red filled histogram. The blue dotted histogram for absorption line objects, on the other hand, refers to 3$\sigma$ upper limit in \ld. Right: Same as left but with \ld~plotted in Eddington units.}\label{fig:disk}
\end{figure*}

\subsection{Literature Collection}
There are 10 blazars in our sample whose optical spectra could not be retrieved, however, their emission line parameters or \mbh~values have been published \citep[e.g.,][]{1981ApJ...243...76B,2015MNRAS.451.4193C}. We were able to collect/estimate both \mbh~and \ld~values for 10 such objects and report them in Table~\ref{tab:em_mbh_ld}. These objects are: 
4FGL J0243.2$-$0550, 4FGL J0509.4+0542, 4FGL J0501.2$-$0158, 4FGL J0525.4$-$4600, 4FGL J0836.5$-$2026, 4FGL J1214.6$-$1926, 4FGL J1557.9$-$0001, 4FGL J1626.0$-$2950, 4FGL J2056.2$-$4714, and 4FGL J2323.5$-$0317. 

Finally, we tabulate references of all the research articles used to collect optical spectra or spectral parameters for 1077 blazars in Appendix (Table~\ref{tab:ref}).

\section{Properties of the Central Engine}\label{sec:central_engine}

In the left panel of Figure~\ref{fig:bh_mass}, we show the distribution of the \mbh~values estimated from different methods described above. The average \mbh~for the whole population is $\langle \log~M_{{\rm BH,all}~\msun} \rangle=8.60$. Considering emission and absorption line systems separately, we get $\langle \log~M_{{\rm BH,emi}~\msun} \rangle=8.48$ and $\langle \log~M_{{\rm BH,abs}~\msun} \rangle=8.80$, respectively. The dispersion for all three distributions is similar, $\sim$0.6, when fitted with a Gaussian function. Though the spread is large, there are tentative evidences hinting the association of more massive black holes with the absorption line systems, i.e., objects lacking broad emission lines or BL Lacs. To understand it further, we compared \mbh~for a sub-sample of blazars with $0.3\leq z \leq0.7$. This was done to avoid any redshift dependent selection effect and to compare a similar number of sources from emission (139) and absorption line (110) blazar samples. In the middle panel of Figure~\ref{fig:bh_mass}, we show the histograms of \mbh~values derived for the two populations in the opted redshift range. The average masses for emission and absorption line blazars are $\langle \log~M_{{\rm BH},~\msun} \rangle=8.13~{\rm and}~8.70$, respectively. Though the spread is large ($\sigma \sim$0.55), this finding hints that BL Lac objects do tend to host more massive black holes than broad line FSRQs. A Kolmogorov-Smirnov test was carried out to determine whether two populations are distinctively different. The derived $p$-value is $1.9\times10^{-8}$, suggesting that the null-hypothesis of both samples belonging to the same source population can be rejected at $>$5$\sigma$ confidence level. 

The observation of more massive black holes residing in BL Lac objects is likely to be connected with the cosmic evolution of blazars \citep[][]{2002ApJ...564...86B,2002ApJ...571..226C,2014ApJ...780...73A}. According to the proposed evolutionary sequence, high-power ($L_{\rm bol.} >10^{46}$ \lum) blazars, primarily FSRQs, evolve to their low-power counterparts, mainly BL Lacs, over cosmological timescales. Such a transition is likely caused by a gradual depletion of the central environment by accretion onto the black hole. In other words, luminous broad emission line blazars evolve to low-luminosity objects exhibiting quasi-featureless spectra as their accretion mode changes from being radiatively efficient to advection dominated accretion flow \citep[][]{1997ApJ...478L..79N}. If so, one would expect the black holes hosted in the latter to be more massive than that found in the former since they would keep growing as more and more mass is dumped by accretion.

We show the variation of \mbh~as a function of the blazar redshift in the right panel of Figure~\ref{fig:bh_mass}. There is a little overlap since all of the absorption line objects are located below $z=0.85$, whereas, a major fraction of the emission line sources are above it (see also Figure~\ref{fig:redshift}). There is a trend of more massive black holes being located at higher redshifts among the broad line sources. This is likely a selection effect since we expect to detect only the most luminous systems, hence the most massive black holes, at cosmological distances. 

The left panel of Figure~\ref{fig:disk} shows the \ld~distribution for broad line objects. For a comparison, we also plot the histogram of the 3$\sigma$ upper limits on the \ld~measured for blazars lacking emission lines. As expected, strong emission line systems have more luminous accretion disks. The difference between the two populations becomes clearer when we derive the \ld~values in Eddington units. The results are shown in the right panel of Figure~\ref{fig:disk} and reveal a bi-modality. A major fraction of broad line objects have \ld/\ledd~$\gtrsim 0.01$, whereas, \ld~of absorption line systems have values $\lesssim$1\% of \ledd~which is expected to be even lower since the plotted quantity corresponds to 3$\sigma$ luminosity upper limit. Interestingly, a low-level of accretion activity (\ld/\ledd~$\lesssim 0.01$) is noticed from a few emission line blazars. These are objects typically classified as BL Lacs, however, have revealed faint broad emission lines in their optical spectra taken during a low jet activity state \citep[cf.][]{1995ApJ...452L...5V}.

\begin{figure*}[t!]
\hbox{
\includegraphics[scale=0.4]{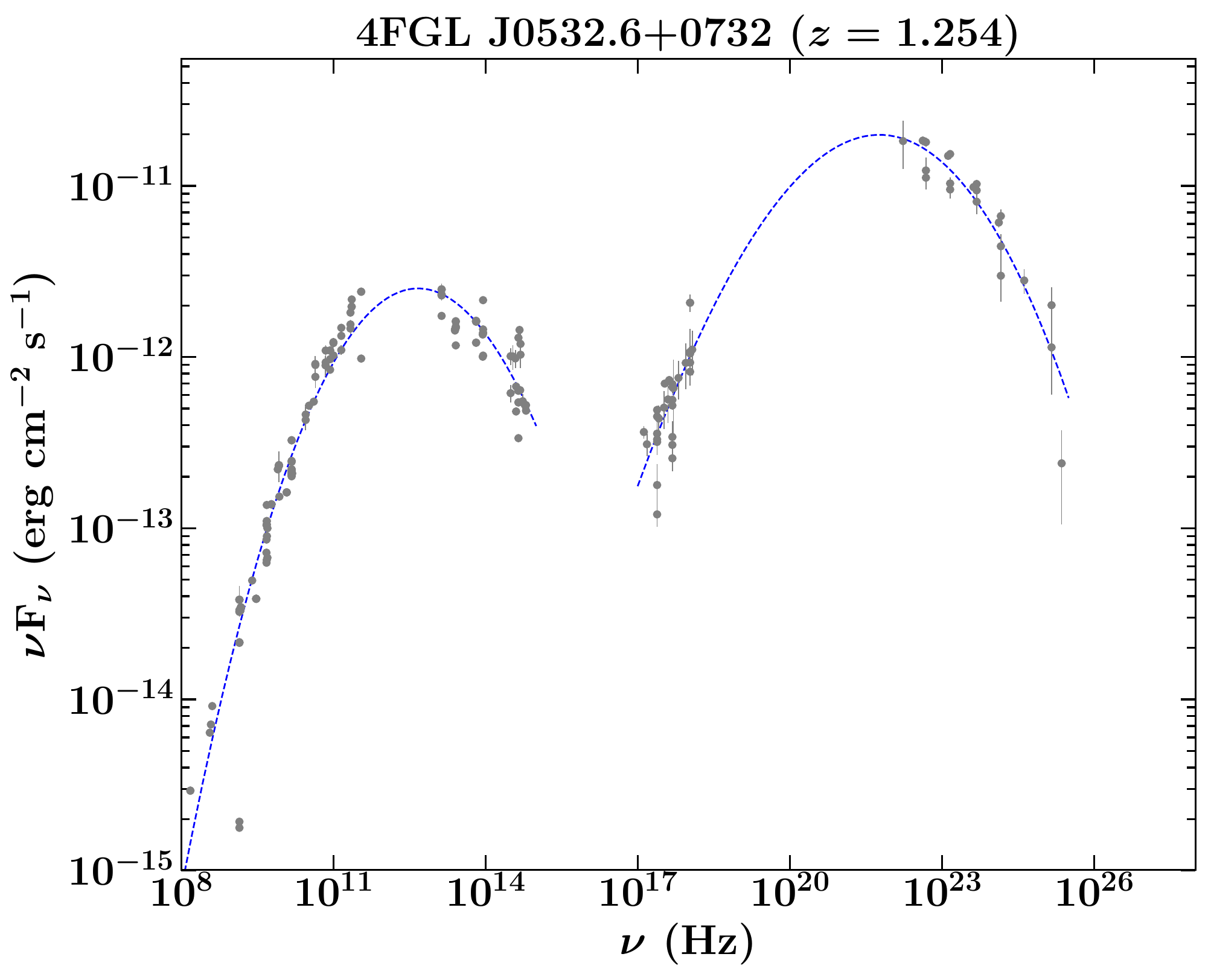} 
\includegraphics[scale=0.4]{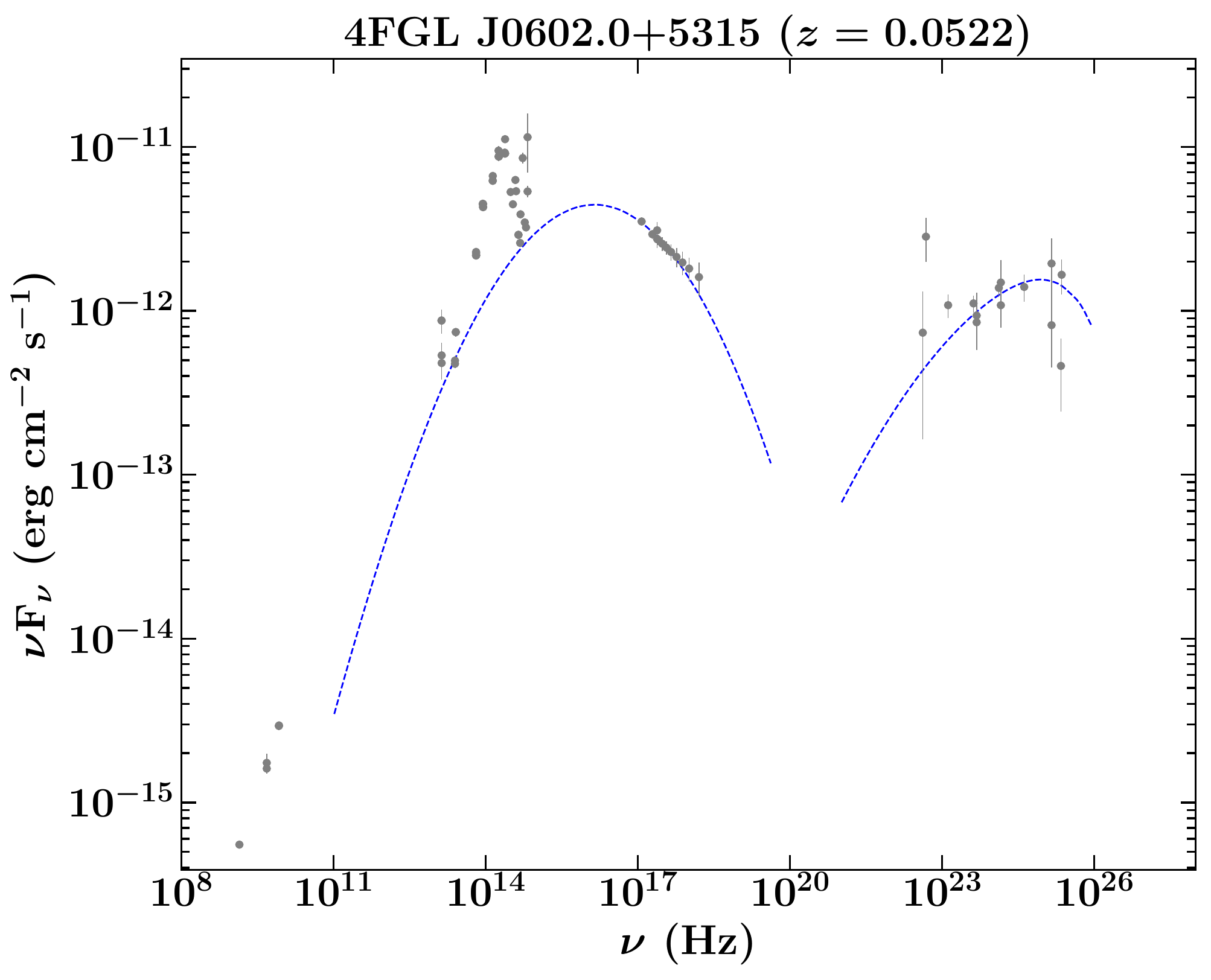} 
}
\caption{The multi-frequency SEDs of two \fermi-LAT detected blazars (grey data points) when fitted with second-degree polynomials (left) and a synchrotron-SSC model (right). The fitted functions are shown with the blue dotted lines. In the right panel, the bump seen at IR-optical wavelengths arises from the host galaxy.}
\label{fig:cd}
\end{figure*}

\section{A Compton Dominance Based Classification}\label{sec:discussion}

A blazar classification based on \ld~in units of \ledd~is physically intuitive \citep[][]{2011MNRAS.414.2674G}, however, extending it to lineless sources remained a challenge. If the optical spectrum of a FSRQ is dominated by the non-thermal jetted radiation which swamps out broad emission lines, it may not be possible to characterize its state of accretion even if it hosts a luminous accretion disk. Therefore, it is necessary to identify some other observational features which can indicate the state of accretion or, in other words, show a correlation with \ld/\ledd. Below we show that Compton dominance or CD can be considered as one such parameter to reveal the physics of the central engine in beamed AGN.

According to the canonical picture of the blazar leptonic emission models, radio-to-optical-UV radiation is dominated by the synchrotron emission, though the big blue bump arising from the accretion disk has also been observed. If the accretion process is radiatively efficient, it will illuminate the BLR and dusty torus enabling a photon-rich environment surrounding the jet. In such objects, the high-energy X- and \gm-ray emission is mainly powered by the external Compton mechanism \citep[see, e.g.,][]{1994ApJ...421..153S}. Due to an additional beaming factor associated with the external Compton process \citep[][]{1995ApJ...446L..63D}, the bolometric jet emission will be dominated by the high-energy \gm-ray emission leading to the observation of a Compton dominated SED \citep[e.g.,][]{2015ApJ...804...74P}. On the other hand, the overall SED of a blazar with an intrinsically low-level of accretion activity (\ld/\ledd~$<0.01$), will be synchrotron dominated. This is because, the \gm-ray emission in such objects originates primarily via synchrotron self Compton mechanism \citep[SSC; cf.][]{2008ApJ...686..181F} since the environment surrounding the jet lacks the external photons needed for the external Compton process. To summarize, one can get a fair idea of the accretion level of a blazar by estimating CD from its multi-frequency SED. Having measured the \ld~in Eddington units for a large sample of blazars, we, therefore, next attempted determining their CD as described below.

\begin{figure}[t!]
\hbox{
\includegraphics[width=\linewidth]{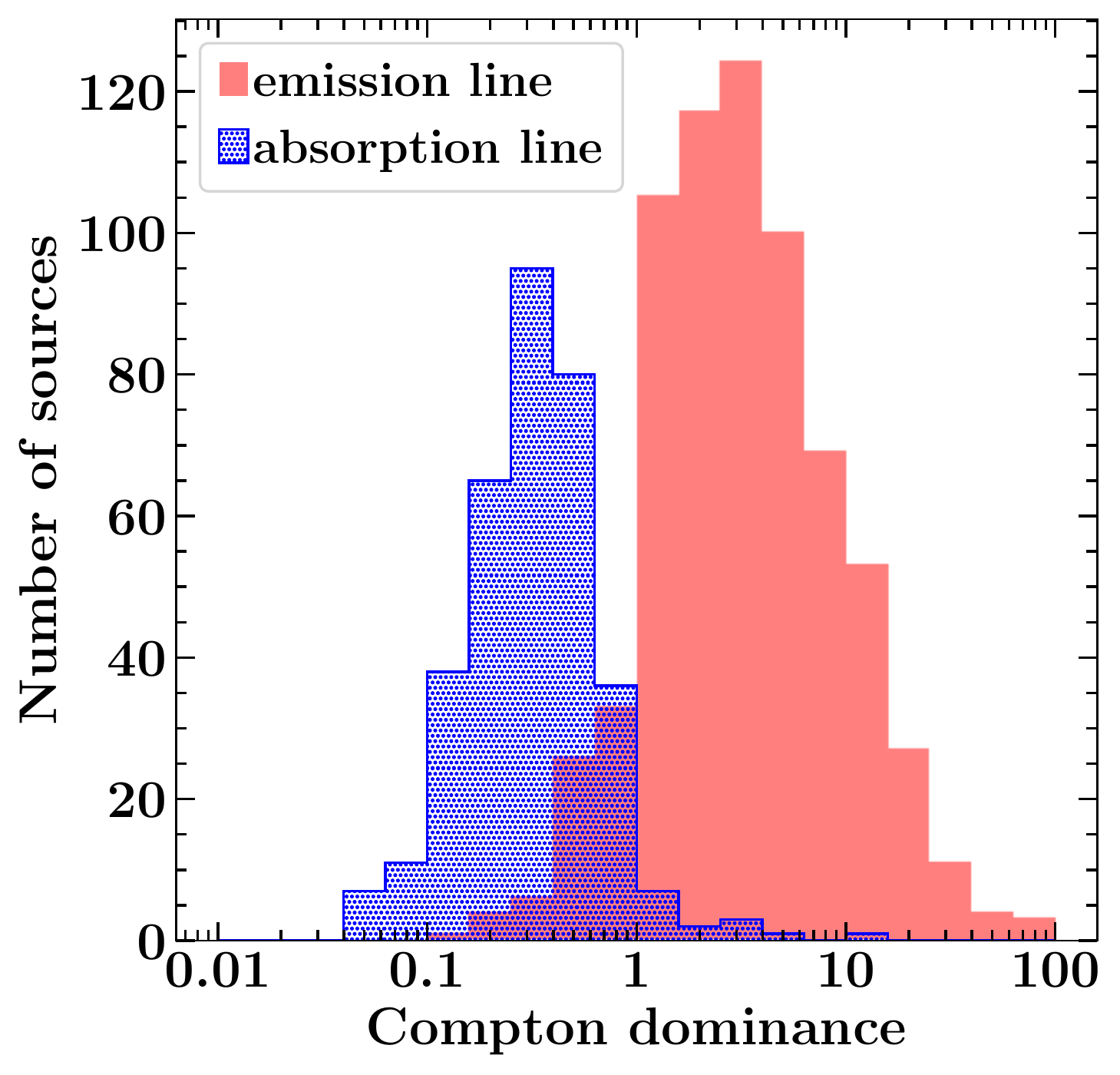} 
}
\caption{The Compton dominance histograms of the emission (red filled) and absorption line (blue dotted) blazars.}
\label{fig:cd_histo}
\end{figure}
\begin{figure*}[t!]
\hbox{
\includegraphics[width=\linewidth]{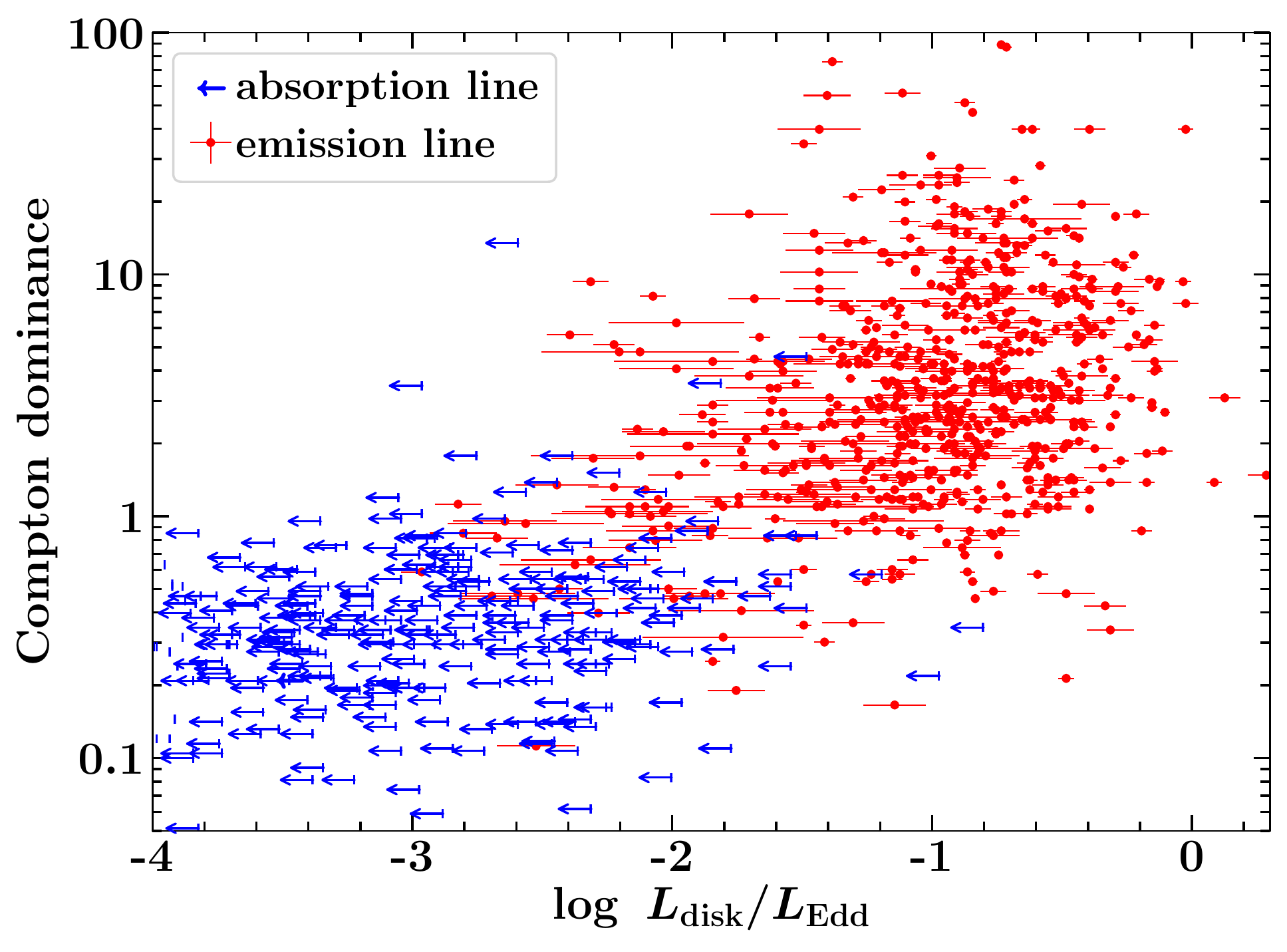} 
}
\caption{The Compton dominance as a function of the \ld~in Eddington units. For the sake of clarity, uncertainties associated with the \mbh~values were not propagated while calculating \ld/\ledd.}\label{fig:cd_ledd}
\end{figure*}
\subsection{Compton Dominance Measurement}

We collected broadband spectral data of blazars present in our sample from Space Science Data Center (SSDC) SED builder tool\footnote{\url{https://tools.ssdc.asi.it/SED/}}. While doing so, we did not consider United States Naval Observatory data and flux values reported by the Catalina Real-Time Transient Survey since their fluxes are often outside the range of the other data taken at the same frequency. To improve the data coverage at X- and \gm-ray bands, we included the flux measurements from the 2$^{\rm nd}$ \swift~X-ray Point Source catalog \citep[2SXPS;][]{2020ApJS..247...54E} and the 4FGL-DR2 catalog \citep[][]{2020ApJS..247...33A}. The synchrotron and inverse Compton peak frequencies and corresponding flux values were then estimated by fitting a second-degree polynomial to both SED peaks using the built-in function provided in the SSDC SED builder tool. In many blazars, the soft X-ray spectrum is still dominated by the synchrotron emission and forms the tail-end of the low-energy peak. In such objects, it may not be possible to accurately constrain the high-energy peak with only \fermi-LAT data. The multi-wavelength SEDs of such blazars, therefore, were fitted with a synchrotron-SSC model assuming a log-parabolic electron energy distribution \citep[][]{2009A&A...501..879T}. Furthermore, the SEDs of powerful FSRQs often exhibit the accretion disk bump at optical-UV energies and many high-frequency peaked BL Lacs show a host galaxy emission at IR-optical wavelengths. Such features were avoided during the fit. The examples of this analysis are illustrated in Figure~\ref{fig:cd}. The CD was derived by taking the ratio of the high- and low-energy peak luminosities, which is equivalent to the ratio of their peak fluxes since it is essentially a redshift independent quantity. The estimated values are provided in Table~\ref{tab:em_mbh_ld} and \ref{tab:abs_mbh_ld} and we plot the CD histograms for both emission and absorption line blazars in Figure~\ref{fig:cd_histo}. As can be seen, a bi-modality appears with the broad emission line blazars have an average CD $>$1 ($\langle {\rm CD_{emi}} \rangle = 5.85$), whereas, absorption line objects have less Compton dominated SED ($\langle {\rm CD_{abs}} \rangle = 0.46$). This finding is in line with the hypothesis of broad line blazars being more Compton dominated sources compared to lineless BL Lac objects as discussed above.

\subsection{Compton Dominance and \ld/\ledd}

In Figure~\ref{fig:cd_ledd}, we show the distribution of CD as a function of \ld~in Eddington units. A positive correlation is apparent with highly accreting objects tend to have larger CD. To quantify the strength of the correlation, we adopted the astronomical survival analysis package \citep[ASURV;][]{1992BAAS...24..839L} which takes into account the upper/lower limits in the data \citep[][]{1986ApJ...306..490I}. The derived Spearman's correlation coefficient is $\rho=0.76$ with probability of no correlation or PNC $<1\times10^{-10}$ indicating a positive correlation with high ($>$5$\sigma$) confidence. This strengthens our argument that CD can be considered as a good proxy for the accretion rate in blazars. Furthermore, a linear regression analysis carried out using ASURV software resulted in the following empirical relation connecting the two variables

\begin{equation}
\log~{\rm CD} = (1.16\pm0.03) + (0.74\pm0.02)\log\left({L_{\rm disk}} \over {L_{\rm Edd}}\right).
\end{equation}

Figure~\ref{fig:cd_ledd} reveals a clear distinction of two groups in regions defined by ${\rm CD} >1$, \ld/\ledd $>$0.01 for broad line blazars and ${\rm CD} <1$, \ld/\ledd$<$0.01 for absorption line systems. The two populations overlap smoothly for intermediate values of CD and \ld~in Eddington units. Based on this finding, we propose that blazars can be classified as high-Compton dominated (HCD) with CD $>$1 and low-Compton dominated (LCD) sources for CD $<$1. Moreover, blazars having ${\rm CD} >1$, \ld/\ledd $>$0.01 should be identified as FSRQs and those with ${\rm CD} <1$, \ld/\ledd$<$0.01 as BL Lacs. One good example is the blazar 4FGL~J0238.6+1637 or AO~0235+164 ($z=0.94$) which is historically classified as a BL Lac object \citep[][]{1975ApJ...201..275S}. For this object, we have found \ld/\ledd~= 0.04 and CD = 2.69, thus indicating an underlying radiatively efficient accretion. Therefore, this source can be identified as a FSRQ and as a HCD blazar \citep[see also][]{2011MNRAS.414.2674G}.

The above mentioned classification scheme is analogous to that based on $L_{\rm BLR}$/\ledd~proposed by \citet[][]{2011MNRAS.414.2674G}. However, it overcomes the limitation imposed by the difficulty in measuring $L_{\rm BLR}$ and \mbh~for objects with quasi-featureless spectra. Another advantage of a CD based classification is that one can easily get an idea about the intrinsic physical nature of blazars just by measuring the relative dominance of the high-energy peak in their multi-frequency SEDs. The proposed scheme is useful keeping in mind the ongoing and next-generation surveys, e.g., Very Large Array Sky Survey, Dark Energy Survey, {\it e-ROSITA}, All-sky MeV Energy Gamma-ray Observatory, and Cherenkov Telescope Array, which will provide unprecedented broadband coverage of the blazar SEDs.

Based on Figure~\ref{fig:cd_ledd}, we argue that the region of  ${\rm CD} \lesssim1$, \ld/\ledd $>$0.01 could be populated with strong line AGN having misaligned jets, e.g., broad line radio galaxies. This is due to the fact that the external Compton mechanism producing X- to \gm-ray emission is highly sensitive to the Doppler boosting. Therefore, as the viewing angle increases, the decrease in the high-energy peak flux is much more rapid than the synchrotron peak flux, thereby effectively reducing CD. However, since the BLR emission is likely to be isotropic, the \ld~of the source does not change. At the same time, the region of ${\rm CD} \gtrsim1$, \ld/\ledd $<$0.01 may remain largely forbidden due to low-level of accretion and hence a jet environment starved of the seed photons needed for the external Compton process. 

Recently, a minority of sources, so-called masquerading BL Lacs, have been identified challenging our current understanding of the FSRQ/BL Lac division. It has been proposed that these objects are intrinsically FSRQs, i.e., radiatively efficient accreting systems. However, since their optical spectra lack emission lines due to strongly amplified jetted radiation contamination, they are classified as BL Lac sources \citep[][]{2013MNRAS.431.1914G,2019MNRAS.484L.104P}. Furthermore, many of such objects exhibit high-frequency peaked SEDs \citep[][]{2012MNRAS.422L..48P}. This was explained by arguing the \gm-ray emitting region to be located outside BLR where a relatively weak cooling environment allows jet electrons to attain very high-energies \citep[][]{2012MNRAS.425.1371G}. In the context of our work, physics of such blue FSRQs can be explained as follows: the Doppler boosting effect that swamps out broad emission lines by enhancing synchrotron radiation should amplify the inverse Compton emission by an even larger factor if the primary mechanism to produce \gm-rays is external Compton. If so, the ratio of the SED peak luminosities, i.e., CD, cannot be significantly lower than one \citep[see, e.g.,][]{2012MNRAS.425.1371G}. This supports our claim that radiatively efficient accreting systems should have a Compton dominated SED. Based on Figure~\ref{fig:cd_ledd}, it can be understood that sources lying around CD $\approx$1 and \ld/\ledd~$\approx$0.01 may belong to the family of masquerading BL Lacs.

We stress that our physical interpretation of the obtained CD and \ld/\ledd~correlation may not be extendable to blazar flares which can be extremely complex and diverse \citep[e.g.,][]{2020MNRAS.498.5128R}.  This is because we used multi-wavelength SEDs generated using all archival observations to compute CD hence reflect the average activity state of sources. Similarly, optical spectra too were mostly taken during average/low activity state of sources and may not be during any specific flaring episodes. Therefore, the derived correlation of CD and \ld/\ledd~refers to the average physical properties of the blazar population. Extending this work to consider blazar flares is beyond the scope of this paper since even one source behaves differently during different flares \citep[e.g.,][]{2013ApJ...763L..11C,2015ApJ...803...15P,2016ApJ...817...61P}.

\section{A Sequence of \ld/\ledd}\label{sec:sequence}
The observed anti-correlation between the synchrotron peak luminosity and corresponding peak frequency, i.e., blazar sequence, has remained one of the crucial topics of research in jet physics. This phenomenon has been argued to have a physical origin \citep[e.g.,][]{1998MNRAS.301..451G} and also criticized due to possible selection effects of missing high-frequency peaked, luminous blazars \citep[cf.][]{2007Ap&SS.309...63P}. A few other works have proposed the Doppler boosting/viewing angle to be the driving factor of the observed sequence \citep[e.g.,][]{2008A&A...488..867N,2011ApJ...740...98M,2017ApJ...835L..38F,2020arXiv200712661K}. Recently, the identification of the first $z>3$ BL Lac object \citep[][]{2020ApJ...903L...8P} has also suggested that a population of luminous, high-frequency peaked blazars may exist which is against the prediction of the blazar sequence \citep[][]{2017MNRAS.469..255G}.

\begin{figure}[t!]
\hbox{
\includegraphics[width=\linewidth]{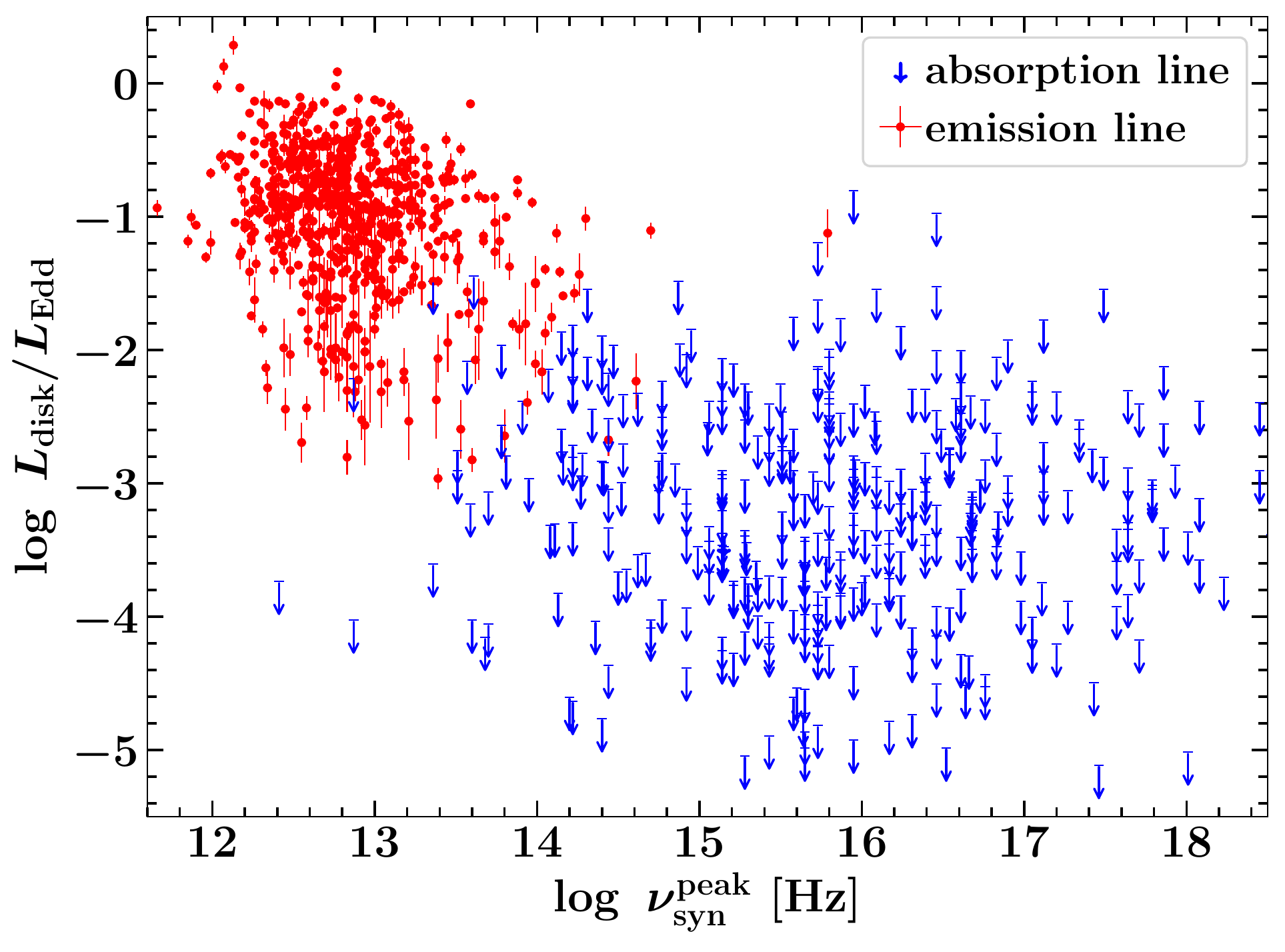} 
}
\caption{This diagram illustrates the variation of the \ld~in Eddington units as a function of the synchrotron peak frequency for \fermi~blazars studied in this work.}\label{fig:nu_ledd}
\end{figure}

In Figure~\ref{fig:nu_ledd}, we show the variation of the \mbh~normalized \ld~as a function of the synchrotron peak frequency for objects studied in this work. Note that \ld~values are derived from the optical spectroscopic emission line fluxes or upper limits and therefore are free from the Doppler boosting effects. In other words, Figure~\ref{fig:nu_ledd} provides a glimpse of the intrinsic physical behavior of beamed AGN population. A strong anti-correlation of the plotted quantities is evident which is confirmed from the Spearmann's test done using ASURV giving $\rho=-0.75$ with PNC $<1\times10^{-10}$. This anti-correlation can be understood as follows. 

Population studies focused on the average activity state of blazars have revealed the blazar-zone to lie at a distance of a few hundreds/thousands of Schwarzschild radii from the central black hole \citep[e.g.,][]{2017ApJ...851...33P}. Considering the BLR/torus radius-\ld~relationship \citep[e.g.,][]{2008MNRAS.386..945T}, strong line blazars hosting a luminous accretion disk have a large BLR/torus and therefore the emission region is typically located within it. The relativistic electrons present in the jet would loose energy mainly by interacting with ambient photons before they could reach to high-energies, leading to the observation of a low-frequency peaked SED. In radiatively inefficient systems, on the other hand, \ld~has a low-value implying a relatively small BLR/torus radius and hence the emission region would be located farther out from it. In the absence of a strong photon field, jet electrons can attain high-energies making the observed SED to be high-frequency peaked. Moreover, Figure~\ref{fig:nu_ledd} also suggests that the shift of the SED peaks to higher frequencies occurs with the depletion of the central environment by accretion, i.e., with the evolution of HCD blazars to their LCD counterparts \citep[][]{2002ApJ...564...86B,2002ApJ...571..226C}. 

Finally, since \ld/\ledd~positively correlates with CD (Figure~\ref{fig:cd_ledd}), one would also expect an anti-correlation of the latter with the synchrotron peak frequency similar to Figure~\ref{fig:nu_ledd}. Such an anti-correlation has already been reported by \citet[][]{2013ApJ...763..134F}.

\section{Summary}\label{sec:summary}

In this work, we have presented a catalog of the central engine properties, i.e., \mbh~and \ld, for a sample of 1077 \gm-ray detected blazars. We summarize our results below.

\begin{enumerate}
\item The average \mbh~for the whole blazar population is $\langle \log~M_{{\rm BH,all}~\msun} \rangle=8.60$. There are evidences indicating black holes residing in absorption line systems to be more massive compared to emission line blazars.

\item The distribution of \ld/\ledd~reveals a clear bi-modality with emission line sources tend to have a larger accretion rate (\ld/\ledd$>$0.01) compared to blazars whose optical spectra are dominated by the absorption lines arising from the host galaxy. Similar results were obtained for CD where emission line blazars exhibit a more Compton dominated SED than absorption line sources.

\item There is strong positive correlation between \ld/\ledd~and CD, suggesting that the latter can be used to determine the state of accretion activity in beamed AGNs.

\item Based on their position in \ld/\ledd-CD plane, blazars can be classified as High-Compton Dominated (HCD, CD$>$1) and Low-Compton Dominated (LCD, CD$<$1) objects. Our results suggests that this scheme is analogous to that predicted based on $L_{\rm BLR}$/\ledd, however, can be extended to objects lacking high-quality optical spectroscopic measurements. We also propose that blazars having ${\rm CD} >1$, \ld/\ledd$>$0.01 should be identified as FSRQs and ${\rm CD} <1$, \ld/\ledd$<$0.01 as BL Lacs.

\item \fermi~blazars show a significant anti-correlation between \ld/\ledd~and synchrotron peak frequency. Being free from Doppler boosting related effects, the observed trend has a physical origin.

\item The overall findings reported in this work directly points to a scenario in which the average physical properties of blazars are mainly controlled by the accretion rate in Eddington units.
\end{enumerate}

We make the catalog publicly available online at Zenodo (doi: 10.5281/zenodo.4456365)\footnote{The fits file is also available at \url{http://www.ucm.es/blazars/engines}}.

\software{ASURV \citep[][]{1992BAAS...24..839L}, Astropy \citep[][]{2013A&A...558A..33A,2018AJ....156..123A}, pPXF \citep[][]{2004PASP..116..138C}, PyQSOFit \citep[][]{2018ascl.soft09008G}.}

\bibliographystyle{aasjournal}
\bibliography{Master}

\begin{thebibliography}{}
\expandafter\ifx\csname natexlab\endcsname\relax\def\natexlab#1{#1}\fi
\providecommand{\url}[1]{\href{#1}{#1}}
\providecommand{\dodoi}[1]{doi:~\href{http://doi.org/#1}{\nolinkurl{#1}}}
\providecommand{\doeprint}[1]{\href{http://ascl.net/#1}{\nolinkurl{http://ascl.net/#1}}}
\providecommand{\doarXiv}[1]{\href{https://arxiv.org/abs/#1}{\nolinkurl{https://arxiv.org/abs/#1}}}

\bibitem[{{Abdollahi} {et~al.}(2020){Abdollahi}, {Acero}, {Ackermann},
  {Ajello}, {Atwood}, {Axelsson}, {Baldini}, {Ballet}, {Barbiellini},
  {Bastieri}, {Becerra Gonzalez}, {Bellazzini}, {Berretta}, {Bissaldi},
  {Blandford}, {Bloom}, {Bonino}, {Bottacini}, {Brandt}, {Bregeon}, {Bruel},
  {Buehler}, {Burnett}, {Buson}, {Cameron}, {Caputo}, {Caraveo}, {Casandjian},
  {Castro}, {Cavazzuti}, {Charles}, {Chaty}, {Chen}, {Cheung}, {Chiaro},
  {Ciprini}, {Cohen-Tanugi}, {Cominsky}, {Coronado-Bl{\'a}zquez}, {Costantin},
  {Cuoco}, {Cutini}, {D'Ammando}, {DeKlotz}, {de la Torre Luque}, {de Palma},
  {Desai}, {Digel}, {Di Lalla}, {Di Mauro}, {Di Venere}, {Dom{\'\i}nguez},
  {Dumora}, {Fana Dirirsa}, {Fegan}, {Ferrara}, {Franckowiak}, {Fukazawa},
  {Funk}, {Fusco}, {Gargano}, {Gasparrini}, {Giglietto}, {Giommi}, {Giordano},
  {Giroletti}, {Glanzman}, {Green}, {Grenier}, {Griffin}, {Grondin}, {Grove},
  {Guiriec}, {Harding}, {Hayashi}, {Hays}, {Hewitt}, {Horan},
  {J{\'o}hannesson}, {Johnson}, {Kamae}, {Kerr}, {Kocevski}, {Kovac'evic'},
  {Kuss}, {Landriu}, {Larsson}, {Latronico}, {Lemoine-Goumard}, {Li},
  {Liodakis}, {Longo}, {Loparco}, {Lott}, {Lovellette}, {Lubrano}, {Madejski},
  {Maldera}, {Malyshev}, {Manfreda}, {Marchesini}, {Marcotulli},
  {Mart{\'\i}-Devesa}, {Martin}, {Massaro}, {Mazziotta}, {McEnery}, {Mereu},
  {Meyer}, {Michelson}, {Mirabal}, {Mizuno}, {Monzani}, {Morselli},
  {Moskalenko}, {Negro}, {Nuss}, {Ojha}, {Omodei}, {Orienti}, {Orlando},
  {Ormes}, {Palatiello}, {Paliya}, {Paneque}, {Pei}, {Pe{\~n}a-Herazo},
  {Perkins}, {Persic}, {Pesce-Rollins}, {Petrosian}, {Petrov}, {Piron}, {Poon},
  {Porter}, {Principe}, {Rain{\`o}}, {Rando}, {Razzano}, {Razzaque}, {Reimer},
  {Reimer}, {Remy}, {Reposeur}, {Romani}, {Saz Parkinson}, {Schinzel},
  {Serini}, {Sgr{\`o}}, {Siskind}, {Smith}, {Spandre}, {Spinelli}, {Strong},
  {Suson}, {Tajima}, {Takahashi}, {Tak}, {Thayer}, {Thompson}, {Tibaldo},
  {Torres}, {Torresi}, {Valverde}, {Van Klaveren}, {van Zyl}, {Wood},
  {Yassine}, \& {Zaharijas}}]{2020ApJS..247...33A}
{Abdollahi}, S., {Acero}, F., {Ackermann}, M., {et~al.} 2020, \apjs, 247, 33,
  \dodoi{10.3847/1538-4365/ab6bcb}

\bibitem[{{Acosta-Pulido} {et~al.}(2010){Acosta-Pulido}, {Agudo}, {Barrena},
  {Ramos Almeida}, {Manchado}, \& {Rodr{\'\i}guez-Gil}}]{2010ANA...519A...5A}
{Acosta-Pulido}, J.~A., {Agudo}, I., {Barrena}, R., {et~al.} 2010, \aap, 519,
  A5, \dodoi{10.1051/0004-6361/200913953}

\bibitem[{{Afanas'Ev} {et~al.}(2005){Afanas'Ev}, {Dodonov}, {Moiseev},
  {Gorshkov}, {Konnikova}, \& {Mingaliev}}]{2005ARep...49..374A}
{Afanas'Ev}, V.~L., {Dodonov}, S.~N., {Moiseev}, A.~V., {et~al.} 2005,
  Astronomy Reports, 49, 374, \dodoi{10.1134/1.1923547}

\bibitem[{{Afanas'Ev} {et~al.}(2006){Afanas'Ev}, {Dodonov}, {Moiseev},
  {Gorshkov}, {Konnikova}, \& {Mingaliev}}]{2006ARep...50..255A}
---. 2006, Astronomy Reports, 50, 255, \dodoi{10.1134/S1063772906040019}

\bibitem[{{Ahumada} {et~al.}(2020){Ahumada}, {Allende Prieto}, {Almeida},
  {Anders}, {Anderson}, {Andrews}, {Anguiano}, {Arcodia}, {Armengaud},
  {Aubert}, {Avila}, {Avila-Reese}, {Badenes}, {Balland }, {Barger},
  {Barrera-Ballesteros}, {Basu}, {Bautista}, {Beaton}, {Beers}, {Benavides},
  {Bender}, {Bernardi}, {Bershady}, {Beutler}, {Bidin}, {Bird}, {Bizyaev},
  {Blanc}, {Blanton}, {Boquien}, {Borissova}, {Bovy}, {Brand t}, {Brinkmann},
  {Brownstein}, {Bundy}, {Bureau}, {Burgasser}, {Burtin}, {Cano-D{\'\i}az},
  {Capasso}, {Cappellari}, {Carrera}, {Chabanier}, {Chaplin}, {Chapman},
  {Cherinka}, {Chiappini}, {Doohyun Choi}, {Chojnowski}, {Chung}, {Clerc},
  {Coffey}, {Comerford}, {Comparat}, {da Costa}, {Cousinou}, {Covey}, {Crane},
  {Cunha}, {da Silva Ilha}, {Dai}, {Damsted}, {Darling}, {Davidson}, {Davies},
  {Dawson}, {De}, {de la Macorra}, {De Lee}, {de Andrade Queiroz}, {Deconto
  Machado}, {de la Torre}, {Dell'Agli}, {du Mas des Bourboux},
  {Diamond-Stanic}, {Dillon}, {Donor}, {Drory}, {Duckworth}, {Dwelly},
  {Ebelke}, {Eftekharzadeh}, {Eigenbrot}, {Elsworth}, {Eracleous},
  {Erfanianfar}, {Escoffier}, {Fan}, {Farr}, {Fern{\'a}ndez-Trincado},
  {Feuillet}, {Finoguenov}, {Fofie}, {Fraser-McKelvie}, {Frinchaboy},
  {Fromenteau}, {Fu}, {Galbany}, {Garcia}, {Garc{\'\i}a-Hern{\'a}ndez}, {Garma
  Oehmichen}, {Ge}, {Geimba Maia}, {Geisler}, {Gelfand }, {Goddy},
  {Gonzalez-Perez}, {Grabowski}, {Green}, {Grier}, {Guo}, {Guy}, {Harding},
  {Hasselquist}, {Hawken}, {Hayes}, {Hearty}, {Hekker}, {Hogg}, {Holtzman},
  {Horta}, {Hou}, {Hsieh}, {Huber}, {Hunt}, {Ider Chitham}, {Imig}, {Jaber},
  {Jimenez Angel}, {Johnson}, {Jones}, {J{\"o}nsson}, {Jullo}, {Kim},
  {Kinemuchi}, {Kirkpatrick}, {Kite}, {Klaene}, {Kneib}, {Kollmeier}, {Kong},
  {Kounkel}, {Krishnarao}, {Lacerna}, {Lan}, {Lane}, {Law}, {Le Goff}, {Leung},
  {Lewis}, {Li}, {Lian}, {Lin}, {Long}, {Longa-Pe{\~n}a}, {Lundgren}, {Lyke},
  {Ted Mackereth}, {MacLeod}, {Majewski}, {Manchado}, {Maraston}, {Martini},
  {Masseron}, {Masters}, {Mathur}, {McDermid}, {Merloni}, {Merrifield},
  {M{\'e}sz{\'a}ros}, {Miglio}, {Minniti}, {Minsley}, {Miyaji}, {Mohammad},
  {Mosser}, {Mueller}, {Muna}, {Mu{\~n}oz-Guti{\'e}rrez}, {Myers}, {Nadathur},
  {Nair}, {Nandra}, {do Nascimento}, {Nevin}, {Newman}, {Nidever}, {Nitschelm},
  {Noterdaeme}, {O'Connell}, {Olmstead}, {Oravetz}, {Oravetz}, {Osorio},
  {Pace}, {Padilla}, {Palanque-Delabrouille}, {Palicio}, {Pan}, {Pan},
  {Parker}, {Paviot}, {Peirani}, {Pe{\~n}a Ram{\'r}ez}, {Penny}, {Percival},
  {Perez-Fournon}, {P{\'e}rez-R{\`a}fols}, {Petitjean}, {Pieri},
  {Pinsonneault}, {Poovelil}, {Povick}, {Prakash}, {Price-Whelan}, {Raddick},
  {Raichoor}, {Ray}, {Rembold}, {Rezaie}, {Riffel}, {Riffel}, {Rix}, {Robin},
  {Roman-Lopes}, {Rom{\'a}n-Z{\'u}{\~n}iga}, {Rose}, {Ross}, {Rossi}, {Rowland
  s}, {Rubin}, {Salvato}, {S{\'a}nchez}, {S{\'a}nchez-Menguiano},
  {S{\'a}nchez-Gallego}, {Sayres}, {Schaefer}, {Schiavon}, {Schimoia},
  {Schlafly}, {Schlegel}, {Schneider}, {Schultheis}, {Schwope}, {Seo},
  {Serenelli}, {Shafieloo}, {Shamsi}, {Shao}, {Shen}, {Shetrone}, {Shirley},
  {Silva Aguirre}, {Simon}, {Skrutskie}, {Slosar}, {Smethurst}, {Sobeck},
  {Sodi}, {Souto}, {Stark}, {Stassun}, {Steinmetz}, {Stello}, {Stermer},
  {Storchi-Bergmann}, {Streblyanska}, {Stringfellow}, {Stutz}, {Su{\'a}rez},
  {Sun}, {Taghizadeh-Popp}, {Talbot}, {Tayar}, {Thakar}, {Theriault}, {Thomas},
  {Thomas}, {Tinker}, {Tojeiro}, {Toledo}, {Tremonti}, {Troup}, {Tuttle},
  {Unda-Sanzana}, {Valentini}, {Vargas-Gonz{\'a}lez}, {Vargas-Maga{\~n}a},
  {V{\'a}zquez-Mata}, {Vivek}, {Wake}, {Wang}, {Weaver}, {Weijmans}, {Wild},
  {Wilson}, {Wilson}, {Wolthuis}, {Wood-Vasey}, {Yan}, {Yang}, {Y{\`e}che},
  {Zamora}, {Zarrouk}, {Zasowski}, {Zhang}, {Zhao}, {Zhao}, {Zheng}, {Zheng},
  {Zhu}, \& {Zou}}]{2020ApJS..249....3A}
{Ahumada}, R., {Allende Prieto}, C., {Almeida}, A., {et~al.} 2020, \apjs, 249,
  3, \dodoi{10.3847/1538-4365/ab929e}

\bibitem[{{Ajello} {et~al.}(2014){Ajello}, {Romani}, {Gasparrini}, {Shaw},
  {Bolmer}, {Cotter}, {Finke}, {Greiner}, {Healey}, {King}, {Max-Moerbeck},
  {Michelson}, {Potter}, {Rau}, {Readhead}, {Richards}, \&
  {Schady}}]{2014ApJ...780...73A}
{Ajello}, M., {Romani}, R.~W., {Gasparrini}, D., {et~al.} 2014, \apj, 780, 73,
  \dodoi{10.1088/0004-637X/780/1/73}

\bibitem[{{Ajello} {et~al.}(2016){Ajello}, {Ghisellini}, {Paliya}, {Kocevski},
  {Tagliaferri}, {Madejski}, {Rau}, {Schady}, {Greiner}, {Massaro},
  {Balokovi{\'c}}, {B{\"u}hler}, {Giomi}, {Marcotulli}, {D'Ammando}, {Stern},
  {Boggs}, {Christensen}, {Craig}, {Hailey}, {Harrison}, \&
  {Zhang}}]{2016ApJ...826...76A}
{Ajello}, M., {Ghisellini}, G., {Paliya}, V.~S., {et~al.} 2016, \apj, 826, 76,
  \dodoi{10.3847/0004-637X/826/1/76}

\bibitem[{{Ajello} {et~al.}(2020){Ajello}, {Angioni}, {Axelsson}, {Ballet},
  {Barbiellini}, {Bastieri}, {Becerra Gonzalez}, {Bellazzini}, {Bissaldi},
  {Bloom}, {Bonino}, {Bottacini}, {Bruel}, {Buson}, {Cafardo}, {Cameron},
  {Cavazzuti}, {Chen}, {Cheung}, {Ciprini}, {Costantin}, {Cutini}, {D'Ammando},
  {de la Torre Luque}, {de Menezes}, {de Palma}, {Desai}, {Di Lalla}, {Di
  Venere}, {Dom{\'\i}nguez}, {Dirirsa}, {Ferrara}, {Finke}, {Franckowiak},
  {Fukazawa}, {Funk}, {Fusco}, {Gargano}, {Garrappa}, {Gasparrini},
  {Giglietto}, {Giordano}, {Giroletti}, {Green}, {Grenier}, {Guiriec},
  {Harita}, {Hays}, {Horan}, {Itoh}, {J{\'o}hannesson}, {Kovac'evic'},
  {Krauss}, {Kreter}, {Kuss}, {Larsson}, {Leto}, {Li}, {Liodakis}, {Longo},
  {Loparco}, {Lott}, {Lovellette}, {Lubrano}, {Madejski}, {Maldera},
  {Manfreda}, {Mart{\'\i}-Devesa}, {Massaro}, {Mazziotta}, {Mereu}, {Meyer},
  {Migliori}, {Mirabal}, {Mizuno}, {Monzani}, {Morselli}, {Moskalenko},
  {Negro}, {Nemmen}, {Nuss}, {Ojha}, {Ojha}, {Omodei}, {Orienti}, {Orlando},
  {Ormes}, {Paliya}, {Pei}, {Pe{\~n}a-Herazo}, {Persic}, {Pesce-Rollins},
  {Petrov}, {Piron}, {Poon}, {Principe}, {Rain{\`o}}, {Rando}, {Rani},
  {Razzano}, {Razzaque}, {Reimer}, {Reimer}, {Schinzel}, {Serini}, {Sgr{\`o}},
  {Siskind}, {Spandre}, {Spinelli}, {Suson}, {Tachibana}, {Thompson}, {Torres},
  {Torresi}, {Troja}, {Valverde}, {van Zyl}, \&
  {Yassine}}]{2020ApJ...892..105A}
{Ajello}, M., {Angioni}, R., {Axelsson}, M., {et~al.} 2020, \apj, 892, 105,
  \dodoi{10.3847/1538-4357/ab791e}

\bibitem[{{Aliu} {et~al.}(2011){Aliu}, {Aune}, {Beilicke}, {Benbow},
  {B{\"o}ttcher}, {Bouvier}, {Bradbury}, {Buckley}, {Bugaev}, {Cannon},
  {Cesarini}, {Ciupik}, {Connolly}, {Cui}, {Decerprit}, {Dickherber}, {Duke},
  {Errando}, {Falcone}, {Feng}, {Finnegan}, {Fortson}, {Furniss}, {Galante},
  {Gall}, {Gillanders}, {Godambe}, {Griffin}, {Grube}, {Gyuk}, {Hanna},
  {Hivick}, {Holder}, {Huan}, {Hughes}, {Hui}, {Humensky}, {Kaaret},
  {Karlsson}, {Kertzman}, {Kieda}, {Krawczynski}, {Krennrich}, {Maier},
  {Majumdar}, {McArthur}, {McCann}, {Moriarty}, {Mukherjee}, {Nelson}, {Ong},
  {Orr}, {Otte}, {Park}, {Perkins}, {Pichel}, {Pohl}, {Prokoph}, {Quinn},
  {Ragan}, {Reyes}, {Reynolds}, {Roache}, {Rose}, {Ruppel}, {Saxon},
  {Sembroski}, {Skole}, {Smith}, {Staszak}, {Te{\v{s}}i{\'c}}, {Theiling},
  {Thibadeau}, {Tsurusaki}, {Tyler}, {Varlotta}, {Vassiliev}, {Wakely},
  {Weekes}, {Weinstein}, {Williams}, {Zitzer}, {VERITAS Collaboration},
  {Ciprini}, {Fumagalli}, {Kaplan}, {Paneque}, \&
  {Prochaska}}]{2011ApJ...742..127A}
{Aliu}, E., {Aune}, T., {Beilicke}, M., {et~al.} 2011, \apj, 742, 127,
  \dodoi{10.1088/0004-637X/742/2/127}

\bibitem[{{{\'A}lvarez Crespo} {et~al.}(2016{\natexlab{a}}){{\'A}lvarez
  Crespo}, {Massaro}, {Milisavljevic}, {Landoni}, {Chavushyan},
  {Pati{\~n}o-{\'A}lvarez}, {Masetti}, {Jim{\'e}nez-Bail{\'o}n}, {Strader},
  {Chomiuk}, {Katagiri}, {Kagaya}, {Cheung}, {Paggi}, {D'Abrusco}, {Ricci}, {La
  Franca}, {Smith}, \& {Tosti}}]{2016AJ....151...95A}
{{\'A}lvarez Crespo}, N., {Massaro}, F., {Milisavljevic}, D., {et~al.}
  2016{\natexlab{a}}, \aj, 151, 95, \dodoi{10.3847/0004-6256/151/4/95}

\bibitem[{{{\'A}lvarez Crespo} {et~al.}(2016{\natexlab{b}}){{\'A}lvarez
  Crespo}, {Masetti}, {Ricci}, {Land oni}, {Pati{\~n}o-{\'A}lvarez}, {Massaro},
  {D'Abrusco}, {Paggi}, {Chavushyan}, {Jim{\'e}nez-Bail{\'o}n}, {Torrealba},
  {Latronico}, {La Franca}, {Smith}, \& {Tosti}}]{2016AJ....151...32A}
{{\'A}lvarez Crespo}, N., {Masetti}, N., {Ricci}, F., {et~al.}
  2016{\natexlab{b}}, \aj, 151, 32, \dodoi{10.3847/0004-6256/151/2/32}

\bibitem[{{Astropy Collaboration} {et~al.}(2013){Astropy Collaboration},
  {Robitaille}, {Tollerud}, {Greenfield}, {Droettboom}, {Bray}, {Aldcroft},
  {Davis}, {Ginsburg}, {Price-Whelan}, {Kerzendorf}, {Conley}, {Crighton},
  {Barbary}, {Muna}, {Ferguson}, {Grollier}, {Parikh}, {Nair}, {Unther},
  {Deil}, {Woillez}, {Conseil}, {Kramer}, {Turner}, {Singer}, {Fox}, {Weaver},
  {Zabalza}, {Edwards}, {Azalee Bostroem}, {Burke}, {Casey}, {Crawford},
  {Dencheva}, {Ely}, {Jenness}, {Labrie}, {Lim}, {Pierfederici}, {Pontzen},
  {Ptak}, {Refsdal}, {Servillat}, \& {Streicher}}]{2013A&A...558A..33A}
{Astropy Collaboration}, {Robitaille}, T.~P., {Tollerud}, E.~J., {et~al.} 2013,
  \aap, 558, A33, \dodoi{10.1051/0004-6361/201322068}

\bibitem[{{Astropy Collaboration} {et~al.}(2018){Astropy Collaboration},
  {Price-Whelan}, {Sip{\H{o}}cz}, {G{\"u}nther}, {Lim}, {Crawford}, {Conseil},
  {Shupe}, {Craig}, {Dencheva}, {Ginsburg}, {Vand erPlas}, {Bradley},
  {P{\'e}rez-Su{\'a}rez}, {de Val-Borro}, {Aldcroft}, {Cruz}, {Robitaille},
  {Tollerud}, {Ardelean}, {Babej}, {Bach}, {Bachetti}, {Bakanov}, {Bamford},
  {Barentsen}, {Barmby}, {Baumbach}, {Berry}, {Biscani}, {Boquien}, {Bostroem},
  {Bouma}, {Brammer}, {Bray}, {Breytenbach}, {Buddelmeijer}, {Burke},
  {Calderone}, {Cano Rodr{\'\i}guez}, {Cara}, {Cardoso}, {Cheedella}, {Copin},
  {Corrales}, {Crichton}, {D'Avella}, {Deil}, {Depagne}, {Dietrich}, {Donath},
  {Droettboom}, {Earl}, {Erben}, {Fabbro}, {Ferreira}, {Finethy}, {Fox},
  {Garrison}, {Gibbons}, {Goldstein}, {Gommers}, {Greco}, {Greenfield},
  {Groener}, {Grollier}, {Hagen}, {Hirst}, {Homeier}, {Horton}, {Hosseinzadeh},
  {Hu}, {Hunkeler}, {Ivezi{\'c}}, {Jain}, {Jenness}, {Kanarek}, {Kendrew},
  {Kern}, {Kerzendorf}, {Khvalko}, {King}, {Kirkby}, {Kulkarni}, {Kumar},
  {Lee}, {Lenz}, {Littlefair}, {Ma}, {Macleod}, {Mastropietro}, {McCully},
  {Montagnac}, {Morris}, {Mueller}, {Mumford}, {Muna}, {Murphy}, {Nelson},
  {Nguyen}, {Ninan}, {N{\"o}the}, {Ogaz}, {Oh}, {Parejko}, {Parley}, {Pascual},
  {Patil}, {Patil}, {Plunkett}, {Prochaska}, {Rastogi}, {Reddy Janga},
  {Sabater}, {Sakurikar}, {Seifert}, {Sherbert}, {Sherwood-Taylor}, {Shih},
  {Sick}, {Silbiger}, {Singanamalla}, {Singer}, {Sladen}, {Sooley},
  {Sornarajah}, {Streicher}, {Teuben}, {Thomas}, {Tremblay}, {Turner},
  {Terr{\'o}n}, {van Kerkwijk}, {de la Vega}, {Watkins}, {Weaver}, {Whitmore},
  {Woillez}, {Zabalza}, \& {Astropy Contributors}}]{2018AJ....156..123A}
{Astropy Collaboration}, {Price-Whelan}, A.~M., {Sip{\H{o}}cz}, B.~M., {et~al.}
  2018, \aj, 156, 123, \dodoi{10.3847/1538-3881/aabc4f}

\bibitem[{{Bade} {et~al.}(1998){Bade}, {Beckmann}, {Douglas}, {Barthel},
  {Engels}, {Cordis}, {Nass}, \& {Voges}}]{1998ANA...334..459B}
{Bade}, N., {Beckmann}, V., {Douglas}, N.~G., {et~al.} 1998, \aap, 334, 459.
\newblock \doarXiv{astro-ph/9803204}

\bibitem[{{Bade} {et~al.}(1995){Bade}, {Fink}, {Engels}, {Voges}, {Hagen},
  {Wisotzki}, \& {Reimers}}]{1995ANAS..110..469B}
{Bade}, N., {Fink}, H.~H., {Engels}, D., {et~al.} 1995, \aaps, 110, 469

\bibitem[{{Baker} {et~al.}(1999){Baker}, {Hunstead}, {Kapahi}, \&
  {Subrahmanya}}]{1999ApJS..122...29B}
{Baker}, J.~C., {Hunstead}, R.~W., {Kapahi}, V.~K., \& {Subrahmanya}, C.~R.
  1999, \apjs, 122, 29, \dodoi{10.1086/313209}

\bibitem[{{Baldwin} {et~al.}(1981){Baldwin}, {Wampler}, \&
  {Burbidge}}]{1981ApJ...243...76B}
{Baldwin}, J.~A., {Wampler}, E.~J., \& {Burbidge}, E.~M. 1981, \apj, 243, 76,
  \dodoi{10.1086/158568}

\bibitem[{{Baldwin} {et~al.}(1989){Baldwin}, {Wampler}, \&
  {Gaskell}}]{1989ApJ...338..630B}
{Baldwin}, J.~A., {Wampler}, E.~J., \& {Gaskell}, C.~M. 1989, \apj, 338, 630,
  \dodoi{10.1086/167224}

\bibitem[{{Becerra Gonz{\'a}lez} {et~al.}(2020){Becerra Gonz{\'a}lez},
  {Acosta-Pulido}, \& {Clavero}}]{2020MNRAS.494.6036B}
{Becerra Gonz{\'a}lez}, J., {Acosta-Pulido}, J.~A., \& {Clavero}, R. 2020,
  \mnras, 494, 6036, \dodoi{10.1093/mnras/staa1144}

\bibitem[{{Bechtold} {et~al.}(2002){Bechtold}, {Dobrzycki}, {Wilden}, {Morita},
  {Scott}, {Dobrzycka}, {Tran}, \& {Aldcroft}}]{2002ApJS..140..143B}
{Bechtold}, J., {Dobrzycki}, A., {Wilden}, B., {et~al.} 2002, \apjs, 140, 143,
  \dodoi{10.1086/342489}

\bibitem[{{Boisse} \& {Bergeron}(1988)}]{1988ANA...192....1B}
{Boisse}, P., \& {Bergeron}, J. 1988, \aap, 192, 1

\bibitem[{{Boroson} \& {Green}(1992)}]{1992ApJS...80..109B}
{Boroson}, T.~A., \& {Green}, R.~F. 1992, \apjs, 80, 109,
  \dodoi{10.1086/191661}

\bibitem[{{B{\"o}ttcher} \& {Dermer}(2002)}]{2002ApJ...564...86B}
{B{\"o}ttcher}, M., \& {Dermer}, C.~D. 2002, \apj, 564, 86,
  \dodoi{10.1086/324134}

\bibitem[{{Bruni} {et~al.}(2018){Bruni}, {Panessa}, {Ghisellini}, {Chavushyan},
  {Pe{\~n}a-Herazo}, {Hern{\'a}ndez-Garc{\'\i}a}, {Bazzano}, {Ubertini}, \&
  {Kraus}}]{2018ApJ...854L..23B}
{Bruni}, G., {Panessa}, F., {Ghisellini}, G., {et~al.} 2018, \apjl, 854, L23,
  \dodoi{10.3847/2041-8213/aaacfb}

\bibitem[{{Buttiglione} {et~al.}(2009){Buttiglione}, {Capetti}, {Celotti},
  {Axon}, {Chiaberge}, {Macchetto}, \& {Sparks}}]{2009ANA...495.1033B}
{Buttiglione}, S., {Capetti}, A., {Celotti}, A., {et~al.} 2009, \aap, 495,
  1033, \dodoi{10.1051/0004-6361:200811102}

\bibitem[{{Cappellari} \& {Emsellem}(2004)}]{2004PASP..116..138C}
{Cappellari}, M., \& {Emsellem}, E. 2004, \pasp, 116, 138,
  \dodoi{10.1086/381875}

\bibitem[{{Carangelo} {et~al.}(2003){Carangelo}, {Falomo}, {Kotilainen},
  {Treves}, \& {Ulrich}}]{2003ANA...412..651C}
{Carangelo}, N., {Falomo}, R., {Kotilainen}, J., {Treves}, A., \& {Ulrich},
  M.~H. 2003, \aap, 412, 651, \dodoi{10.1051/0004-6361:20031519}

\bibitem[{{Cardelli} {et~al.}(1989){Cardelli}, {Clayton}, \&
  {Mathis}}]{1989ApJ...345..245C}
{Cardelli}, J.~A., {Clayton}, G.~C., \& {Mathis}, J.~S. 1989, \apj, 345, 245,
  \dodoi{10.1086/167900}

\bibitem[{{Cavaliere} \& {D'Elia}(2002)}]{2002ApJ...571..226C}
{Cavaliere}, A., \& {D'Elia}, V. 2002, \apj, 571, 226, \dodoi{10.1086/339778}

\bibitem[{{Celotti} {et~al.}(1997){Celotti}, {Padovani}, \&
  {Ghisellini}}]{1997MNRAS.286..415C}
{Celotti}, A., {Padovani}, P., \& {Ghisellini}, G. 1997, \mnras, 286, 415

\bibitem[{{Chai} {et~al.}(2012){Chai}, {Cao}, \& {Gu}}]{2012ApJ...759..114C}
{Chai}, B., {Cao}, X., \& {Gu}, M. 2012, \apj, 759, 114,
  \dodoi{10.1088/0004-637X/759/2/114}

\bibitem[{{Chatterjee} {et~al.}(2013){Chatterjee}, {Fossati}, {Urry}, {Bailyn},
  {Maraschi}, {Buxton}, {Bonning}, {Isler}, \& {Coppi}}]{2013ApJ...763L..11C}
{Chatterjee}, R., {Fossati}, G., {Urry}, C.~M., {et~al.} 2013, \apjl, 763, L11,
  \dodoi{10.1088/2041-8205/763/1/L11}

\bibitem[{{Chen} {et~al.}(2018){Chen}, {Berton}, {La Mura}, {Congiu}, {Cracco},
  {Foschini}, {Fan}, {Ciroi}, {Rafanelli}, \& {Bastieri}}]{2018A&A...615A.167C}
{Chen}, S., {Berton}, M., {La Mura}, G., {et~al.} 2018, \aap, 615, A167,
  \dodoi{10.1051/0004-6361/201832678}

\bibitem[{{Chen} {et~al.}(2015){Chen}, {Zhang}, {Zhang}, \&
  {Yu}}]{2015MNRAS.451.4193C}
{Chen}, Y.~Y., {Zhang}, X., {Zhang}, H.~J., \& {Yu}, X.~L. 2015, \mnras, 451,
  4193, \dodoi{10.1093/mnras/stv658}

\bibitem[{{Chomiuk} {et~al.}(2013){Chomiuk}, {Strader}, {Landt}, \&
  {Cheung}}]{2013ATel.4777....1C}
{Chomiuk}, L., {Strader}, J., {Landt}, H., \& {Cheung}, C.~C. 2013, The
  Astronomer's Telegram, 4777, 1

\bibitem[{{Chu} {et~al.}(1986){Chu}, {Zhu}, \& {Butcher}}]{1986ChANA..10..196C}
{Chu}, Y.-q., {Zhu}, X.-f., \& {Butcher}, H. 1986, \caa, 10, 196,
  \dodoi{10.1016/0275-1062(86)90005-6}

\bibitem[{{Colless} {et~al.}(2001){Colless}, {Dalton}, {Maddox}, {Sutherland },
  {Norberg}, {Cole}, {Bland -Hawthorn}, {Bridges}, {Cannon}, {Collins},
  {Couch}, {Cross}, {Deeley}, {De Propris}, {Driver}, {Efstathiou}, {Ellis},
  {Frenk}, {Glazebrook}, {Jackson}, {Lahav}, {Lewis}, {Lumsden}, {Madgwick},
  {Peacock}, {Peterson}, {Price}, {Seaborne}, \&
  {Taylor}}]{2001MNRAS.328.1039C}
{Colless}, M., {Dalton}, G., {Maddox}, S., {et~al.} 2001, \mnras, 328, 1039,
  \dodoi{10.1046/j.1365-8711.2001.04902.x}

\bibitem[{{de Menezes} {et~al.}(2020){de Menezes}, {Amaya-Almaz{\'a}n},
  {Marchesini}, {Pe{\~n}a-Herazo}, {Massaro}, {Chavushyan}, {Paggi}, {Landoni},
  {Masetti}, {Ricci}, {D'Abrusco}, {La Franca}, {Smith}, {Milisavljevic},
  {Tosti}, {Jim{\'e}nez-Bail{\'o}n}, \& {Cheung}}]{2020ApNSS.365...12D}
{de Menezes}, R., {Amaya-Almaz{\'a}n}, R.~A., {Marchesini}, E.~J., {et~al.}
  2020, \apss, 365, 12, \dodoi{10.1007/s10509-020-3727-5}

\bibitem[{{Dermer}(1995)}]{1995ApJ...446L..63D}
{Dermer}, C.~D. 1995, \apjl, 446, L63, \dodoi{10.1086/187931}

\bibitem[{{Dermer} {et~al.}(2009){Dermer}, {Finke}, {Krug}, \&
  {B{\"o}ttcher}}]{2009ApJ...692...32D}
{Dermer}, C.~D., {Finke}, J.~D., {Krug}, H., \& {B{\"o}ttcher}, M. 2009, \apj,
  692, 32, \dodoi{10.1088/0004-637X/692/1/32}

\bibitem[{{Desai} {et~al.}(2019){Desai}, {Marchesi}, {Rajagopal}, \&
  {Ajello}}]{2019ApJS..241....5D}
{Desai}, A., {Marchesi}, S., {Rajagopal}, M., \& {Ajello}, M. 2019, \apjs, 241,
  5, \dodoi{10.3847/1538-4365/ab01fc}

\bibitem[{{di Serego Alighieri} {et~al.}(1994){di Serego Alighieri},
  {Danziger}, {Morganti}, \& {Tadhunter}}]{1994MNRAS.269..998D}
{di Serego Alighieri}, S., {Danziger}, I.~J., {Morganti}, R., \& {Tadhunter},
  C.~N. 1994, \mnras, 269, 998, \dodoi{10.1093/mnras/269.4.998}

\bibitem[{{Drinkwater} {et~al.}(1997){Drinkwater}, {Webster}, {Francis},
  {Condon}, {Ellison}, {Jauncey}, {Lovell}, {Peterson}, \&
  {Savage}}]{1997MNRAS.284...85D}
{Drinkwater}, M.~J., {Webster}, R.~L., {Francis}, P.~J., {et~al.} 1997, \mnras,
  284, 85

\bibitem[{{Dunlop} {et~al.}(1989){Dunlop}, {Peacock}, {Savage}, {Lilly},
  {Heasley}, \& {Simon}}]{1989MNRAS.238.1171D}
{Dunlop}, J.~S., {Peacock}, J.~A., {Savage}, A., {et~al.} 1989, \mnras, 238,
  1171, \dodoi{10.1093/mnras/238.4.1171}

\bibitem[{{Evans} \& {Koratkar}(2004)}]{2004ApJS..150...73E}
{Evans}, I.~N., \& {Koratkar}, A.~P. 2004, \apjs, 150, 73,
  \dodoi{10.1086/379649}

\bibitem[{{Evans} {et~al.}(2020){Evans}, {Page}, {Osborne}, {Beardmore},
  {Willingale}, {Burrows}, {Kennea}, {Perri}, {Capalbi}, {Tagliaferri}, \&
  {Cenko}}]{2020ApJS..247...54E}
{Evans}, P.~A., {Page}, K.~L., {Osborne}, J.~P., {et~al.} 2020, \apjs, 247, 54,
  \dodoi{10.3847/1538-4365/ab7db9}

\bibitem[{{Falomo} {et~al.}(1997){Falomo}, {Kotilainen}, {Pursimo},
  {Sillanpaeae}, {Takalo}, \& {Heidt}}]{1997ANA...321..374F}
{Falomo}, R., {Kotilainen}, J., {Pursimo}, T., {et~al.} 1997, \aap, 321, 374

\bibitem[{{Falomo} {et~al.}(2000){Falomo}, {Scarpa}, {Treves}, \&
  {Urry}}]{2000ApJ...542..731F}
{Falomo}, R., {Scarpa}, R., {Treves}, A., \& {Urry}, C.~M. 2000, \apj, 542,
  731, \dodoi{10.1086/317044}

\bibitem[{{Falomo} {et~al.}(2017){Falomo}, {Treves}, {Scarpa}, {Paiano}, \&
  {Land oni}}]{2017MNRAS.470.2814F}
{Falomo}, R., {Treves}, A., {Scarpa}, R., {Paiano}, S., \& {Land oni}, M. 2017,
  \mnras, 470, 2814, \dodoi{10.1093/mnras/stx1411}

\bibitem[{{Fan} {et~al.}(2017){Fan}, {Yang}, {Xiao}, {Lin}, {Constantin},
  {Luo}, {Pei}, {Hao}, \& {Mao}}]{2017ApJ...835L..38F}
{Fan}, J.~H., {Yang}, J.~H., {Xiao}, H.~B., {et~al.} 2017, \apjl, 835, L38,
  \dodoi{10.3847/2041-8213/835/2/L38}

\bibitem[{{Ferrarese} \& {Merritt}(2000)}]{2000ApJ...539L...9F}
{Ferrarese}, L., \& {Merritt}, D. 2000, \apjl, 539, L9, \dodoi{10.1086/312838}

\bibitem[{{Finke}(2013)}]{2013ApJ...763..134F}
{Finke}, J.~D. 2013, \apj, 763, 134, \dodoi{10.1088/0004-637X/763/2/134}

\bibitem[{{Finke} {et~al.}(2008){Finke}, {Dermer}, \&
  {B{\"o}ttcher}}]{2008ApJ...686..181F}
{Finke}, J.~D., {Dermer}, C.~D., \& {B{\"o}ttcher}, M. 2008, \apj, 686, 181,
  \dodoi{10.1086/590900}

\bibitem[{{Foschini} {et~al.}(2015){Foschini}, {Berton}, {Caccianiga}, {Ciroi},
  {Cracco}, {Peterson}, {Angelakis}, {Braito}, {Fuhrmann}, {Gallo}, {Grupe},
  {J{\"a}rvel{\"a}}, {Kaufmann}, {Komossa}, {Kovalev}, {L{\"a}hteenm{\"a}ki},
  {Lisakov}, {Lister}, {Mathur}, {Richards}, {Romano}, {Sievers},
  {Tagliaferri}, {Tammi}, {Tibolla}, {Tornikoski}, {Vercellone}, {La Mura},
  {Maraschi}, \& {Rafanelli}}]{2015ANA...575A..13F}
{Foschini}, L., {Berton}, M., {Caccianiga}, A., {et~al.} 2015, \aap, 575, A13,
  \dodoi{10.1051/0004-6361/201424972}

\bibitem[{{Francis} {et~al.}(1991){Francis}, {Hewett}, {Foltz}, {Chaffee},
  {Weymann}, \& {Morris}}]{1991ApJ...373..465F}
{Francis}, P.~J., {Hewett}, P.~C., {Foltz}, C.~B., {et~al.} 1991, \apj, 373,
  465, \dodoi{10.1086/170066}

\bibitem[{{Fricke} {et~al.}(1983){Fricke}, {Kollatschny}, \&
  {Witzel}}]{1983ANA...117...60F}
{Fricke}, K.~J., {Kollatschny}, W., \& {Witzel}, A. 1983, \aap, 117, 60

\bibitem[{{Ghisellini} {et~al.}(1998){Ghisellini}, {Celotti}, {Fossati},
  {Maraschi}, \& {Comastri}}]{1998MNRAS.301..451G}
{Ghisellini}, G., {Celotti}, A., {Fossati}, G., {Maraschi}, L., \& {Comastri},
  A. 1998, \mnras, 301, 451, \dodoi{10.1046/j.1365-8711.1998.02032.x}

\bibitem[{{Ghisellini} {et~al.}(2017){Ghisellini}, {Righi}, {Costamante}, \&
  {Tavecchio}}]{2017MNRAS.469..255G}
{Ghisellini}, G., {Righi}, C., {Costamante}, L., \& {Tavecchio}, F. 2017,
  \mnras, 469, 255, \dodoi{10.1093/mnras/stx806}

\bibitem[{{Ghisellini} {et~al.}(2011){Ghisellini}, {Tavecchio}, {Foschini}, \&
  {Ghirlanda}}]{2011MNRAS.414.2674G}
{Ghisellini}, G., {Tavecchio}, F., {Foschini}, L., \& {Ghirlanda}, G. 2011,
  \mnras, 414, 2674, \dodoi{10.1111/j.1365-2966.2011.18578.x}

\bibitem[{{Ghisellini} {et~al.}(2012){Ghisellini}, {Tavecchio}, {Foschini},
  {Sbarrato}, {Ghirlanda}, \& {Maraschi}}]{2012MNRAS.425.1371G}
{Ghisellini}, G., {Tavecchio}, F., {Foschini}, L., {et~al.} 2012, \mnras, 425,
  1371, \dodoi{10.1111/j.1365-2966.2012.21554.x}

\bibitem[{{Ghisellini} {et~al.}(2014){Ghisellini}, {Tavecchio}, {Maraschi},
  {Celotti}, \& {Sbarrato}}]{2014Natur.515..376G}
{Ghisellini}, G., {Tavecchio}, F., {Maraschi}, L., {Celotti}, A., \&
  {Sbarrato}, T. 2014, \nat, 515, 376, \dodoi{10.1038/nature13856}

\bibitem[{{Gioia} {et~al.}(2004){Gioia}, {Wolter}, {Mullis}, {Henry},
  {B{\"o}hringer}, \& {Briel}}]{2004ANA...428..867G}
{Gioia}, I.~M., {Wolter}, A., {Mullis}, C.~R., {et~al.} 2004, \aap, 428, 867,
  \dodoi{10.1051/0004-6361:20041426}

\bibitem[{{Giommi} {et~al.}(2013){Giommi}, {Padovani}, \&
  {Polenta}}]{2013MNRAS.431.1914G}
{Giommi}, P., {Padovani}, P., \& {Polenta}, G. 2013, \mnras, 431, 1914,
  \dodoi{10.1093/mnras/stt305}

\bibitem[{{Gorham} {et~al.}(2000){Gorham}, {van Zee}, {Unwin}, \&
  {Jacobs}}]{2000AJ....119.1677G}
{Gorham}, P.~W., {van Zee}, L., {Unwin}, S.~C., \& {Jacobs}, C. 2000, \aj, 119,
  1677, \dodoi{10.1086/301289}

\bibitem[{{Graham}(2007)}]{2007MNRAS.379..711G}
{Graham}, A.~W. 2007, \mnras, 379, 711,
  \dodoi{10.1111/j.1365-2966.2007.11950.x}

\bibitem[{{G{\"u}ltekin} {et~al.}(2009){G{\"u}ltekin}, {Richstone}, {Gebhardt},
  {Lauer}, {Tremaine}, {Aller}, {Bender}, {Dressler}, {Faber}, {Filippenko},
  {Green}, {Ho}, {Kormendy}, {Magorrian}, {Pinkney}, \&
  {Siopis}}]{2009ApJ...698..198G}
{G{\"u}ltekin}, K., {Richstone}, D.~O., {Gebhardt}, K., {et~al.} 2009, \apj,
  698, 198, \dodoi{10.1088/0004-637X/698/1/198}

\bibitem[{{Guo} {et~al.}(2019){Guo}, {Liu}, {Shen}, {Loeb}, {Monroe}, \&
  {Prochaska}}]{2019MNRAS.482.3288G}
{Guo}, H., {Liu}, X., {Shen}, Y., {et~al.} 2019, \mnras, 482, 3288,
  \dodoi{10.1093/mnras/sty2920}

\bibitem[{{Guo} {et~al.}(2018){Guo}, {Shen}, \& {Wang}}]{2018ascl.soft09008G}
{Guo}, H., {Shen}, Y., \& {Wang}, S. 2018, {PyQSOFit: Python code to fit the
  spectrum of quasars}.
\newblock \doeprint{1809.008}

\bibitem[{{Halpern} {et~al.}(1991){Halpern}, {Chen}, {Madejski}, \&
  {Chanan}}]{1991AJ....101..818H}
{Halpern}, J.~P., {Chen}, V.~S., {Madejski}, G.~M., \& {Chanan}, G.~A. 1991,
  \aj, 101, 818, \dodoi{10.1086/115725}

\bibitem[{{Halpern} {et~al.}(1997){Halpern}, {Eracleous}, \&
  {Forster}}]{1997AJ....114.1736H}
{Halpern}, J.~P., {Eracleous}, M., \& {Forster}, K. 1997, \aj, 114, 1736,
  \dodoi{10.1086/118602}

\bibitem[{{Halpern} {et~al.}(2003){Halpern}, {Eracleous}, \&
  {Mattox}}]{2003AJ....125..572H}
{Halpern}, J.~P., {Eracleous}, M., \& {Mattox}, J.~R. 2003, \aj, 125, 572,
  \dodoi{10.1086/345796}

\bibitem[{{Hao} {et~al.}(2005){Hao}, {Strauss}, {Tremonti}, {Schlegel},
  {Heckman}, {Kauffmann}, {Blanton}, {Fan}, {Gunn}, {Hall}, {Ivezi{\'c}},
  {Knapp}, {Krolik}, {Lupton}, {Richards}, {Schneider}, {Strateva}, {Zakamska},
  {Brinkmann}, {Brunner}, \& {Szokoly}}]{2005AJ....129.1783H}
{Hao}, L., {Strauss}, M.~A., {Tremonti}, C.~A., {et~al.} 2005, \aj, 129, 1783,
  \dodoi{10.1086/428485}

\bibitem[{{Healey} {et~al.}(2008){Healey}, {Romani}, {Cotter}, {Michelson},
  {Schlafly}, {Readhead}, {Giommi}, {Chaty}, {Grenier}, \&
  {Weintraub}}]{2008ApJS..175...97H}
{Healey}, S.~E., {Romani}, R.~W., {Cotter}, G., {et~al.} 2008, \apjs, 175, 97,
  \dodoi{10.1086/523302}

\bibitem[{{Henstock} {et~al.}(1997){Henstock}, {Browne}, {Wilkinson}, \&
  {McMahon}}]{1997MNRAS.290..380H}
{Henstock}, D.~R., {Browne}, I.~W.~A., {Wilkinson}, P.~N., \& {McMahon}, R.~G.
  1997, \mnras, 290, 380, \dodoi{10.1093/mnras/290.2.380}

\bibitem[{{Isobe} {et~al.}(1986){Isobe}, {Feigelson}, \&
  {Nelson}}]{1986ApJ...306..490I}
{Isobe}, T., {Feigelson}, E.~D., \& {Nelson}, P.~I. 1986, \apj, 306, 490,
  \dodoi{10.1086/164359}

\bibitem[{{Jones} {et~al.}(2009){Jones}, {Read}, {Saunders}, {Colless},
  {Jarrett}, {Parker}, {Fairall}, {Mauch}, {Sadler}, {Watson}, {Burton},
  {Campbell}, {Cass}, {Croom}, {Dawe}, {Fiegert}, {Frankcombe}, {Hartley},
  {Huchra}, {James}, {Kirby}, {Lahav}, {Lucey}, {Mamon}, {Moore}, {Peterson},
  {Prior}, {Proust}, {Russell}, {Safouris}, {Wakamatsu}, {Westra}, \&
  {Williams}}]{2009MNRAS.399..683J}
{Jones}, D.~H., {Read}, M.~A., {Saunders}, W., {et~al.} 2009, \mnras, 399, 683,
  \dodoi{10.1111/j.1365-2966.2009.15338.x}

\bibitem[{{Junkkarinen}(1984)}]{1984PASP...96..539J}
{Junkkarinen}, V. 1984, \pasp, 96, 539, \dodoi{10.1086/131374}

\bibitem[{{Keenan} {et~al.}(2020){Keenan}, {Meyer}, {Georganopoulos}, {Reddy},
  \& {French}}]{2020arXiv200712661K}
{Keenan}, M., {Meyer}, E.~T., {Georganopoulos}, M., {Reddy}, K., \& {French},
  O.~J. 2020, arXiv e-prints, arXiv:2007.12661.
\newblock \doarXiv{2007.12661}

\bibitem[{{Klindt} {et~al.}(2017){Klindt}, {van Soelen}, {Meintjes}, \&
  {V{\"a}is{\"a}nen}}]{2017MNRAS.467.2537K}
{Klindt}, L., {van Soelen}, B., {Meintjes}, P.~J., \& {V{\"a}is{\"a}nen}, P.
  2017, \mnras, 467, 2537, \dodoi{10.1093/mnras/stx218}

\bibitem[{{Kormendy} \& {Ho}(2013)}]{2013ARA&A..51..511K}
{Kormendy}, J., \& {Ho}, L.~C. 2013, \araa, 51, 511,
  \dodoi{10.1146/annurev-astro-082708-101811}

\bibitem[{{Koss} {et~al.}(2017){Koss}, {Trakhtenbrot}, {Ricci}, {Lamperti},
  {Oh}, {Berney}, {Schawinski}, {Balokovi{\'c}}, {Baronchelli}, {Crenshaw},
  {Fischer}, {Gehrels}, {Harrison}, {Hashimoto}, {Hogg}, {Ichikawa}, {Masetti},
  {Mushotzky}, {Sartori}, {Stern}, {Treister}, {Ueda}, {Veilleux}, \&
  {Winter}}]{2017ApJ...850...74K}
{Koss}, M., {Trakhtenbrot}, B., {Ricci}, C., {et~al.} 2017, \apj, 850, 74,
  \dodoi{10.3847/1538-4357/aa8ec9}

\bibitem[{{Kotilainen} {et~al.}(2005){Kotilainen}, {Hyv{\"o}nen}, \&
  {Falomo}}]{2005ANA...440..831K}
{Kotilainen}, J.~K., {Hyv{\"o}nen}, T., \& {Falomo}, R. 2005, \aap, 440, 831,
  \dodoi{10.1051/0004-6361:20042548}

\bibitem[{{Landoni} {et~al.}(2013){Landoni}, {Falomo}, {Treves}, {Sbarufatti},
  {Barattini}, {Decarli}, \& {Kotilainen}}]{2013AJ....145..114L}
{Landoni}, M., {Falomo}, R., {Treves}, A., {et~al.} 2013, \aj, 145, 114,
  \dodoi{10.1088/0004-6256/145/4/114}

\bibitem[{{Landoni} {et~al.}(2018){Landoni}, {Paiano}, {Falomo}, {Scarpa}, \&
  {Treves}}]{2018ApJ...861..130L}
{Landoni}, M., {Paiano}, S., {Falomo}, R., {Scarpa}, R., \& {Treves}, A. 2018,
  \apj, 861, 130, \dodoi{10.3847/1538-4357/aac77c}

\bibitem[{{Landt} {et~al.}(2001){Landt}, {Padovani}, {Perlman}, {Giommi},
  {Bignall}, \& {Tzioumis}}]{2001MNRAS.323..757L}
{Landt}, H., {Padovani}, P., {Perlman}, E.~S., {et~al.} 2001, \mnras, 323, 757,
  \dodoi{10.1046/j.1365-8711.2001.04269.x}

\bibitem[{{Laurent-Muehleisen} {et~al.}(1998){Laurent-Muehleisen}, {Kollgaard},
  {Ciardullo}, {Feigelson}, {Brinkmann}, \& {Siebert}}]{1998ApJS..118..127L}
{Laurent-Muehleisen}, S.~A., {Kollgaard}, R.~I., {Ciardullo}, R., {et~al.}
  1998, \apjs, 118, 127, \dodoi{10.1086/313134}

\bibitem[{{Lavalley} {et~al.}(1992){Lavalley}, {Isobe}, \&
  {Feigelson}}]{1992BAAS...24..839L}
{Lavalley}, M.~P., {Isobe}, T., \& {Feigelson}, E.~D. 1992, in Bulletin of the
  American Astronomical Society, Vol.~24, Bulletin of the American Astronomical
  Society, 839--840

\bibitem[{{Lawrence} {et~al.}(1996){Lawrence}, {Zucker}, {Readhead}, {Unwin},
  {Pearson}, \& {Xu}}]{1996ApJS..107..541L}
{Lawrence}, C.~R., {Zucker}, J.~R., {Readhead}, A.~C.~S., {et~al.} 1996, \apjs,
  107, 541, \dodoi{10.1086/192375}

\bibitem[{{Londish} {et~al.}(2007){Londish}, {Croom}, {Heidt}, {Boyle},
  {Sadler}, {Whiting}, {Rector}, {Pursimo}, \&
  {Chynoweth}}]{2007MNRAS.374..556L}
{Londish}, D., {Croom}, S.~M., {Heidt}, J., {et~al.} 2007, \mnras, 374, 556,
  \dodoi{10.1111/j.1365-2966.2006.11165.x}

\bibitem[{{Marcha} {et~al.}(1996){Marcha}, {Browne}, {Impey}, \&
  {Smith}}]{1996MNRAS.281..425M}
{Marcha}, M.~J.~M., {Browne}, I.~W.~A., {Impey}, C.~D., \& {Smith}, P.~S. 1996,
  \mnras, 281, 425, \dodoi{10.1093/mnras/281.2.425}

\bibitem[{{Marchesini} {et~al.}(2016){Marchesini}, {Andruchow}, {Cellone},
  {Combi}, {Zibecchi}, {Mart{\'{\i}}}, {Romero}, {Mu{\~n}oz-Arjonilla},
  {Luque-Escamilla}, \& {S{\'a}nchez-Sutil}}]{2016ANA...591A..21M}
{Marchesini}, E.~J., {Andruchow}, I., {Cellone}, S.~A., {et~al.} 2016, \aap,
  591, A21, \dodoi{10.1051/0004-6361/201527632}

\bibitem[{{Marchesini} {et~al.}(2019){Marchesini}, {Pe{\~n}a-Herazo},
  {{\'A}lvarez Crespo}, {Ricci}, {Negro}, {Milisavljevic}, {Massaro},
  {Masetti}, {Land oni}, {Chavushyan}, {D'Abrusco}, {Jim{\'e}nez-Bail{\'o}n},
  {La Franca}, {Paggi}, {Smith}, \& {Tosti}}]{2019ApNSS.364....5M}
{Marchesini}, E.~J., {Pe{\~n}a-Herazo}, H.~A., {{\'A}lvarez Crespo}, N.,
  {et~al.} 2019, \apss, 364, 5, \dodoi{10.1007/s10509-018-3490-z}

\bibitem[{{Marziani} {et~al.}(2003){Marziani}, {Sulentic}, {Zamanov},
  {Calvani}, {Dultzin-Hacyan}, {Bachev}, \& {Zwitter}}]{2003ApJS..145..199M}
{Marziani}, P., {Sulentic}, J.~W., {Zamanov}, R., {et~al.} 2003, \apjs, 145,
  199, \dodoi{10.1086/346025}

\bibitem[{{Masetti} {et~al.}(2008){Masetti}, {Mason}, {Landi}, {Giommi},
  {Bassani}, {Malizia}, {Bird}, {Bazzano}, {Dean}, {Gehrels}, {Palazzi}, \&
  {Ubertini}}]{2008ANA...480..715M}
{Masetti}, N., {Mason}, E., {Landi}, R., {et~al.} 2008, \aap, 480, 715,
  \dodoi{10.1051/0004-6361:20078901}

\bibitem[{{Massaro} {et~al.}(2015){Massaro}, {Landoni}, {D'Abrusco},
  {Milisavljevic}, {Paggi}, {Masetti}, {Smith}, \&
  {Tosti}}]{2015ANA...575A.124M}
{Massaro}, F., {Landoni}, M., {D'Abrusco}, R., {et~al.} 2015, \aap, 575, A124,
  \dodoi{10.1051/0004-6361/201425119}

\bibitem[{{McLure} \& {Dunlop}(2004)}]{2004MNRAS.352.1390M}
{McLure}, R.~J., \& {Dunlop}, J.~S. 2004, \mnras, 352, 1390,
  \dodoi{10.1111/j.1365-2966.2004.08034.x}

\bibitem[{{Merritt}(1997)}]{1997AJ....114..228M}
{Merritt}, D. 1997, \aj, 114, 228, \dodoi{10.1086/118467}

\bibitem[{{Meyer} {et~al.}(2011){Meyer}, {Fossati}, {Georganopoulos}, \&
  {Lister}}]{2011ApJ...740...98M}
{Meyer}, E.~T., {Fossati}, G., {Georganopoulos}, M., \& {Lister}, M.~L. 2011,
  \apj, 740, 98, \dodoi{10.1088/0004-637X/740/2/98}

\bibitem[{{Mezcua} {et~al.}(2012){Mezcua}, {Chavushyan}, {Lobanov}, \&
  {Le{\'o}n-Tavares}}]{2012ANA...544A..36M}
{Mezcua}, M., {Chavushyan}, V.~H., {Lobanov}, A.~P., \& {Le{\'o}n-Tavares}, J.
  2012, \aap, 544, A36, \dodoi{10.1051/0004-6361/201117724}

\bibitem[{{Morton} {et~al.}(1978){Morton}, {Savage}, \&
  {Bolton}}]{1978MNRAS.185..735M}
{Morton}, D.~C., {Savage}, A., \& {Bolton}, J.~G. 1978, \mnras, 185, 735,
  \dodoi{10.1093/mnras/185.4.735}

\bibitem[{{Murdoch} {et~al.}(1984){Murdoch}, {Hunstead}, \&
  {White}}]{1984PASAu...5..341M}
{Murdoch}, H.~S., {Hunstead}, R.~W., \& {White}, G.~L. 1984, Proceedings of the
  Astronomical Society of Australia, 5, 341, \dodoi{10.1017/S1323358000017173}

\bibitem[{{Narayan} {et~al.}(1997){Narayan}, {Garcia}, \&
  {McClintock}}]{1997ApJ...478L..79N}
{Narayan}, R., {Garcia}, M.~R., \& {McClintock}, J.~E. 1997, \apjl, 478, L79,
  \dodoi{10.1086/310554}

\bibitem[{{Nieppola} {et~al.}(2008){Nieppola}, {Valtaoja}, {Tornikoski},
  {Hovatta}, \& {Kotiranta}}]{2008A&A...488..867N}
{Nieppola}, E., {Valtaoja}, E., {Tornikoski}, M., {Hovatta}, T., \&
  {Kotiranta}, M. 2008, \aap, 488, 867, \dodoi{10.1051/0004-6361:200809716}

\bibitem[{{Nilsson} {et~al.}(2003){Nilsson}, {Pursimo}, {Heidt}, {Takalo},
  {Sillanp{\"a}{\"a}}, \& {Brinkmann}}]{2003ANA...400...95N}
{Nilsson}, K., {Pursimo}, T., {Heidt}, J., {et~al.} 2003, \aap, 400, 95,
  \dodoi{10.1051/0004-6361:20021861}

\bibitem[{{Nkundabakura} \& {Meintjes}(2012)}]{2012MNRAS.427..859N}
{Nkundabakura}, P., \& {Meintjes}, P.~J. 2012, \mnras, 427, 859,
  \dodoi{10.1111/j.1365-2966.2012.21953.x}

\bibitem[{{Olgu{\'\i}n-Iglesias} {et~al.}(2016){Olgu{\'\i}n-Iglesias},
  {Le{\'o}n-Tavares}, {Kotilainen}, {Chavushyan}, {Tornikoski}, {Valtaoja},
  {A{\~n}orve}, {Vald{\'e}s}, \& {Carrasco}}]{2016MNRAS.460.3202O}
{Olgu{\'\i}n-Iglesias}, A., {Le{\'o}n-Tavares}, J., {Kotilainen}, J.~K.,
  {et~al.} 2016, \mnras, 460, 3202, \dodoi{10.1093/mnras/stw1208}

\bibitem[{{Padovani}(2007)}]{2007Ap&SS.309...63P}
{Padovani}, P. 2007, \apss, 309, 63, \dodoi{10.1007/s10509-007-9455-2}

\bibitem[{{Padovani} {et~al.}(2012){Padovani}, {Giommi}, \&
  {Rau}}]{2012MNRAS.422L..48P}
{Padovani}, P., {Giommi}, P., \& {Rau}, A. 2012, \mnras, 422, L48,
  \dodoi{10.1111/j.1745-3933.2012.01234.x}

\bibitem[{{Padovani} {et~al.}(2019){Padovani}, {Oikonomou}, {Petropoulou},
  {Giommi}, \& {Resconi}}]{2019MNRAS.484L.104P}
{Padovani}, P., {Oikonomou}, F., {Petropoulou}, M., {Giommi}, P., \& {Resconi},
  E. 2019, \mnras, 484, L104, \dodoi{10.1093/mnrasl/slz011}

\bibitem[{{Paiano} {et~al.}(2017{\natexlab{a}}){Paiano}, {Falomo},
  {Franceschini}, {Treves}, \& {Scarpa}}]{2017ApJ...851..135P}
{Paiano}, S., {Falomo}, R., {Franceschini}, A., {Treves}, A., \& {Scarpa}, R.
  2017{\natexlab{a}}, \apj, 851, 135, \dodoi{10.3847/1538-4357/aa9af4}

\bibitem[{{Paiano} {et~al.}(2019){Paiano}, {Falomo}, {Treves}, {Franceschini},
  \& {Scarpa}}]{2019ApJ...871..162P}
{Paiano}, S., {Falomo}, R., {Treves}, A., {Franceschini}, A., \& {Scarpa}, R.
  2019, \apj, 871, 162, \dodoi{10.3847/1538-4357/aaf6e4}

\bibitem[{{Paiano} {et~al.}(2020){Paiano}, {Falomo}, {Treves}, \&
  {Scarpa}}]{2020MNRAS.497...94P}
{Paiano}, S., {Falomo}, R., {Treves}, A., \& {Scarpa}, R. 2020, \mnras, 497,
  94, \dodoi{10.1093/mnras/staa1840}

\bibitem[{{Paiano} {et~al.}(2017{\natexlab{b}}){Paiano}, {Landoni}, {Falomo},
  {Treves}, \& {Scarpa}}]{2017ApJ...844..120P}
{Paiano}, S., {Landoni}, M., {Falomo}, R., {Treves}, A., \& {Scarpa}, R.
  2017{\natexlab{b}}, \apj, 844, 120, \dodoi{10.3847/1538-4357/aa7aac}

\bibitem[{{Paiano} {et~al.}(2017{\natexlab{c}}){Paiano}, {Landoni}, {Falomo},
  {Treves}, {Scarpa}, \& {Righi}}]{2017ApJ...837..144P}
{Paiano}, S., {Landoni}, M., {Falomo}, R., {et~al.} 2017{\natexlab{c}}, \apj,
  837, 144, \dodoi{10.3847/1538-4357/837/2/144}

\bibitem[{{Paliya}(2015)}]{2015ApJ...804...74P}
{Paliya}, V.~S. 2015, \apj, 804, 74, \dodoi{10.1088/0004-637X/804/1/74}

\bibitem[{{Paliya} {et~al.}(2020{\natexlab{a}}){Paliya}, {Ajello}, {Cao},
  {Giroletti}, {Kaur}, {Madejski}, {Lott}, \& {Hartmann}}]{2020ApJ...897..177P}
{Paliya}, V.~S., {Ajello}, M., {Cao}, H.~M., {et~al.} 2020{\natexlab{a}}, \apj,
  897, 177, \dodoi{10.3847/1538-4357/ab9c1a}

\bibitem[{{Paliya} {et~al.}(2020{\natexlab{b}}){Paliya}, {B{\"o}ttcher},
  {Olmo-Garc{\'\i}a}, {Dom{\'\i}nguez}, {Gil de Paz}, {Franckowiak},
  {Garrappa}, \& {Stein}}]{2020ApJ...902...29P}
{Paliya}, V.~S., {B{\"o}ttcher}, M., {Olmo-Garc{\'\i}a}, A., {et~al.}
  2020{\natexlab{b}}, \apj, 902, 29, \dodoi{10.3847/1538-4357/abb46e}

\bibitem[{{Paliya} {et~al.}(2016){Paliya}, {Diltz}, {B{\"o}ttcher}, {Stalin},
  \& {Buckley}}]{2016ApJ...817...61P}
{Paliya}, V.~S., {Diltz}, C., {B{\"o}ttcher}, M., {Stalin}, C.~S., \&
  {Buckley}, D. 2016, \apj, 817, 61, \dodoi{10.3847/0004-637X/817/1/61}

\bibitem[{{Paliya} {et~al.}(2017{\natexlab{a}}){Paliya}, {Marcotulli},
  {Ajello}, {Joshi}, {Sahayanathan}, {Rao}, \&
  {Hartmann}}]{2017ApJ...851...33P}
{Paliya}, V.~S., {Marcotulli}, L., {Ajello}, M., {et~al.} 2017{\natexlab{a}},
  \apj, 851, 33, \dodoi{10.3847/1538-4357/aa98e1}

\bibitem[{{Paliya} {et~al.}(2015){Paliya}, {Sahayanathan}, \&
  {Stalin}}]{2015ApJ...803...15P}
{Paliya}, V.~S., {Sahayanathan}, S., \& {Stalin}, C.~S. 2015, \apj, 803, 15,
  \dodoi{10.1088/0004-637X/803/1/15}

\bibitem[{{Paliya} {et~al.}(2017{\natexlab{b}}){Paliya}, {Stalin}, {Ajello}, \&
  {Kaur}}]{2017ApJ...844...32P}
{Paliya}, V.~S., {Stalin}, C.~S., {Ajello}, M., \& {Kaur}, A.
  2017{\natexlab{b}}, \apj, 844, 32, \dodoi{10.3847/1538-4357/aa77f5}

\bibitem[{{Paliya} {et~al.}(2019){Paliya}, {Koss}, {Trakhtenbrot}, {Ricci},
  {Oh}, {Ajello}, {Stern}, {Powell}, {Urry}, {Harrison}, {Lamperti},
  {Mushotzky}, {Marcotulli}, {Mej{\'\i}a-Restrepo}, \&
  {Hartmann}}]{2019ApJ...881..154P}
{Paliya}, V.~S., {Koss}, M., {Trakhtenbrot}, B., {et~al.} 2019, \apj, 881, 154,
  \dodoi{10.3847/1538-4357/ab2f8b}

\bibitem[{{Paliya} {et~al.}(2020{\natexlab{c}}){Paliya}, {Dom{\'\i}nguez},
  {Cabello}, {Cardiel}, {Gallego}, {Siana}, {Ajello}, {Hartmann}, {Gil de Paz},
  \& {Stalin}}]{2020ApJ...903L...8P}
{Paliya}, V.~S., {Dom{\'\i}nguez}, A., {Cabello}, C., {et~al.}
  2020{\natexlab{c}}, \apjl, 903, L8, \dodoi{10.3847/2041-8213/abbc06}

\bibitem[{{Pe{\~n}a-Herazo} {et~al.}(2017){Pe{\~n}a-Herazo}, {Marchesini},
  {{\'A}lvarez Crespo}, {Ricci}, {Massaro}, {Chavushyan}, {Landoni}, {Strader},
  {Chomiuk}, {Cheung}, {Masetti}, {Jim{\'e}nez-Bail{\'o}n}, {D'Abrusco},
  {Paggi}, {Milisavljevic}, {La Franca}, {Smith}, \&
  {Tosti}}]{2017ApNSS.362..228P}
{Pe{\~n}a-Herazo}, H.~A., {Marchesini}, E.~J., {{\'A}lvarez Crespo}, N.,
  {et~al.} 2017, \apss, 362, 228, \dodoi{10.1007/s10509-017-3208-7}

\bibitem[{{Pe{\~n}a-Herazo} {et~al.}(2019){Pe{\~n}a-Herazo}, {Massaro},
  {Chavushyan}, {Marchesini}, {Paggi}, {Landoni}, {Masetti}, {Ricci},
  {D'Abrusco}, {Milisavljevic}, {Jim{\'e}nez-Bail{\'o}n}, {La Franca}, {Smith},
  \& {Tosti}}]{2019ApNSS.364...85P}
{Pe{\~n}a-Herazo}, H.~A., {Massaro}, F., {Chavushyan}, V., {et~al.} 2019,
  \apss, 364, 85, \dodoi{10.1007/s10509-019-3574-4}

\bibitem[{{Pe{\~n}a-Herazo} {et~al.}(2020){Pe{\~n}a-Herazo},
  {Amaya-Almaz{\'a}n}, {Massaro}, {de Menezes}, {Marchesini}, {Chavushyan},
  {Paggi}, {Landoni}, {Masetti}, {Ricci}, {D'Abrusco}, {Cheung}, {La Franca},
  {Smith}, {Milisavljevic}, {Jim{\'e}nez-Bail{\'o}n}, {Pati{\~n}o-{\'A}lvarez},
  \& {Tosti}}]{2020arXiv200907905P}
{Pe{\~n}a-Herazo}, H.~A., {Amaya-Almaz{\'a}n}, R.~A., {Massaro}, F., {et~al.}
  2020, \aap, 643, A103, \dodoi{10.1051/0004-6361/202037978}

\bibitem[{{Perlman} {et~al.}(1998){Perlman}, {Padovani}, {Giommi}, {Sambruna},
  {Jones}, {Tzioumis}, \& {Reynolds}}]{1998AJ....115.1253P}
{Perlman}, E.~S., {Padovani}, P., {Giommi}, P., {et~al.} 1998, \aj, 115, 1253,
  \dodoi{10.1086/300283}

\bibitem[{{Perlman} {et~al.}(1996){Perlman}, {Stocke}, {Schachter}, {Elvis},
  {Ellingson}, {Urry}, {Potter}, {Impey}, \&
  {Kolchinsky}}]{1996ApJS..104..251P}
{Perlman}, E.~S., {Stocke}, J.~T., {Schachter}, J.~F., {et~al.} 1996, \apjs,
  104, 251, \dodoi{10.1086/192300}

\bibitem[{{Pietsch} {et~al.}(1998){Pietsch}, {Bischoff}, {Boller},
  {Doebereiner}, {Kollatschny}, \& {Zimmermann}}]{1998ANA...333...48P}
{Pietsch}, W., {Bischoff}, K., {Boller}, T., {et~al.} 1998, \aap, 333, 48.
\newblock \doarXiv{astro-ph/9801210}

\bibitem[{{Piranomonte} {et~al.}(2007){Piranomonte}, {Perri}, {Giommi},
  {Landt}, \& {Padovani}}]{2007ANA...470..787P}
{Piranomonte}, S., {Perri}, M., {Giommi}, P., {Landt}, H., \& {Padovani}, P.
  2007, \aap, 470, 787, \dodoi{10.1051/0004-6361:20077086}

\bibitem[{{Pita} {et~al.}(2014){Pita}, {Goldoni}, {Boisson}, {Lenain}, {Punch},
  {G{\'e}rard}, {Hammer}, {Kaper}, \& {Sol}}]{2014ANA...565A..12P}
{Pita}, S., {Goldoni}, P., {Boisson}, C., {et~al.} 2014, \aap, 565, A12,
  \dodoi{10.1051/0004-6361/201323071}

\bibitem[{{Plotkin} {et~al.}(2011){Plotkin}, {Markoff}, {Trager}, \&
  {Anderson}}]{2011MNRAS.413..805P}
{Plotkin}, R.~M., {Markoff}, S., {Trager}, S.~C., \& {Anderson}, S.~F. 2011,
  \mnras, 413, 805, \dodoi{10.1111/j.1365-2966.2010.18172.x}

\bibitem[{{Punsly} \& {Kharb}(2017)}]{2017MNRAS.468L..72P}
{Punsly}, B., \& {Kharb}, P. 2017, \mnras, 468, L72,
  \dodoi{10.1093/mnrasl/slx024}

\bibitem[{{Pursimo} {et~al.}(2013){Pursimo}, {Ojha}, {Jauncey}, {Rickett},
  {Dutka}, {Koay}, {Lovell}, {Bignall}, {Kedziora-Chudczer}, \&
  {Macquart}}]{2013ApJ...767...14P}
{Pursimo}, T., {Ojha}, R., {Jauncey}, D.~L., {et~al.} 2013, \apj, 767, 14,
  \dodoi{10.1088/0004-637X/767/1/14}

\bibitem[{{Raiteri} {et~al.}(2007){Raiteri}, {Villata}, {Capetti}, {Heidt},
  {Arnaboldi}, \& {Magazz{\`u}}}]{2007ANA...464..871R}
{Raiteri}, C.~M., {Villata}, M., {Capetti}, A., {et~al.} 2007, \aap, 464, 871,
  \dodoi{10.1051/0004-6361:20066599}

\bibitem[{{Rajput} {et~al.}(2020){Rajput}, {Stalin}, \&
  {Sahayanathan}}]{2020MNRAS.498.5128R}
{Rajput}, B., {Stalin}, C.~S., \& {Sahayanathan}, S. 2020, \mnras, 498, 5128,
  \dodoi{10.1093/mnras/staa2708}

\bibitem[{{Rector} {et~al.}(2000){Rector}, {Stocke}, {Perlman}, {Morris}, \&
  {Gioia}}]{2000AJ....120.1626R}
{Rector}, T.~A., {Stocke}, J.~T., {Perlman}, E.~S., {Morris}, S.~L., \&
  {Gioia}, I.~M. 2000, \aj, 120, 1626, \dodoi{10.1086/301587}

\bibitem[{{Ricci} {et~al.}(2015){Ricci}, {Massaro}, {Landoni}, {D'Abrusco},
  {Milisavljevic}, {Stern}, {Masetti}, {Paggi}, {Smith}, \&
  {Tosti}}]{2015AJ....149..160R}
{Ricci}, F., {Massaro}, F., {Landoni}, M., {et~al.} 2015, \aj, 149, 160,
  \dodoi{10.1088/0004-6256/149/5/160}

\bibitem[{{Richards} {et~al.}(2009){Richards}, {Myers}, {Gray}, {Riegel},
  {Nichol}, {Brunner}, {Szalay}, {Schneider}, \&
  {Anderson}}]{2009ApJS..180...67R}
{Richards}, G.~T., {Myers}, A.~D., {Gray}, A.~G., {et~al.} 2009, \apjs, 180,
  67, \dodoi{10.1088/0067-0049/180/1/67}

\bibitem[{Rohatgi(2020)}]{Rohatgi2020}
Rohatgi, A. 2020, Webplotdigitizer: Version 4.3.
\newblock \url{https://automeris.io/WebPlotDigitizer}

\bibitem[{{S{\'a}nchez-Bl{\'a}zquez} {et~al.}(2006){S{\'a}nchez-Bl{\'a}zquez},
  {Peletier}, {Jim{\'e}nez-Vicente}, {Cardiel}, {Cenarro},
  {Falc{\'o}n-Barroso}, {Gorgas}, {Selam}, \& {Vazdekis}}]{2006MNRAS.371..703S}
{S{\'a}nchez-Bl{\'a}zquez}, P., {Peletier}, R.~F., {Jim{\'e}nez-Vicente}, J.,
  {et~al.} 2006, \mnras, 371, 703, \dodoi{10.1111/j.1365-2966.2006.10699.x}

\bibitem[{{Sandrinelli} {et~al.}(2013){Sandrinelli}, {Treves}, {Falomo},
  {Farina}, {Foschini}, {Landoni}, \& {Sbarufatti}}]{2013AJ....146..163S}
{Sandrinelli}, A., {Treves}, A., {Falomo}, R., {et~al.} 2013, \aj, 146, 163,
  \dodoi{10.1088/0004-6256/146/6/163}

\bibitem[{{Sbarrato} {et~al.}(2014){Sbarrato}, {Padovani}, \&
  {Ghisellini}}]{2014MNRAS.445...81S}
{Sbarrato}, T., {Padovani}, P., \& {Ghisellini}, G. 2014, \mnras, 445, 81,
  \dodoi{10.1093/mnras/stu1759}

\bibitem[{{Sbarufatti} {et~al.}(2009){Sbarufatti}, {Ciprini}, {Kotilainen},
  {Decarli}, {Treves}, {Veronesi}, \& {Falomo}}]{2009AJ....137..337S}
{Sbarufatti}, B., {Ciprini}, S., {Kotilainen}, J., {et~al.} 2009, \aj, 137,
  337, \dodoi{10.1088/0004-6256/137/1/337}

\bibitem[{{Sbarufatti} {et~al.}(2006{\natexlab{a}}){Sbarufatti}, {Falomo},
  {Treves}, \& {Kotilainen}}]{2006ANA...457...35S}
{Sbarufatti}, B., {Falomo}, R., {Treves}, A., \& {Kotilainen}, J.
  2006{\natexlab{a}}, \aap, 457, 35, \dodoi{10.1051/0004-6361:20065455}

\bibitem[{{Sbarufatti} {et~al.}(2005){Sbarufatti}, {Treves}, {Falomo}, {Heidt},
  {Kotilainen}, \& {Scarpa}}]{2005AJ....129..559S}
{Sbarufatti}, B., {Treves}, A., {Falomo}, R., {et~al.} 2005, \aj, 129, 559,
  \dodoi{10.1086/427138}

\bibitem[{{Sbarufatti} {et~al.}(2006{\natexlab{b}}){Sbarufatti}, {Treves},
  {Falomo}, {Heidt}, {Kotilainen}, \& {Scarpa}}]{2006AJ....132....1S}
---. 2006{\natexlab{b}}, \aj, 132, 1, \dodoi{10.1086/503031}

\bibitem[{{Schlegel} {et~al.}(1998){Schlegel}, {Finkbeiner}, \&
  {Davis}}]{1998ApJ...500..525S}
{Schlegel}, D.~J., {Finkbeiner}, D.~P., \& {Davis}, M. 1998, \apj, 500, 525,
  \dodoi{10.1086/305772}

\bibitem[{{Shaw} {et~al.}(2013{\natexlab{a}}){Shaw}, {Filippenko}, {Romani},
  {Cenko}, \& {Li}}]{2013AJ....146..127S}
{Shaw}, M.~S., {Filippenko}, A.~V., {Romani}, R.~W., {Cenko}, S.~B., \& {Li},
  W. 2013{\natexlab{a}}, \aj, 146, 127, \dodoi{10.1088/0004-6256/146/5/127}

\bibitem[{{Shaw} {et~al.}(2012){Shaw}, {Romani}, {Cotter}, {Healey},
  {Michelson}, {Readhead}, {Richards}, {Max-Moerbeck}, {King}, \&
  {Potter}}]{2012ApJ...748...49S}
{Shaw}, M.~S., {Romani}, R.~W., {Cotter}, G., {et~al.} 2012, \apj, 748, 49,
  \dodoi{10.1088/0004-637X/748/1/49}

\bibitem[{{Shaw} {et~al.}(2013{\natexlab{b}}){Shaw}, {Romani}, {Cotter},
  {Healey}, {Michelson}, {Readhead}, {Richards}, {Max-Moerbeck}, {King}, \&
  {Potter}}]{2013ApJ...764..135S}
---. 2013{\natexlab{b}}, \apj, 764, 135, \dodoi{10.1088/0004-637X/764/2/135}

\bibitem[{{Shen} {et~al.}(2011){Shen}, {Richards}, {Strauss}, {Hall},
  {Schneider}, {Snedden}, {Bizyaev}, {Brewington}, {Malanushenko},
  {Malanushenko}, {Oravetz}, {Pan}, \& {Simmons}}]{2011ApJS..194...45S}
{Shen}, Y., {Richards}, G.~T., {Strauss}, M.~A., {et~al.} 2011, \apjs, 194, 45,
  \dodoi{10.1088/0067-0049/194/2/45}

\bibitem[{{Shen} {et~al.}(2019){Shen}, {Hall}, {Horne}, {Zhu}, {McGreer},
  {Simm}, {Trump}, {Kinemuchi}, {Brandt}, {Green}, {Grier}, {Guo}, {Ho},
  {Homayouni}, {Jiang}, {I-Hsiu Li}, {Morganson}, {Petitjean}, {Richards},
  {Schneider}, {Starkey}, {Wang}, {Chambers}, {Kaiser}, {Kudritzki}, {Magnier},
  \& {Waters}}]{2019ApJS..241...34S}
{Shen}, Y., {Hall}, P.~B., {Horne}, K., {et~al.} 2019, \apjs, 241, 34,
  \dodoi{10.3847/1538-4365/ab074f}

\bibitem[{{Sikora} {et~al.}(1994){Sikora}, {Begelman}, \&
  {Rees}}]{1994ApJ...421..153S}
{Sikora}, M., {Begelman}, M.~C., \& {Rees}, M.~J. 1994, \apj, 421, 153,
  \dodoi{10.1086/173633}

\bibitem[{{Smith} {et~al.}(2009){Smith}, {Montiel}, {Rightley}, {Turner},
  {Schmidt}, \& {Jannuzi}}]{2009arXiv0912.3621S}
{Smith}, P.~S., {Montiel}, E., {Rightley}, S., {et~al.} 2009, arXiv:0912.3621.
\newblock \doarXiv{0912.3621}

\bibitem[{{Sowards-Emmerd} {et~al.}(2003){Sowards-Emmerd}, {Romani}, \&
  {Michelson}}]{2003ApJ...590..109S}
{Sowards-Emmerd}, D., {Romani}, R.~W., \& {Michelson}, P.~F. 2003, \apj, 590,
  109, \dodoi{10.1086/374981}

\bibitem[{{Sowards-Emmerd} {et~al.}(2004){Sowards-Emmerd}, {Romani},
  {Michelson}, \& {Ulvestad}}]{2004ApJ...609..564S}
{Sowards-Emmerd}, D., {Romani}, R.~W., {Michelson}, P.~F., \& {Ulvestad}, J.~S.
  2004, \apj, 609, 564, \dodoi{10.1086/421239}

\bibitem[{{Spinrad} \& {Smith}(1975)}]{1975ApJ...201..275S}
{Spinrad}, H., \& {Smith}, H.~E. 1975, \apj, 201, 275, \dodoi{10.1086/153883}

\bibitem[{{Steidel} \& {Sargent}(1991)}]{1991ApJ...382..433S}
{Steidel}, C.~C., \& {Sargent}, W. L.~W. 1991, \apj, 382, 433,
  \dodoi{10.1086/170732}

\bibitem[{{Stickel} {et~al.}(1988){Stickel}, {Fried}, \&
  {Kuehr}}]{1988ANA...191L..16S}
{Stickel}, M., {Fried}, J.~W., \& {Kuehr}, H. 1988, \aap, 191, L16

\bibitem[{{Stickel} {et~al.}(1989){Stickel}, {Fried}, \&
  {Kuehr}}]{1989ANAS...80..103S}
---. 1989, \aaps, 80, 103

\bibitem[{{Stickel} {et~al.}(1993{\natexlab{a}}){Stickel}, {Fried}, \&
  {Kuehr}}]{1993ANAS...98..393S}
---. 1993{\natexlab{a}}, \aaps, 98, 393

\bibitem[{{Stickel} \& {Kuehr}(1993)}]{1993ANAS..100..395S}
{Stickel}, M., \& {Kuehr}, H. 1993, \aaps, 100, 395

\bibitem[{{Stickel} \& {Kuehr}(1994)}]{1994ANAS..105...67S}
---. 1994, \aaps, 105, 67

\bibitem[{{Stickel} {et~al.}(1993{\natexlab{b}}){Stickel}, {Kuehr}, \&
  {Fried}}]{1993ANAS...97..483S}
{Stickel}, M., {Kuehr}, H., \& {Fried}, J.~W. 1993{\natexlab{b}}, \aaps, 97,
  483

\bibitem[{{Stickel} \& {Kuhr}(1993)}]{1993ANAS..101..521S}
{Stickel}, M., \& {Kuhr}, H. 1993, \aaps, 101, 521

\bibitem[{{Stickel} {et~al.}(1991){Stickel}, {Padovani}, {Urry}, {Fried}, \&
  {Kuehr}}]{1991ApJ...374..431S}
{Stickel}, M., {Padovani}, P., {Urry}, C.~M., {Fried}, J.~W., \& {Kuehr}, H.
  1991, \apj, 374, 431, \dodoi{10.1086/170133}

\bibitem[{{Tavecchio} \& {Ghisellini}(2008)}]{2008MNRAS.386..945T}
{Tavecchio}, F., \& {Ghisellini}, G. 2008, \mnras, 386, 945,
  \dodoi{10.1111/j.1365-2966.2008.13072.x}

\bibitem[{{Thompson} {et~al.}(1990){Thompson}, {Djorgovski}, \& {de
  Carvalho}}]{1990PASP..102.1235T}
{Thompson}, D.~J., {Djorgovski}, S., \& {de Carvalho}, R. 1990, \pasp, 102,
  1235, \dodoi{10.1086/132758}

\bibitem[{{Titov} {et~al.}(2011){Titov}, {Jauncey}, {Johnston}, {Hunstead}, \&
  {Christensen}}]{2011AJ....142..165T}
{Titov}, O., {Jauncey}, D.~L., {Johnston}, H.~M., {Hunstead}, R.~W., \&
  {Christensen}, L. 2011, \aj, 142, 165, \dodoi{10.1088/0004-6256/142/5/165}

\bibitem[{{Titov} {et~al.}(2017){Titov}, {Pursimo}, {Johnston}, {Stanford},
  {Hunstead}, {Jauncey}, \& {Zenere}}]{2017AJ....153..157T}
{Titov}, O., {Pursimo}, T., {Johnston}, H.~M., {et~al.} 2017, \aj, 153, 157,
  \dodoi{10.3847/1538-3881/aa61fd}

\bibitem[{{Titov} {et~al.}(2013){Titov}, {Stanford}, {Johnston}, {Pursimo},
  {Hunstead}, {Jauncey}, {Maslennikov}, \& {Boldycheva}}]{2013AJ....146...10T}
{Titov}, O., {Stanford}, L.~M., {Johnston}, H.~M., {et~al.} 2013, \aj, 146, 10,
  \dodoi{10.1088/0004-6256/146/1/10}

\bibitem[{{Torrealba} {et~al.}(2012){Torrealba}, {Chavushyan},
  {Cruz-Gonz{\'a}lez}, {Arshakian}, {Bertone}, \&
  {Rosa-Gonz{\'a}lez}}]{2012RMxAA..48....9T}
{Torrealba}, J., {Chavushyan}, V., {Cruz-Gonz{\'a}lez}, I., {et~al.} 2012,
  \rmxaa, 48, 9.
\newblock \doarXiv{1107.3416}

\bibitem[{{Tramacere} {et~al.}(2009){Tramacere}, {Giommi}, {Perri},
  {Verrecchia}, \& {Tosti}}]{2009A&A...501..879T}
{Tramacere}, A., {Giommi}, P., {Perri}, M., {Verrecchia}, F., \& {Tosti}, G.
  2009, \aap, 501, 879, \dodoi{10.1051/0004-6361/200810865}

\bibitem[{{Urry} {et~al.}(2000){Urry}, {Scarpa}, {O'Dowd}, {Falomo}, {Pesce},
  \& {Treves}}]{2000ApJ...532..816U}
{Urry}, C.~M., {Scarpa}, R., {O'Dowd}, M., {et~al.} 2000, \apj, 532, 816,
  \dodoi{10.1086/308616}

\bibitem[{{van den Berg} {et~al.}(2019){van den Berg}, {B{\"o}ttcher},
  {Dom{\'\i}nguez}, \& {L{\'o}pez-Moya}}]{2019ApJ...874...47V}
{van den Berg}, J.~P., {B{\"o}ttcher}, M., {Dom{\'\i}nguez}, A., \&
  {L{\'o}pez-Moya}, M. 2019, \apj, 874, 47, \dodoi{10.3847/1538-4357/aafdfd}

\bibitem[{{Vandenbroucke} {et~al.}(2010){Vandenbroucke}, {Buehler}, {Ajello},
  {Bechtol}, {Bellini}, {Bolte}, {Cheung}, {Civano}, {Donato}, {Fuhrmann},
  {Funk}, {Healey}, {Hill}, {Knigge}, {Madejski}, {Romani},
  {Santander-Garc{\'\i}a}, {Shaw}, {Steeghs}, {Torres}, {Van Etten}, \&
  {Williams}}]{2010ApJ...718L.166V}
{Vandenbroucke}, J., {Buehler}, R., {Ajello}, M., {et~al.} 2010, \apjl, 718,
  L166, \dodoi{10.1088/2041-8205/718/2/L166}

\bibitem[{{Vazdekis} {et~al.}(2010){Vazdekis}, {S{\'a}nchez-Bl{\'a}zquez},
  {Falc{\'o}n-Barroso}, {Cenarro}, {Beasley}, {Cardiel}, {Gorgas}, \&
  {Peletier}}]{2010MNRAS.404.1639V}
{Vazdekis}, A., {S{\'a}nchez-Bl{\'a}zquez}, P., {Falc{\'o}n-Barroso}, J.,
  {et~al.} 2010, \mnras, 404, 1639, \dodoi{10.1111/j.1365-2966.2010.16407.x}

\bibitem[{{Vermeulen} {et~al.}(1995){Vermeulen}, {Ogle}, {Tran}, {Browne},
  {Cohen}, {Readhead}, {Taylor}, \& {Goodrich}}]{1995ApJ...452L...5V}
{Vermeulen}, R.~C., {Ogle}, P.~M., {Tran}, H.~D., {et~al.} 1995, \apjl, 452,
  L5, \dodoi{10.1086/309716}

\bibitem[{{Vermeulen} {et~al.}(1996){Vermeulen}, {Taylor}, {Readhead}, \&
  {Browne}}]{1996AJ....111.1013V}
{Vermeulen}, R.~C., {Taylor}, G.~B., {Readhead}, A.~C.~S., \& {Browne},
  I.~W.~A. 1996, \aj, 111, 1013, \dodoi{10.1086/117847}

\bibitem[{{Vestergaard} \& {Peterson}(2006)}]{2006ApJ...641..689V}
{Vestergaard}, M., \& {Peterson}, B.~M. 2006, \apj, 641, 689,
  \dodoi{10.1086/500572}

\bibitem[{{Vestergaard} \& {Wilkes}(2001)}]{2001ApJS..134....1V}
{Vestergaard}, M., \& {Wilkes}, B.~J. 2001, \apjs, 134, 1,
  \dodoi{10.1086/320357}

\bibitem[{{Wallace} {et~al.}(2002){Wallace}, {Halpern}, {Magalh{\~a}es}, \&
  {Thompson}}]{2002ApJ...569...36W}
{Wallace}, P.~M., {Halpern}, J.~P., {Magalh{\~a}es}, A.~M., \& {Thompson},
  D.~J. 2002, \apj, 569, 36, \dodoi{10.1086/339286}

\bibitem[{{White} {et~al.}(2000){White}, {Becker}, {Gregg},
  {Laurent-Muehleisen}, {Brotherton}, {Impey}, {Petry}, {Foltz}, {Chaffee},
  {Richards}, {Oegerle}, {Helfand}, {McMahon}, \&
  {Cabanela}}]{2000ApJS..126..133W}
{White}, R.~L., {Becker}, R.~H., {Gregg}, M.~D., {et~al.} 2000, \apjs, 126,
  133, \dodoi{10.1086/313300}

\bibitem[{{Wilkes} {et~al.}(1983){Wilkes}, {Wright}, {Jauncey}, \&
  {Peterson}}]{1983PASAu...5....2W}
{Wilkes}, B.~J., {Wright}, A.~E., {Jauncey}, D.~L., \& {Peterson}, B.~A. 1983,
  Proceedings of the Astronomical Society of Australia, 5, 2

\bibitem[{{Wolter} {et~al.}(1998){Wolter}, {Ruscica}, \&
  {Caccianiga}}]{1998MNRAS.299.1047W}
{Wolter}, A., {Ruscica}, C., \& {Caccianiga}, A. 1998, \mnras, 299, 1047,
  \dodoi{10.1046/j.1365-8711.1998.01834.x}

\bibitem[{{Yip} {et~al.}(2004{\natexlab{a}}){Yip}, {Connolly}, {Szalay},
  {Budav{\'a}ri}, {SubbaRao}, {Frieman}, {Nichol}, {Hopkins}, {York},
  {Okamura}, {Brinkmann}, {Csabai}, {Thakar}, {Fukugita}, \&
  {Ivezi{\'c}}}]{2004AJ....128..585Y}
{Yip}, C.~W., {Connolly}, A.~J., {Szalay}, A.~S., {et~al.} 2004{\natexlab{a}},
  \aj, 128, 585, \dodoi{10.1086/422429}

\bibitem[{{Yip} {et~al.}(2004{\natexlab{b}}){Yip}, {Connolly}, {Vanden Berk},
  {Ma}, {Frieman}, {SubbaRao}, {Szalay}, {Richards}, {Hall}, {Schneider},
  {Hopkins}, {Trump}, \& {Brinkmann}}]{2004AJ....128.2603Y}
{Yip}, C.~W., {Connolly}, A.~J., {Vanden Berk}, D.~E., {et~al.}
  2004{\natexlab{b}}, \aj, 128, 2603, \dodoi{10.1086/425626}

\end{thebibliography}

\appendix
\begin{deluxetable*}{ll|ll|ll}
\tabletypesize{\normalsize}
\tablecaption{References used in this work.\label{tab:ref}}
\tablewidth{0pt}
\tablehead{
\colhead{4FGL name} & \colhead{Reference} & \colhead{4FGL name} & \colhead{Reference}  & \colhead{4FGL name} & \colhead{Reference}}
\startdata
  J0001.5+2113  &  \citet{2020ApJS..249....3A} &  J0833.4-0458  &  \citet{2020ApJ...897..177P} &  J1417.9+2543  &  \citet{2020ApJS..249....3A}  \\
  J0003.2+2207  &  \citet{2020ApJS..249....3A} &  J0833.9+4223  &  \citet{2020ApJS..249....3A} &  J1417.9+4613  &  \citet{2020ApJS..249....3A}  \\
  J0004.3+4614  &  \citet{2003ApJ...590..109S} &  J0836.5-2026  &  \citet{1983ANA...117...60F} &  J1418.4+3543  &  \citet{2020ApJS..249....3A}  \\
  J0004.4-4737  &  \citet{2012ApJ...748...49S} &  J0836.9+5833  &  \citet{2020ApJS..249....3A} &  J1419.5+3821  &  \citet{2020ApJS..249....3A}  \\
  J0006.3-0620  &  \citet{1989ANAS...80..103S} &  J0837.3+1458  &  \citet{2020ApJS..249....3A} &  J1419.8+5423  &  \citet{2020ApJS..249....3A}  \\
\enddata
\tablecomments{(This table is available in its entirety in a machine-readable form in the online journal. A portion is shown here for guidance regarding its form and content.)}
\tablerefs{\citet{2010ANA...519A...5A,2005ARep...49..374A,2006ARep...50..255A,2020ApJS..249....3A,2016ApJ...826...76A,2011ApJ...742..127A,2016AJ....151...95A,2016AJ....151...32A,1995ANAS..110..469B,1998ANA...334..459B,1999ApJS..122...29B,1981ApJ...243...76B,1989ApJ...338..630B,2020MNRAS.494.6036B,2002ApJS..140..143B,1988ANA...192....1B,2018ApJ...854L..23B,2009ANA...495.1033B,2003ANA...412..651C,2012ApJ...759..114C,2013ATel.4777....1C,1986ChANA..10..196C,2001MNRAS.328.1039C,2019ApJS..241....5D,1997MNRAS.284...85D,1989MNRAS.238.1171D,2004ApJS..150...73E,1997ANA...321..374F,2000ApJ...542..731F,2017MNRAS.470.2814F,2015ANA...575A..13F,1983ANA...117...60F,2004ANA...428..867G,2000AJ....119.1677G,1991AJ....101..818H,1997AJ....114.1736H,2003AJ....125..572H,2008ApJS..175...97H,1997MNRAS.290..380H,2009MNRAS.399..683J,1984PASP...96..539J,2017MNRAS.467.2537K,2017ApJ...850...74K,2005ANA...440..831K,2013AJ....145..114L,2018ApJ...861..130L,2001MNRAS.323..757L,1998ApJS..118..127L,1996ApJS..107..541L,2007MNRAS.374..556L,1996MNRAS.281..425M,2016ANA...591A..21M,2019ApNSS.364....5M,2003ApJS..145..199M,2008ANA...480..715M,2015ANA...575A.124M,2012ANA...544A..36M,1978MNRAS.185..735M,1984PASAu...5..341M,2003ANA...400...95N,2012MNRAS.427..859N,2016MNRAS.460.3202O,2019MNRAS.484L.104P,2017ApJ...851..135P,2017ApJ...844..120P,2017ApJ...837..144P,2019ApJ...871..162P,2020MNRAS.497...94P,2019ApJ...881..154P,2020ApJ...897..177P,2020ApJ...902...29P,2017ApNSS.362..228P,2019ApNSS.364...85P,2020arXiv200907905P,1996ApJS..104..251P,1998AJ....115.1253P,1998ANA...333...48P,2007ANA...470..787P,2014ANA...565A..12P,2017MNRAS.468L..72P,2013ApJ...767...14P,2007ANA...464..871R,2000AJ....120.1626R,2015AJ....149..160R,2013AJ....146..163S,2005AJ....129..559S,2006ANA...457...35S,2006AJ....132....1S,2009AJ....137..337S,2012ApJ...748...49S,2013AJ....146..127S,2013ApJ...764..135S,2009arXiv0912.3621S,2003ApJ...590..109S,2004ApJ...609..564S,1991ApJ...382..433S,1993ANAS..100..395S,1994ANAS..105...67S,1993ANAS..101..521S,1988ANA...191L..16S,1989ANAS...80..103S,1993ANAS...98..393S,1993ANAS...97..483S,1990PASP..102.1235T,2011AJ....142..165T,2013AJ....146...10T,2017AJ....153..157T,2012RMxAA..48....9T,2000ApJ...532..816U,2010ApJ...718L.166V,1995ApJ...452L...5V,1996AJ....111.1013V,2002ApJ...569...36W,2000ApJS..126..133W,1983PASAu...5....2W,1998MNRAS.299.1047W,2020ApNSS.365...12D,1994MNRAS.269..998D}
}

\end{deluxetable*}

\acknowledgments
We are grateful to the journal referee for constructive criticism. VSP is thankful to A. Desai, R. de Menezes, T. Pursimo, and J. Strader for providing optical spectra of a few blazars in tabular format and H. Guo and M. Cappellari for useful discussions over the application of {\tt PyQSOFit} and {\tt pPXF} tools, respectively. VSP's work was supported by the Initiative and Networking Fund of the Helmholtz Association. A.D. acknowledges the support of the Ram{\'o}n y Cajal program from the Spanish MINECO. A.O.G. acknowledges financial support from the Spanish Ministry of Science, Innovation and Universities (MCIUN) under grant numbers RTI2018-096188-B-I00, and from the Comunidad de Madrid Tec2Space project S2018/NMT-4291.

Funding for the Sloan Digital Sky Survey IV has been provided by the Alfred P. Sloan Foundation, the U.S. Department of Energy Office of Science, and the Participating Institutions. SDSS-IV acknowledges support and resources from the Center for High-Performance Computing at the University of Utah. The SDSS web site is www.sdss.org.

SDSS-IV is managed by the Astrophysical Research Consortium for the Participating Institutions of the SDSS Collaboration including the Brazilian Participation Group, the Carnegie Institution for Science, Carnegie Mellon University, the Chilean Participation Group, the French Participation Group, Harvard-Smithsonian Center for Astrophysics, Instituto de Astrof\'isica de Canarias, The Johns Hopkins University, Kavli Institute for the Physics and Mathematics of the Universe (IPMU) / University of Tokyo, the Korean Participation Group, Lawrence Berkeley National Laboratory, Leibniz Institut f\"ur Astrophysik Potsdam (AIP),  Max-Planck-Institut f\"ur Astronomie (MPIA Heidelberg), Max-Planck-Institut f\"ur Astrophysik (MPA Garching), Max-Planck-Institut f\"ur Extraterrestrische Physik (MPE), National Astronomical Observatories of China, New Mexico State University, New York University, University of Notre Dame, Observat\'ario Nacional / MCTI, The Ohio State University, Pennsylvania State University, Shanghai Astronomical Observatory, United Kingdom Participation Group,Universidad Nacional Aut\'onoma de M\'exico, University of Arizona, University of Colorado Boulder, University of Oxford, University of Portsmouth, University of Utah, University of Virginia, University of Washington, University of Wisconsin, Vanderbilt University, and Yale University.

This research has made use of data obtained through the High Energy Astrophysics Science Archive Research Center Online Service, provided by the NASA/Goddard Space Flight Center. This research has made use of the NASA/IPAC Extragalactic Database (NED), which is operated by the Jet Propulsion Laboratory, California Institute of Technology, under contract with the National Aeronautics and Space Administration. Part of this work is based on archival data, software or online services provided by the Space Science Data Center (SSDC). This research has made use of NASA's Astrophysics Data System Bibliographic Services.

This research made use of Astropy,\footnote{http://www.astropy.org} a community-developed core Python package for Astronomy \citep{2013A&A...558A..33A,2018AJ....156..123A}.

\end{document}